\RequirePackage{tikz}
\documentclass{interact}

\usepackage{epstopdf}
\usepackage[caption=false]{subfig}

\usepackage[numbers,sort&compress]{natbib}
\bibpunct[, ]{[}{]}{,}{n}{,}{,}

\usepackage[utf8]{inputenc}
\usepackage{comment,enumitem, amsmath, mathrsfs,  dsfont, booktabs,lipsum, parskip, multirow,amsfonts, amssymb, makecell, siunitx, booktabs, tikz, fix-cm, mathtools, multicol, bigints,  float, mathtools, parskip, tabstackengine, tabulary, tabularx, adjustbox, booktabs, longtable, lscape, graphics, lipsum, xr-hyper, color, colortbl,  threeparttable}

\usepackage{comment,enumitem, amsmath, mathrsfs, amsthm,   dsfont,booktabs,lipsum,parskip,multirow,amssymb, makecell,siunitx,booktabs, tikz, fix-cm,  mathtools, multicol, bigints,  float, mathtools, parskip, tabstackengine, tabulary,tabularx, adjustbox, booktabs, longtable, lscape, graphics, lipsum, xr-hyper, MnSymbol, multirow, array, amsmath, diagbox, setspace, anyfontsize}

\usepackage[myheadings]{fullpage}
\usepackage[draft=false, citecolor=blue]{hyperref}
\hypersetup{
    colorlinks=true,
    linkcolor=blue,
    filecolor=magenta,      
    urlcolor=cyan,
    pdftitle={Overleaf Example},
    pdfpagemode=FullScreen,
    }
    
\numberwithin{equation}{section}
\newlist{enumthm}{enumerate}{1}
\setlist[enumthm]{label=(\alph*)}

\newtheorem{Definition}{Definition}[section]
\newtheorem{Lemma}{Lemma}[section]
\newtheorem{Theorem}{Theorem}[section]

\newtheorem{Remark}{Remark}[section]
\newtheorem{Corollary}{Corollary}[section]
\newtheorem{Proposition}{Proposition}[section]

\newtheorem{Example}{Example}[section]
\newtheorem{Algorithm}{Algorithm}[section]

\counterwithin{table}{section}
\counterwithin{figure}{section}

\newcounter{desccount}
\newcommand{\descitem}[1]{%
  \item[#1] \refstepcounter{desccount}\label{#1}
}
\newcommand{\descref}[1]{\hyperref[#1]{#1}}

\begin{document}


\title{Robust Nonparametric Two-Sample Tests via Mutual Information using Extended Bregman Divergence}

\author{
 Arijit Pyne \\
  Novartis AG\\
  Hyderabad, India\\
  \texttt{arijitpynestat@gmail.com} \\
}

\maketitle

\begin{abstract}
We introduce a generalized formulation of mutual information (MI) based on the extended Bregman divergence, a framework that subsumes the generalized S-Bregman (GSB) divergence family. The GSB divergence unifies two important classes of statistical distances, namely the S-divergence and the Bregman exponential divergence (BED), thereby encompassing several widely used subfamilies, including the power divergence (PD), density power divergence (DPD), and S-Hellinger distance (S-HD). In parametric inference, minimum divergence estimators are well known to balance robustness with high asymptotic efficiency relative to the maximum likelihood estimator. However, nonparametric tests based on such statistical distances have been relatively less explored. In this paper, we construct a class of consistent and robust nonparametric two-sample tests for the equality of two absolutely continuous distributions using the generalized MI. We establish the asymptotic normality of the proposed test statistics under the null and contiguous alternatives. The robustness properties of the generalized MI are rigorously studied through the influence function and the breakdown point, demonstrating that stability of the generalized MI translates into stability of the associated tests. Extensive simulation studies show that divergences beyond the PD family often yield superior robustness under contamination while retaining high asymptotic power. A data-driven scheme for selecting optimal tuning parameters is also proposed. Finally, the methodology is illustrated with applications to real data.
\end{abstract}

\begin{keywords}
Extended Bregman Divergence, Mutual Information, Nonparametric Two-Sample Test, Consistency, Robustness, Influence Function, Asymptotic Breakdown Point
\end{keywords}

\section{ Introduction }
\label{Introduction}

Since the beginning of the twentieth century, researchers have applied statistical methods to analyze scientific problems across diverse disciplines. Statistical inference, which forms the core of statistical methods, evolved along the way to its present form. Ever since, likelihood-based methods have emerged to take the center stage in every subsequent development. Early uses of the maximum likelihood estimator (MLE) trace back to the time of Gauss, Laplace, and many others. Nevertheless, its widespread importance in statistical inference began to unfold primarily due to Sir R.A. Fisher, whose pioneering works in the theory of statistical estimation built a strong foundation for all subsequent breakthroughs. The use of the likelihood function in parametric inference is preceded only by the choice of an “appropriate” mathematical model, indexed by an unknown parameter (possibly a vector) for the probability distribution underpinning the data set at hand. The parameter of the model is then estimated by maximizing the likelihood function, which is further used to test statistical hypotheses. Maximization of the likelihood function can be equivalently carried out by minimizing the likelihood disparity (LD) between the model and the data. Simultaneously with the rise of MLE, the likelihood ratio test (LRT), Wald test, and score test received wide acclaim, among many other methods for testing statistical hypotheses. Parametric models are often hard to formulate in real-life situations; in such scenarios, nonparametric procedures have come to statisticians’ rescue time and again.

The wide popularity of likelihood-based inferential methods is primarily due to their full asymptotic optimality under certain regularity conditions when an underlying parametric model truly generates the data. Such conditions are rarely met in most real-life situations. In these cases, likelihood-based methods generally become highly sensitive even to mild misspecification of the parametric models. As we obtain data from increasingly diverse sources of studies, the question of the robustness of statistical methods becomes increasingly pertinent. This quest has led to the development of robust inferential procedures, which also served as a precursor to the development of M-estimation theory (see Huber (2011) \cite{huber2011robust}; Hampel et al. (1986) \cite{hampel1986robust}). Although the MLE belongs to the class of M-estimators, so do many other well-known robust alternatives. Higher robustness often comes at the expense of asymptotic efficiency. In the minimum-distance estimation method, a statistical divergence between the data and a model is constructed, which is then minimized to choose an estimator. Minimum distance estimators form an important class of M-estimators. Subsequently, divergence deviation tests (DDTs) are constructed on the basis of minimum distance estimators. Minimum distance methods have been widely used in the statistical literature due to their high (sometimes even full) asymptotic efficiencies at the model, along with the additional advantage of exhibiting strong robustness features against model misspecifications.

Although Karl Pearson was the first person to use statistical distance by introducing the $\chi^{2}$ measure to fit a parametric model, it took a long time to be imbibed into mainstream research until the late $70$s, when Beran (1977) \cite{beran1977minimum, beran1977robust} introduced a robust and efficient estimator as a minimizer of the Hellinger distance between the data and a parametric model. Such estimators are called minimum Hellinger distance estimators. The Hellinger distance (HD) is a member of the celebrated power divergence (PD) family, developed by Cressie \& Read (1984) \cite{cressie1984multinomial}. However, the power divergence family belongs to a class of more generalized distances, namely the $\phi$-divergence of Csis\'{z}ar (1963)  \cite{csiszar1963informationstheoretische}; Ali \& Silvey (1966) \cite{ali1966general}. All members of the $\phi$-divergence family lead to fully asymptotically efficient inference at the true model, with some members having additional robustness properties. The Bregman divergence (see Bregman (1967) \cite{bregman1967relaxation}) is another important class of statistical distances, which includes the density power divergence (DPD) of Basu et al. (1998) \cite{basu1998robust}. The latter distance is arguably the most important family in the Bregman divergence class, since it does not require kernel smoothing for parameter estimation, even for continuous models. Moreover, it smoothly connects the LD (likelihood disparity, which leads to the MLE) to the squared $L_{2}$-distance (which produces very robust but somewhat inefficient estimators). The Bregman exponential divergence (BED) of Mukherjee et al. (2019) \cite{mukherjee2019b} constitutes another important family in the Bregman divergence class. Apparently, other members of the $\phi$-divergence family, except the likelihood disparity, do not belong to the class of Bregman divergences. However, Ghosh et al. (2017) \cite{ghosh2017generalized} defined the $S$-divergence family, which smoothly connects the power divergence family to the density power divergence family. Finally, the generalized S-Bregman divergence of Basak and Basu (2022) \cite{basak2022extended} unifies both the S-divergence and the BED family through an extension of what is now called the extended Bregman divergence. See Pardo (2006) \cite{pardo2018statistical} and Basu et al. (2011) \cite{basu2011statistical} for more details regarding minimum distance statistical inference.

As mentioned earlier, it is not always possible to formulate a parametric model in real-life situations. In such cases, it is better to use a nonparametric model, which requires only a minimum number of statistical assumptions. Empirical likelihood in the context of nonparametric inference gained popularity mainly through the works of Owen. See Owen (2001) \cite{owen2001empirical} for a nice exposition. Since then, it has been extended to the $\phi$-divergence family; see Morales et al. (2001) \cite{morales2001likelihood}; Balakrishnan et al. (2015, 2017) \cite{balakrishnan2015empirical, balakrishnan2017empirical}. Micheas \& Zografos (2006) \cite{micheas2006measuring} used the $\phi$-divergence to measure stochastic dependence among a set of random variables.

In the context of two-sample testing problems, $\phi$-divergence based on the empirical likelihood function has been used by Jing (1995) \cite{jing1995two}; Qin \& Zhao (2000) \cite{yong2000empirical}; Liu et al. (2008) \cite{liu2008empirical}; Wu \& Yan (2012) \cite{wu2012empirical}; Balakrishnan et al. (2017) \cite{balakrishnan2017empirical}, among many others. In most of these earlier articles, the authors considered the null hypothesis of equality in some summary statistics (viz., mean, etc.). Guha \& Chothia (2014) \cite{guha2014two} used MI (likelihood disparity) to test the equality between two completely unstructured absolutely continuous distributions based on kernel density estimates. When the underlying distributions are continuous, several well-known nonparametric tests are available, many of which are based on the empirical distribution function. Popular examples include the Kolmogorov-Smirnov (KS) test, the Wilcoxon test, the Anderson-Darling (AD) test, and the Cram\'{e}r-von Mises (CVM) test. A two-sample t-test and the Wilcoxon test are commonly used to detect differences in location. The t-test and the Wilcoxon test are less effective for distributions that are matched in location but differ in shape, skewness, or kurtosis. While the AD, CVM, and KS tests perform reasonably well in these situations, Guha and Chothia (2014) \cite{guha2014two} demonstrated that tests based on mutual information (MI) often outperform them. Guha et al. (2021) \cite{guha2021nonparametric} improved the robustness of two-sample nonparametric tests using the $\phi$-divergence, particularly the power divergence family. However, we see further that the asymptotic null distribution of the empirical mutual information based on the power divergence family requires the support of the continuous random variables to be bounded. This stringent requirement, in a way, becomes a serious impediment to the asymptotic null distribution of the power divergence family being used in more general cases where the supports of the continuous random variables are unbounded. To get around this limitation, Guha et al. (2021) \cite{guha2021nonparametric} suggested using a permutation test, which is computationally quite expensive.

In this paper, we propose a two-sample nonparametric test for equality between two completely unstructured continuous distributions based on the $\phi$-generated extended Bregman divergence. The key contributions of this paper are outlined as follows.

\begin{itemize}
\item[(a)] First, we propose a general definition of mutual information based on the extended Bregman divergence, or, in particular, the generalized S-Bregman divergence. Needless to say, this definition extends the scope of using MI for many other families beyond the power divergence family.

\item[(b)] A class of consistent nonparametric tests based on the generalized MI is proposed to test equality between two completely unstructured absolutely continuous distributions.

\item[(c)] We establish the asymptotic normality of the test statistics under the null hypothesis and its contiguous alternatives. The members of the GSB family, except for $\alpha=0$, do not require the support of continuous random variables to be bounded to use the asymptotic null distribution.

\item[(d)] The infinitesimal stability behavior of the generalized MI functional at a point is studied through the influence function. Also, the {\it level influence function} ($\mathcal{LIF}$) and {\it power influence function} ($\mathcal{PIF}$) of the test functional are calculated.

\item[(e)] Under certain regularity conditions, we compute the asymptotic breakdown point of the generalized mutual information functional.

\item[(f)] Simulation results empirically suggest that tests based on power divergence families are generally less stable under higher data contamination compared to the other members of the GSB families. However, all these tests produce higher asymptotic powers, except sometimes in the cases when $\beta < 0$. Nonetheless, a more detailed study reveals that sometimes the rate of convergence to asymptotic normality may become slower for $\beta < 0$.

\item[(g)] We propose an algorithm for choosing {\it optimal} tuning parameters given a data set. This scheme is further applied to a couple of real data examples.
\end{itemize}

Going forward, this article is organized as follows. In Section \ref{Generalized Mutual Information via Extended Bregman Divergence}, or more specifically in Subsection \ref{The Class of Extended Bregman Divergences}, we introduce the $\phi$-generated extended Bregman divergence, and one of its most popular subfamilies, namely the generalized S-Bregman (GSB) divergence. In Subsection \ref{Generalized Mutual Information and Its Properties}, we proceed to define mutual information ($\mathscr{B}$-MI) based on the $\phi$-generated extended Bregman divergence. Although this generalisation does not include all the members of the $\phi$-divergence family, it does include the most well-studied one—namely the power divergence (PD) family. Along with that, we discuss some of its important properties. In Section \ref{A Two-sample Test Based on the MI based on Extended Begman Divergence}, the two-sample testing problem is stated in terms of generalized MI. A nonparametric estimate of generalized MI using a kernel density estimate is discussed, and its asymptotic properties are studied in Subsection \ref{Nonparametric Estimation of BMI}. In Subsection \ref{Asymptotic Normality under Independence}, we establish the asymptotic normality of the test statistics (normalized version of generalized MI) under the null hypothesis. The consistency and power of the test statistics at contiguous alternatives are studied in Subsection \ref{Consistency and Power at Contiguous Alternatives}. We study the robustness behavior of the test statistics in Section \ref{Robustness Studies}. In particular, the influence function ($\mathcal{IF}$) of the test functional, as well as its level and power influence functions, are computed in Subsections \ref{Influence Function analysis} and \ref{Level and Power Influence Functions of Wald-type test}. The asymptotic breakdown point of the mutual information functional is computed in Subsection \ref{Asymptotic Breakdown Point Analysis}. Extensive simulation results are presented in Section \ref{Numeric Studies}. We discuss the choice of tuning parameter selection in Subsections \ref{Tuning Parameter Selection}, and apply it to a couple of real data examples in Subsection \ref{Real Data Examples}. Finally, concluding remarks are made in Section \ref{Concluding Remarks}. The proofs of all the theoretical results are provided in the Appendix \ref{appendix}.

\section{Generalized Mutual Information via Extended Bregman Divergence}
\label{Generalized Mutual Information via Extended Bregman Divergence}

\subsection{The Class of Extended Bregman Divergences}
\label{The Class of Extended Bregman Divergences}

The Bregman divergence (see Bregman (1967)  \cite{bregman1967relaxation}) between two points $x, y \in \mathcal{S}$, generated by a strictly convex and differentiable function $\phi : \mathcal{S} \to \mathds{R}$, is defined as the difference between $\phi(x)$ and its first-order Taylor series approximation at $y$ as
\begin{align}
\label{bregman divergence distance between points}
D_{\phi}(x, y)
=\phi(x)-\phi(y)-\langle \phi'(y), x-y \rangle,
\end{align}
where $\phi'(y)$ is the gradient of $\phi$ evaluated at $y$, and $\mathcal{S}$ is a convex subset of $\mathds{R}^{p}$. The convexity of $\phi$ ensures that $D_{\phi}$ is non-negative and $D_{\phi}(x,y)=0$ if and only if $x=y$. Bregman divergences are similar to a metric, but they satisfy neither the triangle inequality nor symmetry in general. However, they satisfy a generalization of the Pythagorean theorem.

When points are interpreted as probability density functions $g, f$ (defined on a common support $\chi$ with respect to a common $\sigma$-finite measure $\mu$), the resulting Bregman divergence becomes
\begin{align}
\label{bregman divergence}
D_{\phi}(g, f)=\int_{\chi}
\Big\{\phi(g)- \phi(f)
-(g-f) \phi'(f)
\Big\} d\mu.
\end{align}
Later, the symbols $\chi, \mu$ will be made implicit when they are clear from the context. Likelihood disparity (LD) and squared $L_{2}$-distance are two well-known divergences belonging to this class, generated respectively by $\phi(t)=t \log t$ and $\phi(t)=t^{2}$.

An important family in this class is the density power divergence (DPD) of Basu et al. (1998) \cite{basu1998robust}. It is generated by $\phi(t)=\frac{t^{1+\alpha}-t}{\alpha}$ with $\alpha >0$. This divergence takes the following form:
\begin{align}
\label{density power divergence}
d_{\alpha}(g, f)=\int \Big\{f^{1+\alpha}-\Big(1+\frac{1}{\alpha}\Big)f^{\alpha}g+\frac{1}{\alpha}g^{1+\alpha} \Big\}
\mbox{ for } 0 < \alpha \le 1.
\end{align}
The DPD at $\alpha=0$ is defined as $d_{0}(g, f)=\underset{\alpha \downarrow 0+}{\lim}d_{\alpha}(g,f)=\int g \ln \frac{g}{f}$. When $\alpha=1$, it becomes the squared $L_{2}$-distance: $d_{1}(g, f)=\int (g-f)^{2}$. Thus, DPD smoothly connects two well-known statistical distances. In the context of parametric estimation, a parametric model $\{f_{\theta}; \theta \in \Theta\}$ and empirical density $g_{n}$ are substituted respectively for $f$ and $g$. The minimum density power divergence estimator (MDPDE) is known to be very efficient for small values of $\alpha$, and its robustness increases as the tuning parameter $\alpha$ becomes larger.

Mukherjee et al. (2019) \cite{mukherjee2019b} define another important subfamily in the Bregman divergence class, known as the Bregman exponential divergence (also called the B-exponential divergence or BED), which is given by
\begin{align}
\label{B-exponential divergence}
BED_{\beta}(g, f)=\frac{2}{\beta}
\int \Big\{e^{\beta f}\Big(f-\frac{1}{\beta}\Big)-e^{\beta f}g
+\frac{e^{\beta g}}{\beta} \Big\}
\mbox{ for } \beta \ne 0.
\end{align}
This family is generated by $\phi(t)=\frac{2(e^{\beta t}-\beta t -1)}{\beta^{2}}$ with $\beta \ne 0$. At $\beta=0$, BED is defined as $BED_{0}(g, f)=\underset{\beta \to 0}{\lim}BED_{\beta}(g, f)$, which is the squared $L_{2}$-distance.

On the other hand, $\phi$-divergence is another important class of distances with the following form:
\begin{align}
\label{phi divergence}
d_{\phi}(g, f)=\int \phi \Big( \frac{g}{f}\Big)f,
\end{align}
where the convex function $\phi$ is additionally required to satisfy the following conditions: $\phi(1)=\phi'(1)=0$, $0\phi(0/0)=0$, and $0\phi(f/0)=f \underset{u \to +\infty}{\lim} \frac{\phi(u)}{u}$. This class includes the well-known power divergence (PD) family, which is defined as
\begin{align}
\label{power divergence}
PD_{\lambda}(g, f)=\frac{1}{\lambda(1+\lambda)}
\int g \Big\{ \Big(\frac{g}{f} \Big)^{\lambda}-1 \Big\}
\mbox{ for } \lambda \ne -1, 0.
\end{align}
The divergences at $\lambda=-1, 0$ are defined as the continuous limits of $PD_{\lambda}(g, f)$ as $\lambda \to -1, 0$. LD is the only common member belonging to both the classes of Bregman and $\phi$-divergences.

Ghosh et al. (2017) \cite{ghosh2017generalized} define the S-divergence, which integrates the power divergence family with the density power divergence in a unified framework. The S-divergence family, involving two tuning parameters $\alpha \in [0, 1]$ and $\lambda \in \mathds{R}$, is defined as
\begin{align}
\label{S-divergence}
SD_{\alpha, \lambda}(g, f)
=\int \Big\{\frac{1}{B}(g^{A+B}-f^{A+B})-(g^{A}-f^{A})\frac{A+B}{AB}f^{B} \Big\}
\mbox{ for } AB \ne 0,
\end{align}
where $A=1+\lambda(1-\alpha)$, $B=\alpha-\lambda(1-\alpha)$. The S-divergence may be generated from (\ref{bregman divergence}) by $\phi(t)= \frac{t^{1+\frac{B}{A}}}{B}$ with $AB \ne 0$. The divergences at $A=0$ and/or $B=0$ are defined as the continuous limits as $A \to 0$ and $B \to 0$. When $\lambda=0$, the S-divergence $SD_{\alpha, \lambda}(g, f)$ in (\ref{S-divergence}) becomes the density power divergence $d_{\alpha}(g, f)$ as in (\ref{density power divergence}), and at $\alpha=0$ it becomes the power divergence $PD_{\lambda}$ in (\ref{power divergence}). An important subfamily of the S-divergence family is the S-Hellinger distance, which is defined as
\begin{align}
\label{S-hellineger divergence}
SD_{\alpha, \lambda=-0.5}(g, f)=\frac{2}{1+\alpha}
\int \Big(g^{\frac{1+\alpha}{2}} - f^{\frac{1+\alpha}{2}} \Big)^{2}
\mbox{ for } \alpha \in [0, 1].
\end{align}
Although the S-divergence family includes some members of the Bregman divergence, it does not include all of them. However, this was achieved by Basak and Basu (2022) \cite{basak2022extended} through the following extension of the Bregman divergence:
\begin{align}
\label{extended bregman divergence}
D^{(k)}_{\phi}(g, f)=\int
\Big\{\phi(g^{k})- \phi(f^{k})
-(g^{k}-f^{k}) \phi'(f^{k})
\Big\}
\mbox{ for some index } k >0.
\end{align}
This is called the $\phi$-generated extended Bregman divergence with a positive index $k$. In our present discussion, we assume that the convex $\phi$ is four times continuously differentiable. Basak and Basu (2022) \cite{basak2022extended} further define the generalized S-Bregman divergence (GSB) by choosing the following convex function:
\begin{align}
\label{convex function of GSB}
\phi(t)=e^{\beta t}+\frac{t^{+\frac{B}{A}}}{B}
\mbox{ for } AB \ne 0,
\end{align}
where $A, B$ are defined as in (\ref{S-divergence}) with $\alpha \ge -1$ and $\lambda, \beta \in \mathds{R}$. The function $\phi$ in (\ref{convex function of GSB}), together with the exponent $k=A>0$, generates the following form of the extended Bregman divergence:
\begin{align}
\label{GSB}
D^{*}(g, f)
=\int
\Big\{e^{\beta f^{A}}(\beta f^{A}- \beta g^{A}-1)
+ e^{\beta g^{A}}+\frac{1}{B}(g^{A+B}-f^{A+B})
-(g^{A}-f^{A}) \frac{A+B}{AB}f^{B}
\Big\}
\end{align}
for $A, B \ne 0$. The GSB divergences at $A=0$ and/or $B=0$ are defined as the continuous limits of the expression in (\ref{GSB}) as $A, B \to 0$. Notice that the GSB divergence becomes the S-divergence for $\beta=0$. A scaled BED family $BED^{s}_{\beta}(g,f)=\frac{\beta^{2}}{2} \times BED_{\beta}(g,f)$ for $\beta \ne 0$ is generated for $\alpha=-1, \lambda=0$. The GSB divergence not only integrates all members of the Bregman divergence into the power divergence family, but it also produces many other stable and highly efficient estimators in the context of parametric inference, outside both the BED and S-divergence families.

\subsection{Generalized Mutual Information and Its Properties}
\label{Generalized Mutual Information and Its Properties}

\begin{Definition}($\mathscr{B}$-MI)
The mutual information based on a $\phi$-generated extended Bregman divergence for a set of random variables $\{X_{1}, \ldots, X_{m}\}$ may be defined as a map $I_{D^{(k)}_{\phi}}: \otimes_{i=1}^{m}\chi_{i} \longrightarrow [0, +\infty]$ such that 
\begin{align}
    \label{extended bregman MI}
    I_{D^{(k)}_{\phi}}(X_{1}, X_{2}, \ldots, X_{m})
    =D^{(k)}_{\phi}(f_{X_{1}, X_{2}, \ldots , X_{m}}, f_{X_{1}} \cdots  f_{X_{m}} ),
\end{align}
where $f$ denotes the density of the indicated random variable(s), and $\chi_{i}$ being the sample space of $X_{i},\ i=1,2,\ldots,m$.     
\end{Definition}
Note that the expression in (\ref{extended bregman MI}) for any set of random variables is uniquely defined up to a set of measure 0. $I_{D^{(k)}_{\phi}}$ may be interpreted as a measure of stochastic association among a set of random variables $\{X_{1}, \ldots, X_{m}\}$. Substituting the $\phi$-function (\ref{convex function of GSB}) in (\ref{extended bregman MI}), we get the explicit form of the MI based on the GSB divergence. Additionally, when $\alpha=\beta=0$, $I_{D^{(k)}_{\phi}}$ becomes $I_{\lambda}$ of Guha et al. (2021) \cite{guha2021nonparametric}. Similarly, the mutual information based on the DPD ($I_{\alpha}$) may be obtained for $\alpha \ge 0$ with $\lambda=\beta=0$. Expressions of MI for several other divergences may be similarly obtained for different choices of the $\phi$-function.  

The power divergence is the only common member belonging to both the $\phi$-divergence and $\phi$-generated extended Bregman divergence families. These common members are denoted by $C_{B\phi}=\Big\{\phi \mbox{ for some } k>0: d_{\phi}(g,f)=D^{(k)}_{\phi}(g,f) \forall f, g  \Big\}$. Micheas \& Zografos (2006) \cite{micheas2006measuring} explore measures of dependence for multivariate data using the $\phi$-divergence between the joint density of random variables and the product of their marginals.   

Similar properties hold for the mutual information based on the $\phi$-generated extended Bregman divergence. First, we define $\phi_{0}=\phi(0)$, $\phi_{2}=\phi_{0}+\underset{u \to +\infty}{\lim} \frac{\phi(u)}{u}$ and $c_{L}=\int_{f > 0} \frac{ k^{2}f^{2k}\phi''(f^{k})}{\phi''(1)}$ where $f=f_{X_{1}}f_{X_{2}}, \ldots, f_{X_{m}}$. Often, $\phi$-divergences are standardized such that $\phi''(1)=1$. Some mathematical properties of the generalized mutual information $I_{D^{(k)}_{\phi}}$ are listed in the following proposition.  

\begin{Proposition}
	\label{proposition: mathematical properties of MI}
	Mutual information based on the $\phi$-generated extended Bregman divergence has the following properties. 
	\begin{description} 
		\descitem{(P1)} $I_{D^{(k)}_{\phi}}(X_{1}, \ldots, X_{m})=0$ if and only if $f_{X_{1}, X_{2}, \ldots , X_{m}}=f_{X_{1}} f_{X_{2}} \cdots f_{X_{m}}$ a.s..
		
		\descitem{(P2)} $I_{D^{(k)}_{\phi}}(X_{1}, \ldots, X_{m})=I_{D^{(k)}_{\phi}}(X_{i_{1}}, X_{i_{2}}, \ldots, X_{i_{m}})$ for any permutation $(X_{i_{1}}, X_{i_{2}}, \ldots, X_{i_{m}})$ of $(X_{1}, X_{2}, \ldots, X_{m})$.  
		
		\descitem{(P3)} $I_{D^{(k)}_{\phi}}(T_{1}(X_{1}), \ldots , T_{m}(X_{m}))=I_{D^{(k)}_{\phi}}(X_{1}, \ldots , X_{m})$ for any one-one onto transformation $(X_{1}, \ldots, X_{m}) \mapsto (T_{1}(X_{1}), T_{2}(X_{2}), \ldots , T_{m}(X_{m}))$. Consequently, $I_{D^{(k)}_{\phi}}$ is invariant under strictly increasing transformations. 
		
		\descitem{(P4)} Let $\phi_{2}=\infty$ and the joint distribution is singular to the product of its marginals, then $I_{D^{(k)}_{\phi}}(X_{1}, \ldots, X_{m})=\infty$. 
		
		\descitem{(P5)} Suppose $\phi$ is strictly convex, twice continuously differentiable such that $\phi(1)=\phi'(1)=0$ and $\phi_{2}, c_{L}< \infty$. Then 
		$I_{D^{(k)}_{\phi}}(X_{1}, \ldots, X_{m}) =c_{L}\phi_{2} $ if and only if the joint distribution is singular to the product of its marginals for all $\phi \notin C_{B\phi}$. 
		
	\end{description}
\end{Proposition}

\descref{(P1)} states that the generalized MI is zero if and only if the random variables are independent. In \descref{(P2)} we show that the generalized MI remains unchanged for any permutation of its arguments. We show in \descref{(P3)} that the generalized mutual information is invariant and cannot be increased by any one-one and onto transformation of $(X_{1}, \ldots, X_{m})$. In \descref{(P4)} we investigate an upper bound of the generalized mutual information when the joint distribution is singular to the product of its marginals. There exists an almost sure relationship among $(X_{1}, \ldots, X_{m})$ when $I_{D^{(k)}_{\phi}}$ attains its upper bound. Micheas \& Zografos (2006) \cite*{micheas2006measuring} show that mutual information based on the $\phi$-divergence satisfies many other important properties which also carry over into the power divergence family, i.e., $\phi \in C_{B\phi}$. In \descref{(P5)} we prove a converse result of \descref{(P4)} under certain conditions.  

Assume that $X$ is a binary $0-1$ variable and $Y$ is a continuous random variable. Then the joint and conditional densities at a point $(x,y) \in \{0, 1\} \times \mathds{R}$ are defined as
\begin{align}
\label{joint and conditional density}
    f_{x, y}dy
    &=\mathds{P}\Big\{X=x, y< Y < y+dy \Big\}, \\
    f_{y|x}dy
    &=\mathds{P} \Big\{ y < Y < y+dy | X=x \Big\}.
\end{align}
The marginal densities of $X$ and $Y$ are respectively denoted by $f_{x}=\mathds{P}[X=x]$ and $f_{y}=\sum_{x=0}^{1}f_{x,y}$. For notational convenience, we denote $f_{x_{0}}=\mathds{P}[X=0]$ and $f_{x_{1}}=\mathds{P}[X=1]$. 

In this hybrid set-up, the $\mathscr{B}$-MI becomes 
\begin{gather}
    I_{D^{*}}(X, Y)=D^{*}(f_{X,Y},f_{X} f_{Y} ) 
    \nonumber \\
    =\sum_{x=0}^{1}\int_{f_{y}>0}
    \Big\{e^{\beta (f_{x}f_{y})^{A}}(\beta (f_{x}f_{y})^{A}- \beta f_{x,y}^{A}-1) 
    + e^{\beta (f_{x,y})^{A}} 
    +\frac{1}{B}(f_{x,y}^{A+B}-(f_{x}f_{y})^{A+B})
    \nonumber \\
    \label{GSB based MI}
    -(f_{x,y}^{A}-(f_{x}f_{y})^{A}) \frac{A+B}{AB}(f_{x}f_{y})^{B}
    \Big\}dy
    \mbox{ for } AB \ne 0.
\end{gather}
If $A=0$ but $B \ne 0$, the expression in (\ref{GSB based MI}) is defined as 
\begin{align}
\label{case A=0}
    \lim_{A \downarrow 0+} I_{D^{*}}(X, Y)
    =\sum_{x=0}^{1}\int_{f_{y}>0}
    \Bigg\{
    (f_{x}f_{y})^{1+\alpha} \ln \Bigg(\frac{f_{x}f_{y}}{f_{x,y}}\Bigg)
    - \frac{(f_{x}f_{y})^{1+\alpha}-f^{1+\alpha}_{x,y}}{1+\alpha}\Bigg\}dy,
\end{align}
and similarly, it is defined for $A \ne 0, B=0$ as 
\begin{gather}
    \lim_{B \to 0} I_{D^{*}}(X,Y)
    =\sum_{x=0}^{1}\int_{f_{y}>0}
    \Bigg\{e^{\beta (f_{x}f_{y})^{1+\alpha}}
    \Big(\beta (f_{x}f_{y})^{1+\alpha}-\beta f_{x,y}^{1+\alpha}
    -1 \Big)+e^{\beta (f_{x,y})^{1+\alpha}} \nonumber \\
    \label{case B=0}
    +f^{1+\alpha}_{x,y}\ln \Bigg( \frac{f_{x,y}}{f_{x}f_{y}}\Bigg)
    -\frac{f^{1+\alpha}_{x,y}-(f_{x}f_{y})^{1+\alpha}}{1+\alpha}
    \Bigg\} dy
\end{gather}
for $\alpha > -1$. In (\ref{case A=0}) and (\ref{case B=0}), the case $\alpha=-1$ is similarly defined as $\alpha \to -1$. This turns out to be 0 in both cases.

\section{A Two-sample Test based on \texorpdfstring{$\mathscr{B}-MI$}{}}

\label{A Two-sample Test Based on the MI based on Extended Begman Divergence}

In this section, we propose a class of tests based on $\phi$-generated extended Bregman divergence for unstructured comparison of two independent random samples $Y_{0}=(Y_{01}, \ldots, Y_{0n_{0}})$ and $Y_{1}=(Y_{11}, Y_{12}, \ldots, Y_{1n_{1}})$. We assume that $Y_{0}$ and $Y_{1}$ are independently drawn from absolutely continuous distribution functions $F_{0}$ and $F_{1}$, respectively. The hypothesis of unstructured comparison between $F_{0}$ and $F_{1}$ is stated as
\begin{align}
\label{unstructred hypothesis}
\mathds{H}_{0}: F_{0}=F_{1}
\mbox{ against }
\mathds{H}_{1}: F_{0} \ne F_{1}.
\end{align}
Since $F_{0}$ and $F_{1}$ are absolutely continuous, their probability density functions $f_{0}$ and $f_{1}$ exist. Following Guha \& Chotia (2014, 2021) \cite{guha2014two, guha2021nonparametric}, we combine these two samples $Y_{0}$ and $Y_{1}$ as $Y:=(Y_{0}, Y_{1} )$. Further, define a 0–1 binary vector $X=(X_{01}, \ldots, X_{0n_{0}}, X_{11}, \ldots, X_{1n_{1}})$. The components of $X$ are defined as
\begin{align}
\label{X vector}
X_{ij}=\begin{cases}
0 &\mbox{ if } Y_{ij} \in \{Y_{01}, \ldots , Y_{0n_{0}}\}, \\
1 &\mbox{ if } Y_{ij} \in \{Y_{11}, \ldots , Y_{1n_{1}}\}
\end{cases}
\end{align}
for $j=1, \ldots, n_{i}$ and $i=0,1$. Observe that $X$ is a vector of $n_{0}$ {\it zeros} followed by $n_{1}$ {\it ones}. However, the following test will not depend on the particular values of the components of $X$ that we assign to these two groups. When $\mathds{H}_{0}$ is true, all the components of the combined vector ${Y}=({Y}_{0}, {Y}_{1})$ come from the same distribution, and hence, they should not depend on any permutation of components of $X$. Thus, the null hypothesis $\mathds{H}_{0}$ is equivalent to stating that $X$ and $Y$ are independent. In terms of the mutual information $I_{D^{(k)}_{\phi}}(X,Y)$ defined in (\ref{extended bregman MI}), the unstructured testing problem in (\ref{unstructred hypothesis}) may be restated as
\begin{align}
\label{unstructred hypothesis; reformulated}
&\mathds{H}: f_{x,y}=f_{x}f_{y}
\mbox{ a.s.} \iff
I_{D^{(k)}_{\phi}}(X, Y)=0 ,
\\
\mbox{ against }
&\mathds{K}: f_{x,y} \ne f_{x}f_{y}
\mbox{ for at least one } (x,y)
\iff I_{D^{(k)}_{\phi}}(X, Y) > 0.
\end{align}
The expression of $I_{D^{(k)}_{\phi}}(X, Y)$ in the hybrid set-up is already given in (\ref{GSB based MI}). To carry out this testing problem, we need to estimate $I_{D^{(k)}_{\phi}}(X, Y)$ and also its asymptotic distribution under independence. Given a data set, we reject $\mathds{H}$ if the estimated MI exceeds the critical value, and accept otherwise. For simplicity, we restrict our discussion to the two-sample problem. A generalization to more than two samples can be achieved with higher-order kernels. For brevity, we shall interchangeably use the notation $I_{D^{(k)}_{\phi}}$ for $I_{D^{(k)}_{\phi}}(X, Y)$. In the next subsection, we provide a plug-in estimate for $I_{D^{(k)}_{\phi}}(X, Y)$ based on kernel density estimates.

\subsection{Nonparametric Estimation of \texorpdfstring{$\mathscr{B}$-MI}{}}
\label{Nonparametric Estimation of BMI}

Suppose $Y=(Y_{0}, Y_{1})$ is a combined sample, and let $X$ be its $0$–$1$ labeling as before. For simplicity, we enumerate the observations of the combined sample as $Y=(Y_{1}, \ldots, Y_{n_{0}}, Y_{n_{0}+1}, \ldots, Y_{n})$ and similarly for $X$ using only one subscript, where $n=n_{0}+n_{1}$. The generalized mutual information $I_{D^{(k)}_{\phi}}$ may be estimated using the density estimates at a point $(x,y) \in \{0,1\} \times \mathds{R}$ as
\begin{gather}
\widehat{f}_{X}(x)=
\frac{1}{n}\sum_{i=1}^{n}\mathds{1}\{X_{i}=x\}
,  \widehat{f}_{Y}(y)=\frac{1}{nh_{n}}\sum_{i=1}^{n}
K\Bigg(\frac{Y_{i}-y}{h_{n}}\Bigg), \\
\widehat{f}_{X,Y}(x,y)=\frac{1}{nh_{n}}\sum_{i=1}^{n}
K\Bigg(\frac{Y_{i}-y}{h_{n}} \Bigg)
\mathds{1}\{X_{i}=x\},
\end{gather}
where $K$ is a {\it suitable} kernel function and ${h_{n}}$ is a bandwidth sequence. The estimated generalized MI is denoted by $\widehat{I}_{D^{(k)}_{\phi}}$. 

Following Fernandes and N\'eri (2009) \cite{fernandes2009nonparametric}, we prove a similar asymptotic representation of $\widehat{I}_{D^{(k)}_{\phi}}$ up to an error term of appropriate order. First, we make the following assumptions.

\begin{description}

\descitem{(A1)} The convex function $\phi(t)$ defined in (\ref{extended bregman divergence}) is four times differentiable. These derivatives are bounded by integrable functions uniformly for all $t$.     

\descitem{(A2)} The probability density functions $f_{x,y}, f_{y}$ are continuously twice differentiable with respect to $y$ for $x=0,1$. The derivatives are assumed to be bounded. 

\descitem{(A3)} 
\begin{description}
    \descitem{(i)} The kernel $K$ is symmetric about {\it zero} and bounded, i.e.,  
\begin{align}
K(u)=K(-u) \mbox{ for all } u \in \mathds{R},
\mbox{ and }
||K||:=\sup_{u \in \mathds{R}}|K(u)| < \infty.
\end{align}

 \descitem{(ii)} The kernel $K$ has bounded support.
\end{description}

\descitem{(A4)} The bandwidth sequence $\{h_{n}\}$ satisfies the following conditions:

\begin{align}
h_{n} \longrightarrow 0,
n h^{2}_{n} \longrightarrow +\infty,
\mbox{ and }
n h^{4}_{n} \longrightarrow 0
\mbox{ as } n \longrightarrow +\infty.
\end{align}

\end{description}
Assumption \descref{(A1)} allows a second-order Taylor series expansion of $\widehat{I}_{D^{(k)}_{\phi}}$ up to a remainder term.
Assumption \descref{(A2)} requires that the joint
and marginal densities of $Y$ are smooth enough to admit functional expansions. Assumptions \descref{(A1)} and \descref{(A2)} together imply that the remainder term in the expansion of $\widehat{I}_{D^{(k)}_{\phi}}$ is bounded in probability. Assumption \descref{(A3)} imposes conditions on kernels to reduce the bias in kernel density estimation. Assumption \descref{(A4)} is a set of technical conditions on the bandwidth sequence that are required to establish the asymptotic normality of $\widehat{I}_{D^{(k)}_{\phi}}$. Assumptions \descref{(A3)}\descref{(i)} and \descref{(A4)} together ensure that the convergences $\widehat{f}_{x,y} \overset{\mathds{P}}{\longrightarrow} f_{x,y}$ and $\widehat{f}_{y} \overset{\mathds{P}}{\longrightarrow} f_{y}$ are uniform in $x,y$ with an appropriate order. Assumption \descref{(A3)}\descref{(ii)} restricts the use of kernels with unbounded support. As kernel density estimates are generally robust with respect to the choice of kernels, and the class of kernels with bounded support is rich enough, this may not pose a serious problem in practical applications. However, Assumption \descref{(A3)}\descref{(ii)} may be alleviated with conditions on the finiteness of complicated integrals involving the kernels, which we defer to further research. The following asymptotic expansion in the lemma derives the asymptotic distribution under the null hypothesis.

\begin{Lemma}
\label{Lemma: Asymptotic representation of MI under independence}
Suppose the Assumptions \descref{(A1)}–\descref{(A3)}\descref{(i)} and \descref{(A4)} hold. Then under the null hypothesis $\mathds{H}$, $\widehat{I}_{D^{(k)}_{\phi}}$ admits the following expansion:
\begin{align}
\widehat{I}_{D^{(k)}_{\phi}}
&=\frac{1}{2}\int_{f_{y}>0} \sum_{x=0}^{1}
\Bigg[k^{2}(f_{x}f_{y})^{2k}\phi''(f^{k}_{x}f^{k}_{y})
\Bigg(\frac{h_{x}}{f_{x}}+\frac{h_{y}}{f_{y}}-\frac{h_{x,y}}{f_{x}f_{y}}\Bigg)^{2}\Bigg]dy +o_{\mathds{P}}\Bigg(\frac{1}{nh^{1/2}_{n}}\Bigg),
\end{align}
where $h_{x}=\widehat{f}_{x}-f_{x}$, $h_{y}=\widehat{f}_{y}-f_{y}$, $h_{x,y}=\widehat{f}_{x,y}-f_{x,y}$ for all $x,y$.
\end{Lemma}

The terms $h_{x}, h_{y}, h_{x,y}$ quantify biases in the estimates. It is clear from Lemma \ref{Lemma: Asymptotic representation of MI under independence} that the limiting distribution of $\widehat{I}_{D^{(k)}_{\phi}}$ (after proper normalization) is driven by the first non-degenerate derivative of $\widehat{I}_{D^{(k)}_{\phi}}$. As it turns out, the first two terms in the expansion are singular, and the asymptotic distribution is determined only by its second derivative.

This result is the key to establishing the asymptotic normality of the test statistic under independence. The requirement of fourth-order differentiability of $\phi$ is rather strong; however, the members of the generalized S-Bregman divergence may satisfy this condition. In that case, the derivatives of $\phi$ consist of an exponential term $e^{\beta (f_{x}f_{y})^{A}}$ added to a polynomial function. In most situations, the exponential term remains bounded for any finite $\beta$.

\begin{Remark}
For GSB divergence, Lemma \ref{Lemma: Asymptotic representation of MI under independence} gives
\begin{gather}
2\widehat{I}_{D^{*}}(X,Y) \nonumber \\
= \int_{f_{y}>0} \sum_{x=0}^{1}
\Bigg[\Bigg\{(f_{x}f_{y})^{2A}
(A\beta)^{2} e^{\beta (f_{x}f_{y})^{A}}
+\Big(1+\alpha\Big)(f_{x}f_{y})^{1+\alpha} \Bigg\}
\Bigg(\frac{h_{x}}{f_{x}}+\frac{h_{y}}{f_{y}}-\frac{h_{x,y}}{f_{x}f_{y}}\Bigg)^{2}\Bigg] dy+o_{\mathds{P}}\Bigg(\frac{1}{nh^{1/2}_{n}}\Bigg).
\end{gather}
\end{Remark}
The important point to note here is that when $\beta=0$, the asymptotic expansion of $\widehat{I}_{D^{*}}$ depends only on $\alpha \ge 0$. Additionally, with $\alpha=0$, it gives the expression for the power divergence family. Later, we shall see that the asymptotic mean and variance of the power divergence family do not depend on its tuning parameter $\lambda \in \mathds{R}$.

\begin{Lemma}
\label{Lemma: consistency of MI}
Suppose the Assumptions \descref{(A1)} – \descref{(A4)} hold. Then $\widehat{I}_{D^{(k)}_{\phi}} \overset{\mathds{P}}{\longrightarrow} I_{D^{(k)}_{\phi}}$ at the true joint density $f_{X,Y}$.
\end{Lemma}

\begin{Remark}
Parzen (1962) \cite{parzen1962estimation} shows that when $nh_{n} \rightarrow +\infty$, the kernel density estimates are weakly consistent. See Wied \& Wei{\ss}bach (2012) \cite{wied2012consistency} for more details about stronger versions of consistency results.
\end{Remark}

\subsection{Asymptotic Normality of \texorpdfstring{$\widehat{I}_{D^{(k)}_{\phi}}$}{} Under Independence}
\label{Asymptotic Normality under Independence}

In this subsection, we shall formally state the asymptotic distribution of $\widehat{I}_{D^{(k)}_{\phi}}$ under the null hypothesis $\mathds{H}$. To do that, we define  
\begin{align}
	\label{ex-breg mean}
	\mu_{\phi}
	&=\frac{1}{2}\int_{u} K^{2}(u)du
	\int_{f_{y}>0}\Bigg[ \sum_{x=0}^{1}k^{2}(f_{x}f_{y})^{2k-1} \phi''(f^{k}_{x}f^{k}_{y})(1-f_{x})\Bigg]dy,
	\\
	\label{ex-breg var}
	\sigma^{2}_{\phi}
	&=\frac{1}{2}\int_{u} \Big( \int_{z} K(z)K(z+u)dz\Big)^{2} du
	\int_{f_{y}>0} \Bigg[\sum_{x=0}^{1} k^{2}(f_{x}f_{y})^{2k-1}
	\phi''\Big(f^{k}_{x}f^{k}_{y}\Big)(1-f_{x}) \Bigg]^{2} dy. 
\end{align}

In connection with that, we make the following assumption. 
\begin{description}
	\descitem{(A5)} Both $\mu_{\phi}$ and $\sigma_{\phi}$ are finite. Moreover, $\sigma_{\phi}$ must be positive.  
\end{description}

Later, we shall see that $\frac{\mu_{\phi}}{nh_{n}}$ and $\frac{\sigma_{\phi}}{nh^{1/2}_{n}}$ will be used, respectively, as the centring and scaling sequences in normalizing $\widehat{I}_{D^{(k)}_{\phi}}$ to derive its asymptotic distribution under independence. 
To see that Assumption \descref{(A5)} is not superfluous, consider the case when  
\begin{align}
	\label{degenerate mean and var}
	\sum_{x=0}^{1}k^{2}(f_{x}f_{y})^{2k-1}\phi''\Big(f^{k}_{x}f^{k}_{y}\Big)(1-f_{x}) 
\end{align}
remains constant as a function of $y$. In addition to that, let us consider that $Y$ has an unbounded support. In this case, Assumptions \descref{(A1)} and \descref{(A3)} alone would not imply that both $\mu_{\phi}$ and $\sigma_{\phi}$ are finite. So Assumption \descref{(A5)} needs to be separately assumed. Note that (\ref{degenerate mean and var}) is satisfied for the power divergence family; therefore, it is required that both $\mu_{\phi}$ and $\sigma^{2}_{\phi}$ be finite to hold the asymptotic normality result under the null hypothesis. In this case, Assumption \descref{(A5)} is equivalent to assuming that the continuous random variables have bounded support, thus restricting the scope of this result in practical applications. However, we can still use that result for the generalized S-Bregman divergence family even with unbounded support when $\alpha \ne 0$. Now we start with the following lemma that essentially verifies the conditions of Hall and Heyde (2014) \cite*{hall2014martingale} (Lemma $3.1$, pp.57), and will be further used in the proof of CLT in this hybrid setup. Let us define $T_{i}=\frac{2}{nh^{3/2}_{n}}\underset{j < i}{\sum} V_{ij}$ where  
\begin{align}
	V_{ij}=\sum_{x=0}^{1}
	\int_{f_{y}>0}
	k^{2}(f_{x}f_{y})^{2k-2}\phi''(f^{k}_{x}f^{k}_{y})
	Z_{ix}Z_{jx}
	K_{h_{n}i}(y)K_{h_{n}j}(y)dy=V_{ji}.
\end{align}
Let $Z_{n}:=\prod_{j=1}^{n}(1+itT_{j})$ be defined on a common probability space $(\Omega, \mathcal{F}, \mathds{P})$ for fixed $t \in \mathds{R}$.

\begin{Lemma}
	\label{conditions of hall and heyde}
	Denote the sum of random variables as $S_{n}=\sum_{i=1}^{n}T_{i}$, and $\mathcal{F}_{m}:=\sigma(T_{1}, \ldots, T_{m})$ be an increasing sequence of $\sigma$-fields generated by $\{(T_{1}, \ldots, T_{m}): m \le n\}$. Under the null hypothesis $\mathds{H}$ and the Assumptions \descref{(A1)}-\descref{(A5)}, the following statements are true:  
	\begin{description}
		\descitem{(a)} $\sum_{i=1}^{n} T^{2}_{i} \overset{\mathds{P}}{\longrightarrow} 4 \sigma^{2}_{\phi}$, 
		
		\descitem{(b)} $\underset{1 \le i \le n}{\max}|T_{i}| \overset{\mathds{P}}{\longrightarrow} 0$,
		
		\descitem{(c)} $Z_{n} \longrightarrow 1$ ({\it weakly in} $L_{1}$), i.e., $\mathds{E}\Big[Z_{n}\mathds{1}(E)\Big] \longrightarrow \mathds{P}(E)$  for all $E \in \mathcal{F}$ as $n \to \infty$.     
	\end{description} 
\end{Lemma}

Now, we are in a position to state the asymptotic normality related to the test statistics. 

\begin{Theorem}
\label{Theorem: Asymptotic null distribution}
Suppose the Assumptions \descref{(A1)}-\descref{(A5)} are true. Then under the null hypothesis $\mathds{H}$, it holds that 
\begin{align}
    nh^{1/2}_{n}\Big(\widehat{I}_{D^{(k)}_{\phi}}-
    \frac{\mu_{\phi}}{nh_{n}}\Big) \overset{\mathcal{L}}{\longrightarrow}
    \mathcal{N}(0, \sigma^{2}_{\phi})
    \mbox{ as } n \to \infty.
\end{align}

\end{Theorem}

The presence of a bias, which grows as the order of $h^{-1/2}_{n}$ in the hybrid setup, is one of the known problems of the MI estimates. It renders the estimation of bias essential to apply Theorem \ref{Theorem: Asymptotic null distribution}. We observe that the rate of convergence to normality is rather slow for the GSB divergence, in particular when $\beta < 0$. However, this can be improved with larger sample sizes. Next, we compute the asymptotic means excluding the factor $\frac{1}{nh_{n}}$ and asymptotic variances for different special cases of the GSB divergence family. 

\begin{Corollary}
For the GSB divergence, the asymptotic mean and variance are obtained as \begin{align}
\label{GSB mean}
    \mu_{\phi}
    &=\frac{1}{2}\int_{u}K^{2}(u)du
    \int \Bigg[\sum_{x=0}^{1}(f_{x}f_{y})^{2A-1}
    \Big[(A\beta)^{2}e^{\beta (f_{x}f_{y})^{A}}+(A+B)(f_{x}f_{y})^{B-A}\Big](1-f_{x})\Bigg] dy, 
    \\
     \label{GSB var}
    \sigma^{2}_{\phi}
    &=\frac{1}{2}\int \Bigg(\int K(u)K(u+z)dz \Bigg)^{2}du
    \int\Bigg[\sum_{x=0}^{1}(f_{x}f_{y})^{2A-1}
    \Big[(A\beta)^{2}e^{\beta (f_{x}f_{y})^{A}}+(A+B)(f_{x}f_{y})^{B-A}\Big](1-f_{x})\Bigg]^{2} dy.
\end{align}
\end{Corollary}

\begin{Corollary}
{\it (Power divergence family)} Here $\beta=0$, $A+B=1$ and 
\begin{align}
\label{power divergence mean}
    \mu_{\phi} &=\frac{1}{2}\int_{u}K^{2}(u)du
    \int_{f_{y}>0}dy,  
    \\
    \label{power divergence var}
    \sigma^{2}_{\phi} &=\frac{1}{2}\int_{u}
    \Big(\int_{z} K(z)K(z+u)dz \Big)^{2}du
       \int_{f_{y}>0}dy.
\end{align}
\end{Corollary}

\begin{Corollary}
{\it (S-divergence, S-Hellinger, DPD)} Here $\beta=0$, $A+B=1+\alpha$ and
\begin{align}
   \label{S-divergence mean}
    \mu_{\phi}&=\frac{(1+\alpha)}{2}\int_{u}K^{2}(u)du
    \Big( f_{x_{0}}^{\alpha}f_{x_{1}} 
    + f_{x_{1}}^{\alpha}f_{x_{0}} \Big)
    \int_{f_{y}>0}f^{\alpha}_{y}dy,
    \\
    \label{S-divergence var}
    \sigma^{2}_{\phi} &=\frac{(1+\alpha)^{2}}{2}\int_{u}
    \Big(\int_{z} K(z)K(z+u)dz \Big)^{2}du
    \Big(f^{\alpha}_{x_{0}}f_{x_{1}}+f^{\alpha}_{x_{1}}f_{x_{0}}    \Big)^{2}
    \int_{f_{y}>0}f^{2\alpha}_{y}
    dy.
\end{align}
\end{Corollary}

\begin{Corollary}
{\it (Squared $L_{2}$-divergence)} Here $\beta=0$, $A+B=2$. So
\begin{align}
\label{L2 mean}
    \mu_{\phi} &=2f_{x_{0}}f_{x_{1}}\int_{u}K^{2}(u)du,   \\
\label{L2 var}    
    \sigma^{2}_{\phi} &=8(f_{x_{0}}f_{x_{1}})^{2}\int_{u}
    \Big(\int_{z} K(z)K(z+u)dz \Big)^{2}du \int_{f_{y}>0}f^{2}_{y}  dy.
\end{align}
\end{Corollary}

\begin{Corollary}
{\it(BED)} Here $\beta\in \mathds{R} \setminus \{0\}$, $A+B=0, A=1$. Finally multiply it by $\frac{2}{\beta^{2}}$ for mean and $\frac{4}{\beta^{4}}$ for variance. So
\begin{gather}
    \mu_{\phi}
    =(f_{x_{0}}f_{x_{1}})\int_{u}K^{2}(u)du
    \int_{f_{y}>0}\Bigg( 
    e^{\beta (f_{x_{0}}f_{y})} +
    e^{\beta (f_{x_{1}}f_{y})}
    \Bigg)f_{y}dy 
\end{gather}
and
\begin{gather}
    \sigma^{2}_{\phi}=2 (f_{x_{0}}f_{x_{1}})^{2}
    \int_{u}\Bigg( \int K(z)K(z+u)dz\Bigg)^{2} du
    \int_{f_{y}>0}
\Bigg(e^{\beta(f_{x_{0}}f_{y})}
+e^{\beta(f_{x_{1}}f_{y})}\Bigg)^{2}f^{2}_{y}dy. 
\end{gather}
For this family, $\beta=0$ is defined as a continuous limit of $\beta \to 0$. 
\end{Corollary}

We see that the asymptotic mean and variance for the $S$-divergence family depend only on the tuning parameter $\alpha \in [0,1]$. Empirical evidence suggests that as $\alpha$ increases towards $1$, the asymptotic mean and variance decrease. Mutual information based on the $S$-divergence also decreases as $\alpha$ increases. This observation is very crucial, and perhaps partly explains the stability behavior of the test statistics for higher $\alpha$. On the contrary, the robustness for the BED family increases as $\beta$ decreases in the real line.

If the support of $Y$ is unbounded, we cannot use Theorem \ref{Theorem: Asymptotic null distribution} for the power divergence family. Therefore, we need to invoke a permutation algorithm for calculating the empirical level and power when $Y$ have unbounded support. We observe that it takes a lot of computational burden to carry out the permutation test, which can be completely avoided for other members in this family, even when $Y$ has unbounded support. Also note that the asymptotic normality result may sometimes be useful only for very large sample sizes.

This fact is obviously a key highlight among many others to advocate the use of the GSB divergence family except for $\alpha=\beta=0$.

\begin{Remark}
MI based on the {\it Itakura-Saito distance} is obtained from extended Bregman divergence for $\phi(t)=-\frac{\log(t)}{2\pi}$ with $k=1$. For this family, we have 
\begin{align}
    \mu_{\phi}
    &=\frac{1}{4\pi}\int_{u}K^{2}(u)du
    \Bigg[\frac{f_{x_{0}}}{f_{x_{1}}}+
    \frac{f_{x_{1}}}{f_{x_{0}}}\Bigg]
    \int_{f_{y}>0}\frac{dy}{f_{y}},
    \\
    \sigma^{2}_{\phi}
    &=\frac{1}{8\pi^{2}}
    \int_{u}\Bigg( \int K(z)K(z+u)dz\Bigg)^{2} du
    \Bigg[ \frac{f_{x_{1}}}{f_{x_{0}}}  +
    \frac{f_{x_{0}}}{f_{x_{1}}}\Bigg]^{2}
     \int_{f_{y}>0} \frac{dy}{f^{2}_{y}}.
\end{align}
\end{Remark}

\subsection{Consistency and Power at Contiguous Alternatives}
\label{Consistency and Power at Contiguous Alternatives}

Suppose the conditions of the CLT for Theorem \ref{Theorem: Asymptotic null distribution} are satisfied, and that $\mu_{\phi}, \sigma_{\phi}$ are known. Then the null hypothesis $\mathds{H}$ given in (\ref{unstructred hypothesis; reformulated}) will be rejected at $100c\%$ nominal level of significance when $\widehat{I}_{D^{(k)}_{\phi}}> t$ satisfying $\mathds{P}_{\mathds{H}}\{\widehat{I}_{D^{(k)}_{\phi}}> t\}=c$. A simple calculation shows that $t=\frac{\tau_{c} \sigma_{\phi}}{nh^{1/2}_{n}}+\frac{\mu_{\phi}}{nh_{n}}$ for sufficiently large $n$, where $\tau_{c}$ is the upper-$100c\%$ point of $\mathcal{N}(0,1)$. Using Lemma \ref{Lemma: consistency of MI}, it can be easily shown that $\mathds{P}_{\mathds{K}}\Big\{\widehat{I}_{D^{(k)}_{\phi}}> \frac{\tau_{c}\sigma_{\phi}}{nh^{1/2}_{n}}+\frac{\mu_{\phi}}{nh_{n}} \Big\} \longrightarrow 1$ as $n \to \infty$. Hence, the class of tests is consistent. 

In practice, $\mu_{\phi}, \sigma_{\phi}$ are generally unknown, which need to be estimated to carry out the testing procedure. However, $\mu_{\phi}$ and $\sigma_{\phi}$ can be estimated using $\widehat{f}_{X}(x)$ and $\widehat{f}_{Y}(y)$. The normalized test statistics and their empirical versions are denoted by 
\begin{align}
\widehat{T}_{D^{(k)}_{\phi}}
=\frac{nh^{1/2}_{n}
(\widehat{I}_{D^{(k)}_{\phi}}-\frac{\widehat{\mu_{\phi}}}{nh_{n}})}{\widehat{\sigma}_{\phi}} 
\mbox{ and }
T_{D^{(k)}_{\phi}}
=\frac{nh^{1/2}_{n}
(\widehat{I}_{D^{(k)}_{\phi}}-\frac{\mu_{\phi}}{nh_{n}})}{\sigma_{\phi}}.
\end{align}
In the next result, we establish that both $\widehat{T}_{D^{(k)}_{\phi}}$ and $T_{D^{(k)}_{\phi}}$ converge to the same limiting distribution under the null hypothesis $\mathds{H}$, also the tests based on $\widehat{T}_{D^{(k)}_{\phi}}$ are consistent. 
\begin{Theorem}
\label{theorem: consistency of tests}
Under the Assumptions \descref{(A1)}-\descref{(A5)}, the following results are true as $n \to \infty$. 
\begin{itemize}
    \item[(i)] $|T_{D^{(k)}_{\phi}}- \widehat{T}_{D^{(k)}_{\phi}}| \overset{\mathds{P}}{\longrightarrow} 0$, consequently, $\widehat{T}_{D^{(k)}_{\phi}}
    \overset{\mathcal{L}}{\longrightarrow} \mathcal{N}(0,1)$ under the null hypothesis $\mathds{H}$.
    \item[(ii)] Tests based on $\widehat{T}_{D^{(k)}_{\phi}}$ are consistent.
\end{itemize}
\end{Theorem}

When the kernel density estimates are uniformly consistent, so are the tests based on them. Then, it is pertinent to investigate the behavior of its power function when the alternatives are contiguous to the null hypothesis. Suppose we want to test the null hypothesis $\mathds{H}$ against the following sequence of contiguous or local alternatives 
 \begin{align}
 \label{contiguous alternatives}
 \mathds{K}_{n}:
  (X, Y) \sim f^{(n)} \mbox{ where } f^{(n)}_{x,y}=f_{x}f_{y}+\frac{d}{\sqrt{nh^{1/2}_{n}}}
  \Delta_{x,y}
  \mbox{ where } n =1, 2, \ldots ,  
\end{align}
for a known constant $d \ge 0$ and $\Delta_{x,y}$ be such that $\sum_{x=0}^{1}\int_{f_{y}>0}\Delta_{x,y}dy=0$. As $f^{(n)}_{x,y}$ is a contiguous sequence to $\mathds{H}$, so will be its marginals with the same sequence of contaminating proportions. In this connection, denote the contaminating sequences of the marginals as 
\begin{align}
    f^{(n)}_{x}
    =\int_{f_{y}>0} f^{(n)}_{x,y}dy
    =f_{x}+\frac{d}{\sqrt{nh^{1/2}_{n}}}\Delta_{x}
    \mbox{ and }
     f^{(n)}_{y}
    =\sum_{x=0}^{1} f^{(n)}_{x,y}
    =f_{y}+\frac{d}{\sqrt{nh^{1/2}_{n}}}\Delta_{y},
\end{align}
where $\Delta_{x}=\int_{f_{y}>0}\Delta_{x,y}dy, \Delta_{y}=\sum_{x=0}^{1}\Delta_{x,y}$. See that the contiguous densities converge to the null density at a rate $\frac{1}{\sqrt{nh^{1/2}_{n}}}$ slower than $\frac{1}{nh^{1/2}_{n}}$. This rate is precisely required to stabilize a bias term that would invariably shift the asymptotic null distribution when $\mathds{K}_{n}$ is true. So the choice of such a contaminating sequence plays a crucial role in deriving the asymptotic distribution of $T_{D^{(k)}_{\phi}}$ under $\mathds{K}_{n}$. Notice that when $d=0$, the solution under $\mathds{K}_{n}$ is $f_{x,y}=f_{x}f_{y}$ for all $x,y$ almost surely, implying that $\mathds{K}_{n}=\mathds{H}$. Before stating our next result, it is useful to establish some notation. Define
\begin{align}
\label{I(n)}
    I^{(n)}_{D^{(k)}_{\phi}}
    =\sum_{x=0}^{1}\int_{f_{y}>0}
    \Bigg[\phi\Big((f^{(n)}_{x,y})^{k}\Big)-
    \phi\Big((f^{(n)}_{x}f^{(n)}_{y})^{k}\Big)
    -\Big\{\Big(f^{(n)}_{x,y}\Big)^{k} -\Big(f^{(n)}_{x}f^{(n)}_{y}\Big)^{k}\Big\}
    \phi'\Big((f^{(n)}_{x}f^{(n)}_{y})^{k}\Big)
    \Bigg]dy. 
\end{align}
Next, we approximate $I^{(n)}_{D^{(k)}_{\phi}}$ up to a second-order term that will be useful to derive the asymptotic distribution of $T_{D^{(k)}_{\phi}}$ under the contiguous alternatives. 

\begin{Lemma}
	\label{lemma: contiguous alternatives}
	Suppose the Assumption \descref{(A1)} is true. Then it holds that    
	\begin{align}
		\label{bias due to contiguous alternatives}
		I^{(n)}_{D^{(k)}_{\phi}}
		=\frac{d^{2}}{2nh^{1/2}_{n}}
		\sum_{x=0}^{1}\int_{f_{y}>0}
		k^{2}(f_{x}f_{y})^{2k}\phi''(f^{k}_{x}f^{k}_{y})
		\Bigg(\frac{\Delta_{x}}{f_{x}}+\frac{\Delta_{y}}{f_{y}}
		-\frac{\Delta_{x,y}}{f_{x}f_{y}}\Bigg)^{2}dy
		+o\Bigg(\frac{d^{2}}{nh^{1/2}_{n}}\Bigg).
	\end{align}
\end{Lemma}

Next, we show that a suitably scaled version of the second-order term in (\ref{bias due to contiguous alternatives}) induces a location shift in the null distribution of $\widehat{I}_{D^{(k)}_{\phi}}$ under the contiguous alternatives $\mathds{K}_{n}$, while the asymptotic variance remains unchanged.  

Now, we state the asymptotic distribution of $T_{D^{(k)}_{\phi}}$ under the contiguous alternatives $\mathds{K}_{n}$. 

\begin{Theorem}
	\label{Theorem: Asymptotic normality of MI under contiguous alternative}
	Suppose that the Assumptions \descref{(A1)}-\descref{(A5)} are true and $0 \le d \le C \cdot \sup_{x,y}|f_{x,y}-f_{x}f_{y}|$ for a constant $C>0$. Then the following result is true: 
	\begin{align}
		T_{D^{(k)}_{\phi}}
		\overset{\mathcal{L}}{\longrightarrow}
		\frac{d^{2}}{2\sigma_{\phi}}\sum_{x=0}^{1}\int_{f_{y}>0}k^{2}(f_{x}f_{y})^{2k}
		\phi''(f^{k}_{x}f^{k}_{y})
		\Bigg(\frac{\Delta_{x}}{f_{x}}+\frac{\Delta_{y}}{f_{y}}- \frac{\Delta_{x,y}}{f_{x}f_{y}}\Bigg)^{2}dy+
		\mathcal{N}(0, 1) 
	\end{align}
	as $n \to \infty$ under the contiguous alternatives $\mathds{K}_{n}$. 
\end{Theorem}

\begin{Remark}	
	In particular, suppose we take $\Delta_{x,y}=(\delta_{x_{0}}(x)\delta_{y_{0}}(y)-f_{x}f_{y})$ where $\delta_{x_{0}}(x)\delta_{y_{0}}(y)$ is a probability density function degenerate at a point $t_{0}=(x_{0}, y_{0})$. Then we obtain 
	\begin{align}
		T_{D^{(k)}_{\phi}} \overset{\mathcal{L}}{\longrightarrow} \frac{d^{2}}{2\sigma_{\phi}}
		\mathcal{IF}_{2}\Big(I_{D^{(k)}_{\phi}}, f_{X} f_{Y}, t_{0}\Big)+\mathcal{N}(0, 1)  
		\mbox{ under }
		\mathds{K}_{n},
	\end{align}
	where $\mathcal{IF}_{2}$ is the second-order influence function of $I_{D^{(k)}_{\phi}}$ under the null distribution at the point $t_{0}$. Later we will see that $\mathcal{IF}_{2}$ measures the stability behavior of $I_{D^{(k)}_{\phi}}$ at a point $t_{0}$ under infinitesimal contamination when the null hypothesis is true. The computation of $\mathcal{IF}_{2}$ is deferred to the next section. The asymptotic local power at contiguous alternatives $\mathds{K}_{n}$ is therefore given by  
	\begin{align}
		\label{eq: slope function}
		\underset{n \to \infty}{\lim}\mathds{P}_{\mathds{K}_{n}}\Big[T_{D^{(k)}_{\phi}} > \tau_{c}\Big]
		= 1-\Phi_{1} (\tau_{c}- d^{2}S_{\phi})
		\mbox{ where }
		S_{\phi}:=\frac{\mathcal{IF}_{2}(I_{D^{(k)}_{\phi}}, f_{X} f_{Y}, t_{0})}{2\sigma_{\phi}}.
	\end{align}
	We may call $S_{\phi}$ a slope function associated with the asymptotic power of $T_{D^{(k)}_{\phi}}$ at local alternatives. Observe that the local asymptotic power of $T_{D^{(k)}_{\phi}}$ varies as the slope function $S_{\phi}$ changes with the $\phi$-function itself. We can study $S_{\phi}$ to compare the power of 
	the standardized test statistic $T_{D^{(k)}_{\phi}}$ at such contiguous alternatives for different members of the $\phi$-generated extended Bregman divergence family. Also, note that $S_{\phi}$ is positive for all strictly convex and at least twice differentiable $\phi$-function. The larger the slope $S_{\phi}$ becomes, the faster the asymptotic contiguous power increases towards one from the nominal level `$c$' with the increment of $d^{2}$. Thus the second-order influence function $\mathcal{IF}_{2}$, which is a measure of robustness for $I_{D^{(k)}_{\phi}}$, is directly linked to the local asymptotic power of $T_{D^{(k)}_{\phi}}$ at contiguous alternatives $\mathds{K}_{n}$ in an interesting way.
\end{Remark}

In the next Section \ref{Robustness Studies}, we shall study the robustness behavior of $I_{D^{(k)}_{\phi}}$. 

\section{Robustness Studies}
\label{Robustness Studies}

One way to analyze the robustness of a functional is to examine its behavior when the underlying data-generating probability density function is perturbed. The influence function ($\mathcal{IF}$) is a widely used notion of a local measure of robustness. However, it may sometimes fail to capture robustness characteristics. In such cases, it becomes necessary to consider higher-order influence functions.

\subsection{Influence Function of \texorpdfstring{$I_{D^{(k)}_{\phi}}$}{} under Independence}
\label{Influence Function analysis}

Suppose that the joint density of $(X,Y)$ is contaminated at a point $t_{0}=(x_{0}, y_{0}) \in \{0,1\} \times \mathds{R}$ as   
\begin{align}
	\label{point-mass contamination}
	f^{\epsilon}_{X,Y}(x,y)=(1-\epsilon)f_{x,y}+\epsilon \delta_{x_{0}}(x)\delta_{y_{0}}(y),
\end{align}
where $\delta_{z_{0}}(z)=\mathds{1}\{z=z_{0}\}$ is the Dirac delta function and $0 \le \epsilon \le 1$. We know that $\int \delta_{z_{0}}(dz)=1$, or $\sum \delta_{z_{0}}(z)=1$ according to $z$ is continuous or discrete. A convenient abuse of notation for the integration of the Dirac delta function is $\int \delta_{z_{0}}(z)dz=1$, which is understood in the former sense. The marginals corresponding to the $\epsilon$-contaminated joint density are given by  
\begin{align}
	f^{\epsilon}_{X}(x)=(1-\epsilon)f_{x}+\delta_{x_{0}}(x)
	\mbox{ and }
	f^{\epsilon}_{Y}(y)=(1-\epsilon)f_{y}+\epsilon \delta_{y_{0}}(y).
\end{align}
The (first-order) influence function of the generalized mutual information functional $I_{D^{(k)}_{\phi}}$ at the point $t_{0}=(x_{0},y_{0})$ is defined by   
\begin{align}
	\label{def: first-order IF}
	\mathcal{IF}_{1}(I_{D^{(k)}_{\phi}}, f_{X,Y}, t_{0})=
	\Bigg[\frac{\partial D^{(k)}_{\phi}(f^{\epsilon}_{X,Y},f^{\epsilon}_{X}  f^{\epsilon}_{Y})}{\partial \epsilon}\Bigg]_{\epsilon=0}.
\end{align}
A simple calculation shows that  
\begin{gather}
	\label{first-order IF}
	\mathcal{IF}_{1}(I_{D^{(k)}_{\phi}}, f_{XY}, t_{0})
	=\sum_{x=0}^{1}\int_{f_{y}>0}\Bigg[k(f_{x,y})^{k-1}
	\Big(\delta_{x_{0}}(x)\delta_{y_{0}}(y)-f_{x,y}\Big)
	\Big\{\phi'\Big(f^{k}_{x,y}\Big)-\phi'\Big(f^{k}_{x}f^{k}_{y}\Big) \Big\} \nonumber \\
	-k(f_{x}f_{y})^{k-1}
	\Big(f^{k}_{x,y} -(f_{x}f_{y})^{k} \Big)
	(f_{x}\Big(\delta_{Y}(y_{0})-f_{y})+f_{y}(\delta_{X}(x_{0})-f_{x})\Big)\phi''\Big(f^{k}_{x}f^{k}_{y}\Big)\Bigg]dy.
\end{gather}
Under the null hypothesis $\mathds{H}$, this turns out to be  
\begin{align}
	\label{first-order IF under null}
	\mathcal{IF}_{1}(I_{D^{(k)}_{\phi}}, f_{X} f_{Y}, t_{0})=0
	\mbox{ for all } t_{0} \in \{0,1\} \times \mathds{R}.
\end{align}
See that the first-order influence function does not reveal any robustness feature of the functional $I_{D^{(k)}_{\phi}}$ under independence because it becomes constant for all the $\phi$ functions that constitute the extended Bregman divergence family. Therefore, to get further insights, we need to look up to the second-order influence function, defined similarly:  
\begin{align}
	\label{def: second-order IF}
	\mathcal{IF}_{2}(I_{D^{(k)}_{\phi}}, f_{X,Y}, t_{0})=
	\Bigg[\frac{\partial^{2} D^{(k)}_{\phi}(f^{\epsilon}_{X,Y},f^{\epsilon}_{X}  f^{\epsilon}_{Y})}{\partial \epsilon^{2}}\Bigg]_{\epsilon=0}.
\end{align}
An explicit expression of the second-order influence function under the null hypothesis is given in the next result. 
\begin{Theorem}
	\label{Theorem: second-order IF under null}
	Under the null hypothesis $\mathds{H}$, the second-order influence function of $I_{D^{(k)}_{\phi}}$ at a point $t_{0}$ is given by
	\begin{align}
		\label{IF: extended Bregman divergence}
		\mathcal{IF}_{2}(I_{D^{(k)}_{\phi}}, f_{X} f_{Y}, t_{0})
		=\sum_{x=0}^{1}\int_{f_{y}>0}k^{2}(f_{x}f_{y})^{2k}
		\phi''\Big(f^{k}_{x}f^{k}_{y}\Big)
		\Bigg[\frac{\Delta_{x}}{f_{x}}
		+\frac{\Delta_{y}}{f_{y}}
		-\frac{\Delta_{x,y}}{f_{x}f_{y}}
		\Bigg]^{2} dy
	\end{align}
	for $t_{0} \in \{0,1\} \times \mathds{R}$, where $\Delta_{x}=\delta_{x_{0}}(x)-f_{x}$, $\Delta_{y}=\delta_{y_{0}}(y)-f_{y}$ and $\Delta_{x,y}=\delta_{x_{0}}(x)\delta_{y_{0}}(y)-f_{x}f_{y}$.
\end{Theorem}
The proof of this result is straightforward and is therefore omitted in the Appendix. It is clear from (\ref{IF: extended Bregman divergence}) that the values of the second-order influence function are always positive. The higher the absolute values of the influence function, the better the stability of the test statistics becomes. For computational simplicity, it is often useful to use the following property of the Dirac delta function: $\int_{-\infty}^{\infty} g(z)\delta_{z_{0}}(dz)=g(z_{0})$ for a ``well-behaved" function $g$. Later, we shall plot the influence functions for different members of the generalized S-Bregman divergence family and discuss their properties in greater detail. 

Next, we compute the influence function of the test functional. Excluding the rate $nh^{1/2}_{n}$ the test functional may be considered as 
\begin{align}
	W_{D^{(k)}_{\phi}}=\Bigg[
	\frac{I_{D^{(k)}_{\phi}}
		-\frac{\widehat{\mu}_{\phi}}{nh_{n}}}{\widehat{\sigma}_{\phi}} \Bigg].
\end{align}
A simple calculation shows that 
\begin{align}
	\mathcal{IF}_{1}(W_{D^{(k)}_{\phi}}, f_{X} f_{Y}, t_{0})
	=0 \mbox{ and }
	\mathcal{IF}_{2}(W_{D^{(k)}_{\phi}}, f_{X} f_{Y}, t_{0})
	=\frac{1}{\widehat{\sigma}_{\phi}}
	\mathcal{IF}_{2}(I_{D^{(k)}_{\phi}}, f_{X} f_{Y}, t_{0})
\end{align}
under the null hypothesis $\mathds{H}$. The stability behavior of the test functional $W_{D^{(k)}_{\phi}}$ is determined by $I_{D^{(k)}_{\phi}}$ as their second-order influence functions are proportional to each other.

Unfortunately, things get much harder when the null hypothesis is not true. In those cases, we do not have simplified expressions for the second-order influence functions as the alternative hypothesis $\mathds{K}$ is a composite one. An alternative approach is to study its stability at the contiguous alternatives using the level and power influence functions, which will be discussed in the next subsection. 

For the GSB family, Theorem \ref{Theorem: second-order IF under null} gives
\begin{align}
  \label{IF: GSB}
  \mathcal{IF}_{2}(I_{D^{(k)}_{\phi}}, f_{X} f_{Y}, t_{0})
  =\sum_{x=0}^{1}\int_{f_{y}>0}
  \Big\{A^{2}\beta^{2}e^{\beta (f_{x}f_{y})^{A}} (f_{x}f_{y})^{2A-1}
  + (A+B) (f_{x}f_{y})^{A+B-1} \Big\}U_{xy}dy,
\end{align}
where $U_{xy}=(f_{x}f_{y})
   \Bigg(\frac{h^{X}_{x_{0}}}{f_{x}}
    +\frac{h^{Y}_{y_{0}}}{f_{y}}
    -\frac{h^{X,Y}_{x_{0}, y_{0}}}{f_{x}f_{y}}
    \Bigg)^{2}$ and $t_{0} \in \{0,1\} \times \mathds{R}$.  
    
In particular,
\begin{align}
\label{IF: special cases}
    \mathcal{IF}_{2}(I_{D^{(k)}_{\phi}}, f_{X} f_{Y}, t_{0})
    =\begin{cases}
    \sum_{x=0}^{1}\int_{f_{y}>0}
      A^{2}\beta^{2}e^{\beta (f_{x}f_{y})^{A}}(f_{x}f_{y})^{2A-1}U_{xy}dy  
    &\mbox{ if } A+B=0, A, \beta \ne 0, \\ \\
    (1+\alpha)\sum_{x=0}^{1}\int_{f_{y}>0}(f_{x}f_{y})^{\alpha} U_{xy} dy
    &\mbox{ if } \beta=0.
     \end{cases}
\end{align}
Note that when $A+B=0, A=1$, we get the expression for a scaled BED family, which needs to be multiplied by $2/\beta^{2}$ (where $\beta \ne 0$) to get the $\mathcal{IF}_{2}$ for the BED family. 

The boundedness of the influence function essentially depends on controlling the term $U_{xy}$ inside the summation and integration in (\ref{IF: GSB}). Since $f_{y}$ is a probability density function of a continuous random variable, its boundedness may be an issue. However, a sufficient condition, such as the density of $Y$ having a continuously bounded derivative, ensures its density to be uniformly bounded. When $f_{y}$ is uniformly bounded, $U_{xy}$ can still be unbounded because the density remains in its denominator. However, in most situations, we see that the factor $(f_{x}f_{y})^{\tau}$ adds stability to $U_{xy}$ for $\tau>0$.
More stability of the influence function to extreme outliers is achieved for higher values of $\tau$.

From the above speculation, see that the first term of $\mathcal{IF}_{2}$ in (\ref{IF: GSB}) will become more stable for $2A-1 > 0$, i.e., $A > \frac{1}{2}$. Similarly, more stability in the second term is achieved for $A+B-1 > 0$, i.e., $\alpha>0$. Further, observe that $e^{\beta (f_{x}f_{y})}$ always stays bounded for finite $\beta$.

Like Basak and Basu (2022) \cite{basak2022extended}, we study the influence function in the following regions.   
\begin{itemize}
    \item[(1)]  We know that $\beta=0$ gives the S-divergence family. Moreover, when $\alpha=0$, we obtain the expression of $\mathcal{IF}_{2}$ for the power divergence family, which turns out to be independent of $\lambda$. From the earlier discussions, we know that $\mathcal{IF}_{2}$ of the S-divergence family will be more stable only when $\alpha>0$ (see (\ref{IF: special cases})). It becomes very unstable at $\alpha=0$, i.e., for the power divergence family. This region is denoted by   
    $\mathds{S}_{1} =\{(\alpha, \lambda, \beta): \alpha > 0, \lambda \in \mathds{R}, \beta=0\}$.
     
     \item[(2)] When $\beta \ne 0, A=0$, the first term in (\ref{IF: GSB}) drops out and the non-vanishing second term is more stable for $\alpha>0$. However $A=0$ implies that $\lambda=\frac{1}{\alpha-1}$, and the entire expression is independent of $\beta$. So the allowable region is 
     $\mathds{S}_{2}=\{(\alpha, \lambda, \beta): \alpha>0, \lambda=\frac{1}{\alpha-1}, \beta \ne 0\}$. As $\alpha>0$ in $\mathds{S}_{2}$, the allowable tuning parameter $\lambda$ should be negative. 
    
    \item[(3)] When $(A+B)=0$, or, $\alpha=-1$, the second term in (\ref{IF: GSB}) vanishes. Then the first term becomes more stable for $A > \frac{1}{2}$, i.e., $\lambda > -\frac{1}{4}$ and $\beta \ne 0$. In this case, $\mathcal{IF}_{2}$ will be more stable in $\mathds{S}_{3}=\{(\alpha, \lambda, \beta): \alpha=-1, \lambda> -\frac{1}{4}, \beta \ne  0\}$. 
    
    \item[(4)] In the fourth case, all of $\beta, A, (A+B)$ should be non-zero. So the allowable region is $\mathds{S}_{4}=\{(\alpha, \lambda, \beta); \alpha>0, \lambda(1-\alpha)>-\frac{1}{2}, \beta \ne 0\}$. 
    
\end{itemize}
Combining all these disjoint regions, we see that the infinitesimal effect of an outlier as measured by $\mathcal{IF}_{2}$ will be clearly downweighted when the tuning parameters belong to $\mathds{S}:=\cup_{i=1}^{4}\mathds{S}_{i}$. Outside this region, the influence function becomes very unstable. The PD family is part of this unstable region.

\begin{figure}
\begin{center}
 \includegraphics[width=0.8\textwidth]{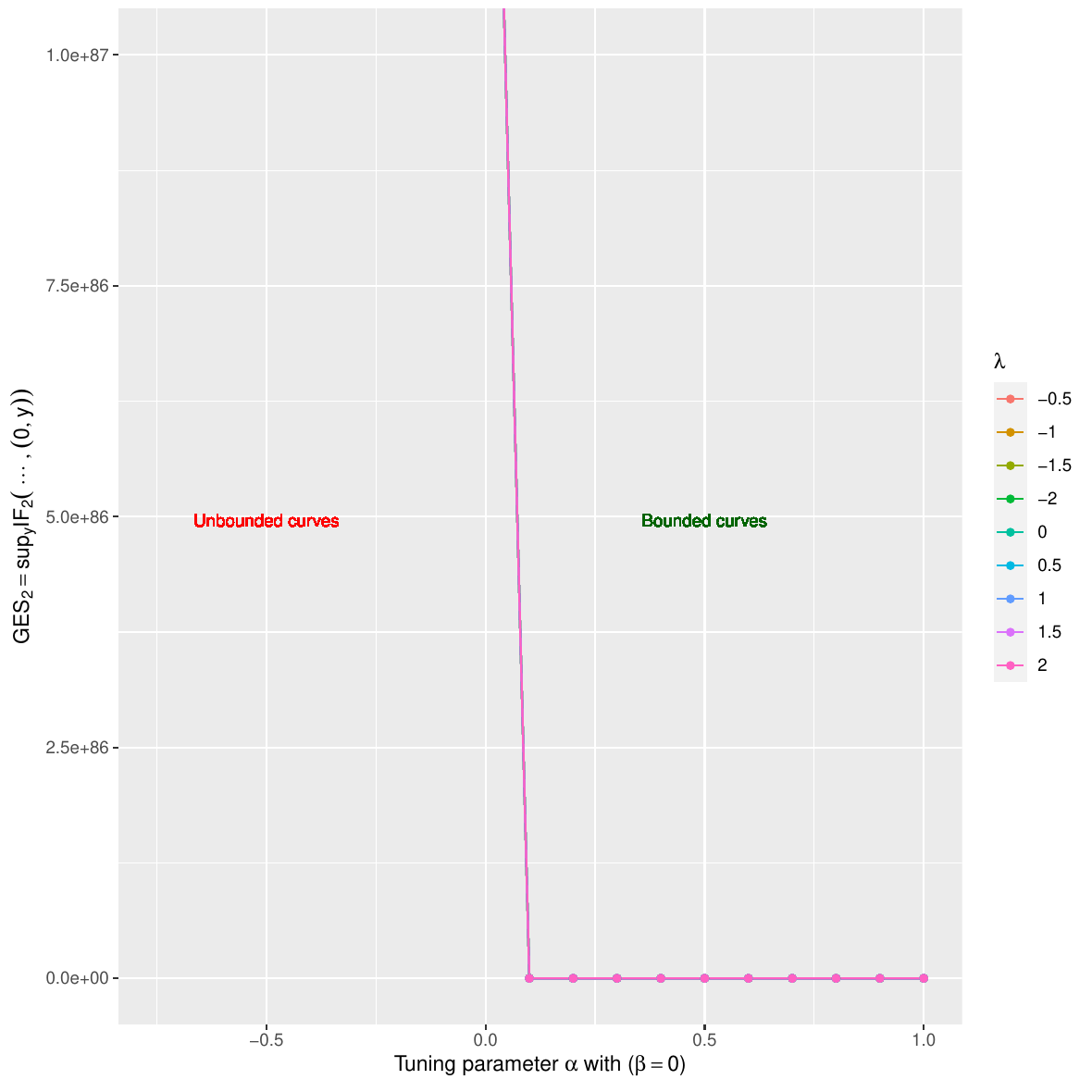}
 \end{center}
   \caption{Bounded GES curves when $(\alpha, \lambda, \beta) \in \mathds{S}_{1}$, and unbounded curves when $\alpha \le 0$.}
    \label{S1 region}
\end{figure}

\begin{figure}
\begin{center}
 \includegraphics[width=0.8\textwidth]{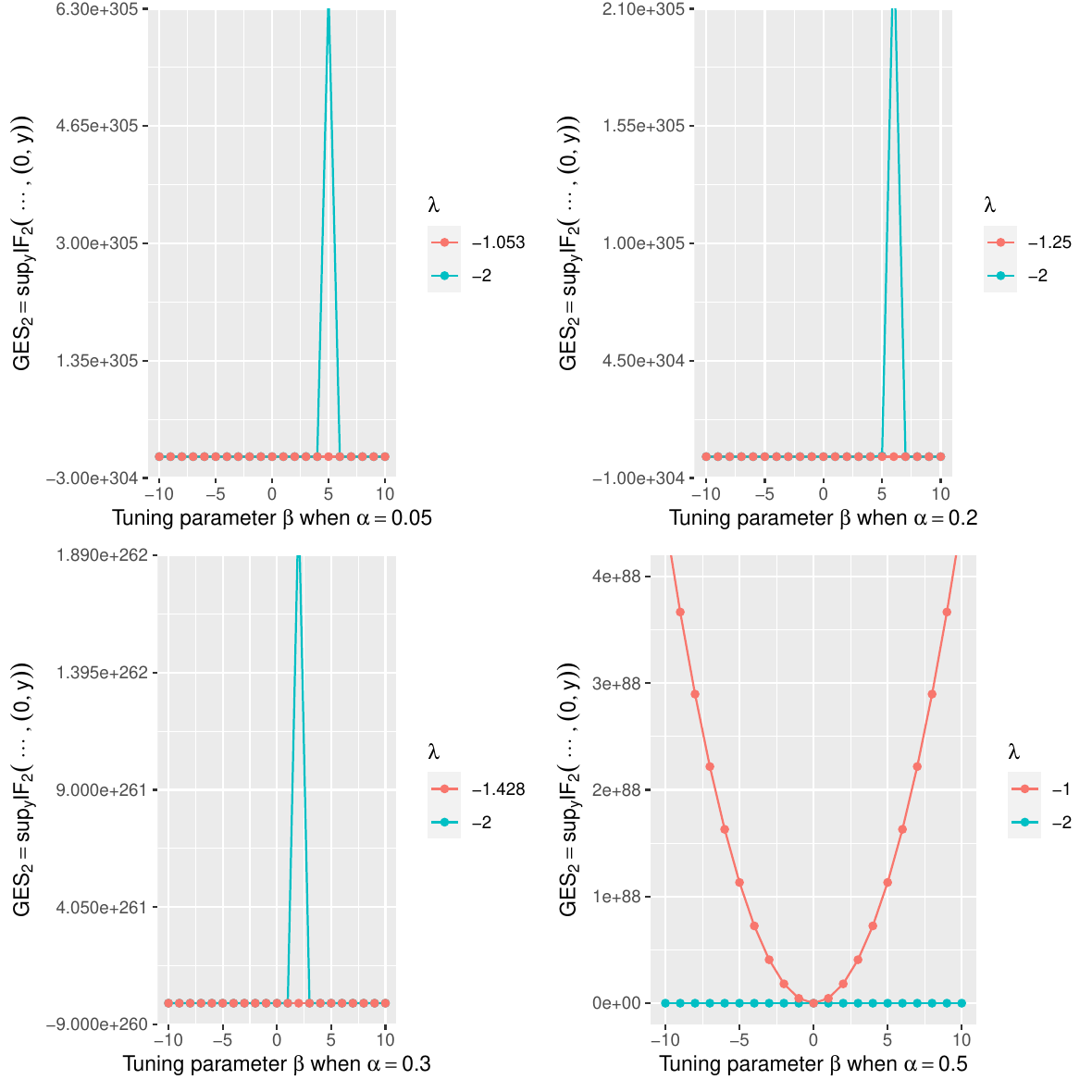}
 \end{center}
   \caption{Bounded GES curves when $(\alpha, \lambda, \beta) \in \mathds{S}_{2}$, and some unbounded curves when $\lambda \ne \frac{1}{\alpha-1}$.}
    \label{S2 region}
\end{figure}

\begin{figure}
\begin{center}
 \includegraphics[width=0.8\textwidth]{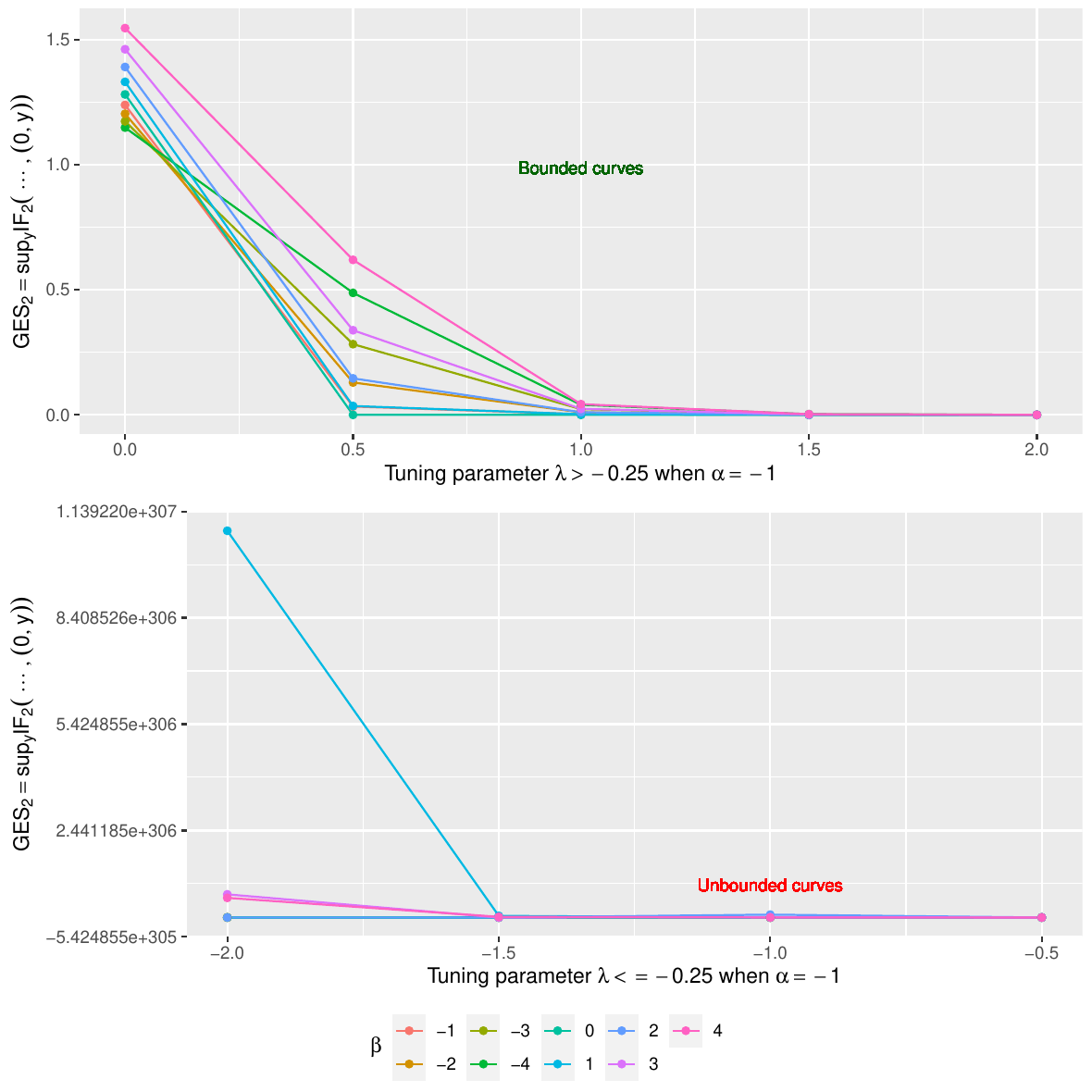}
 \end{center}
   \caption{Bounded GES curves when $(\alpha, \lambda, \beta) \in \mathds{S}_{3}$, and some unbounded curves when $\lambda \le -0.25$.}
    \label{S3 region}
\end{figure}

\begin{figure}
\begin{center}
 \includegraphics[width=0.8\textwidth]{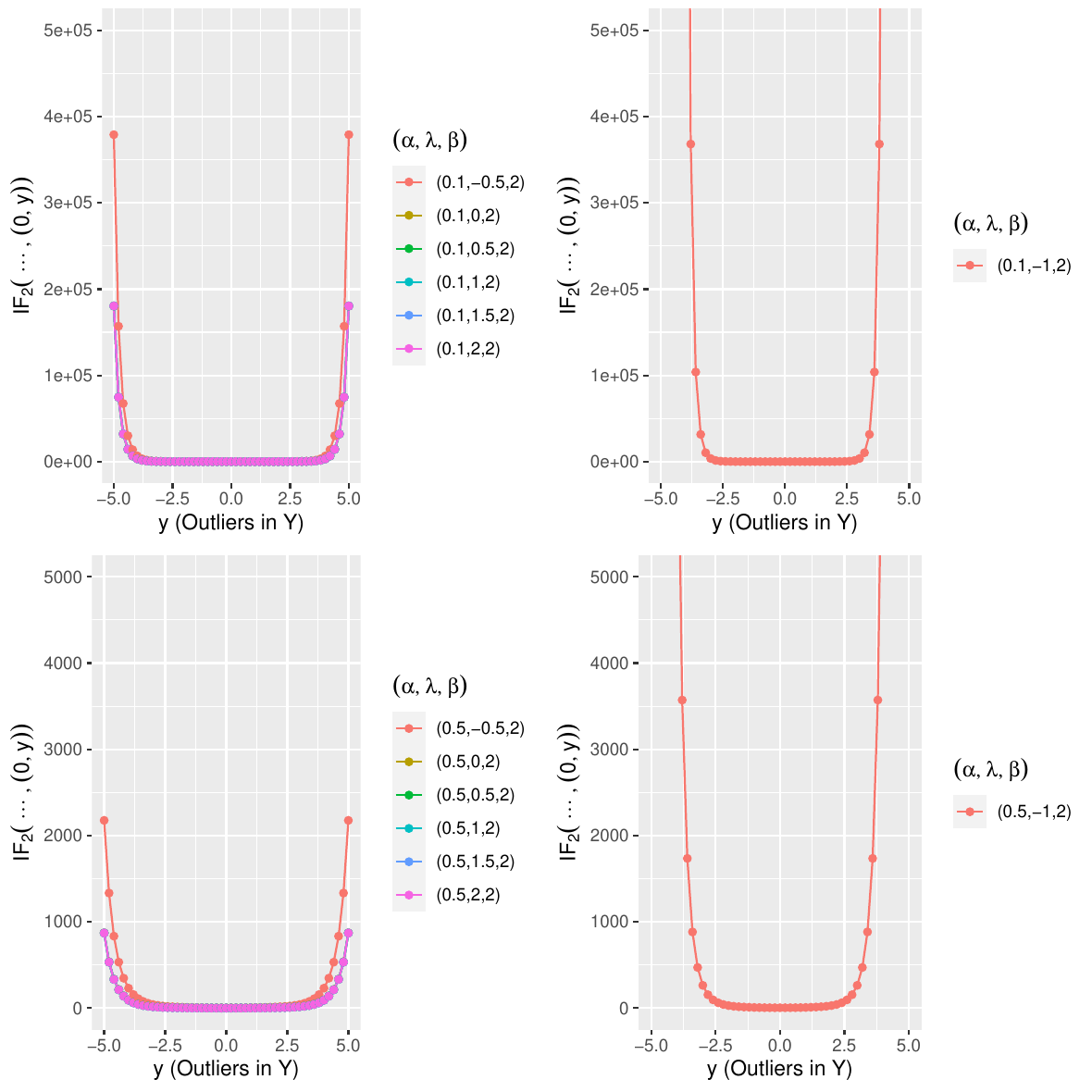}
 \end{center}
   \caption{Bounded $\mathcal{IF}_{2}$ when $(\alpha, \lambda, \beta) \in \mathds{S}_{4}$, and some unbounded curves when $\lambda(1-\alpha) \le -0.5$.}
    \label{S4 region}
\end{figure}

To explore further on this front, we choose $X \sim Bernoulli(0.5)$ and $Y \sim \mathcal{N}(0, 1)$. $X, Y$ are assumed to be independent under the null hypothesis $\mathds{H}$. In Figure \ref{S1 region}, \ref{S2 region} and \ref{S3 region}, we plot the second-order gross error sensitivity (GES) that is defined as $GES_{2}:=\sup_{y}\mathcal{IF}_{2}(\cdots,(0,y))$ fixing $x=0$. For practical purposes, we have taken the supremum over $[-20, 20]$, which is a very large range for outliers to be added to the continuous distribution $\mathcal{N}(0, 1)$. From Figure \ref{S1 region}, it is clear that all members of the power divergence family ($\alpha=\beta=0$) are equivalent in terms of robustness. Thus, outliers coming from the extreme region of the true continuous distribution, or even lying outside the support, make the tests based on the PD family unstable. However, with $\alpha>0$, their robustness increases substantially. In other figures, we plot the bounded GES curves when the tuning parameters belong to $\mathds{S}_{i}, i=2, 3$, and some unbounded curves outside those regions. In Figure \ref{S4 region}, we plot the $\mathcal{IF}_{2}$ and appropriately identify some unbounded curves outside that region.


 Empirical evidence suggests that members outside the PD family generally produce better stability upon extreme data contamination at a point mass. For fixed $\lambda \in \mathds{R}$, better stability is usually achieved for higher values of $\alpha$ and negative $\beta$.

\begin{Remark}
	\label{Remark: outliers in nonparametric setup}	
    \textbf{(Notion of Contamination in This Setup)}
	In parametric estimation, it is generally assumed that a contaminated density should not belong to the parametric family, or it should be geometrically separated from the true distribution. A similar interpretation may be drawn in this case as well. Notice that the null hypothesis $\mathds{H}$ defines the class of densities $\mathcal{H}:=\{f_{x,y}: f_{x,y}=f_{x}f_{y} \text{ for all } x, y\}$. The joint density under the null gets contaminated as 
	\begin{align}
		\label{contaminated null density}	
		f^{\epsilon}_{x,y}=(1-\epsilon)f_{x}f_{y}+\epsilon \delta_{x_{0}}(x)\delta_{y_{0}}(y),
	\end{align} 
	which no longer belongs to the class $\mathcal{H}$. A non-robust distance, such as the Kullback-Leibler divergence, which is very sensitive to contamination proportion, enlarges the dissimilarity between the contaminated and the true density (under null). Consequently, the level gets inflated. However, using robust statistical distances mitigates this dissimilarity. Therefore, the levels become stable. The role of the influence function in revealing the robustness of these nonparametric tests will be further clarified through the plots for specific families.  
\end{Remark}

\subsection{Level and Power Influence Functions}
\label{Level and Power Influence Functions of  Wald-type test}

In the previous subsection, we examined the robustness of $I_{D^{(k)}_{\phi}}$ and $T_{D^{(k)}_{\phi}}$ using the second-order influence function analysis under the null hypothesis. However, an important question remains: how are the level and power of this class of tests affected when the distributions under the null and contiguous alternatives become contaminated? The answer is not straightforward. One possible approach is to evaluate the influence functions of the level and power functionals corresponding to this class of tests. The main challenge, however, lies in the fact that closed-form expressions for the size and power are generally intractable under contaminated distributions. Nevertheless, suitable approximations can be made under certain regularity conditions. In this section, we explore these aspects in greater detail.

Suppose that the densities under the null and contiguous alternatives are subject to contamination in the following manner:
\begin{align}
	\label{ch6:contaminated null}	
	f^{\epsilon}_{x,y}
	=f_{x}f_{y}+ \frac{\epsilon}{\sqrt{nh^{1/2}_{n}}} \Delta_{x,y}
	\mbox{ and }
	f^{(P_{n})}_{x,y}
	=f^{(n)}_{x,y}+\frac{\epsilon}{\sqrt{nh^{1/2}_{n}}}
	\Delta^{'}_{x,y}
	\mbox{ for all } x, y	
\end{align} 
respectively, where $\epsilon \ge 0$ and $\sum_{x=0}^{1}\int_{f_{y}>0}\Delta'_{x,y}dy=0$. Here we allow $\Delta'_{x,y}$ to depend on the sample size $n$. The densities $f^{(n)}_{x,y}$ under the contiguous alternatives are defined in (\ref{contiguous alternatives}). When $\epsilon = 0$, no contamination occurs in either scenario. In particular, when $\Delta_{x,y} = \Delta'_{x,y}$ for all $x,y$, along with $d = 0$, it follows that $f^{(P_{n})}_{x,y} = f^{\epsilon}_{x,y}$ for all $x,y$. Now, consider point-mass contaminations in both cases, defined as $\Delta_{x,y} = (\delta_{x_{0}}(x)\delta_{y_{0}}(y) - f_{x}f_{y})$ and $\Delta'_{x,y} = (\delta_{x_{1}}(x)\delta_{y_{1}}(y) - f^{(n)}_{x,y})$, where $t_{0} = (x_{0}, y_{0})$ and $t_{1} = (x_{1}, y_{1})$. Under these contaminations, the level and power functionals are respectively given by
\begin{align}
	\label{ch6:level and power}
	\alpha(f^{\epsilon}_{X,Y}, t_{0})= \mathds{P}_{f^{\epsilon}_{X,Y}}\Big\{ T_{D^{(k)}_{\phi}} > \tau_{c} \Big\}
	\mbox{ and }
	\pi(f^{(P_{n})}_{X,Y}, t_{1})= \mathds{P}_{f^{(P_{n})}_{X,Y}}  \Big\{ T_{D^{(k)}_{\phi}} > \tau_{c} \Big\}. 
\end{align}

Following Hampel (1986) \cite*{hampel1986robust}, the level and power influence functions are defined as
\begin{align}
\label{ch6:LIF and PIF}
\mathcal{LIF}(\alpha, t_{0}) &= \underset{n \longrightarrow \infty}{\lim} \Bigg[\frac{\partial \alpha(f^{(\epsilon)}_{X,Y}, t{0})}{\partial \epsilon} \Bigg]_{\epsilon=0}, \quad
\mathcal{PIF}(\pi, t_{1}) = \underset{n \longrightarrow \infty}{\lim} \Bigg[\frac{\partial \pi(f^{(P_{n})}_{X,Y}, t_{1})}{\partial \epsilon} \Bigg]_{\epsilon=0}.
\end{align}
To evaluate these level and power influence functions as in (\ref{ch6:LIF and PIF}), we require the asymptotic distributions of $T_{D^{(k)}_{\phi}}$ under both the contaminated null hypothesis and the contaminated contiguous alternatives. The former has already been derived in Theorem \ref{Theorem: Asymptotic normality of MI under contiguous alternative}. Our next objective is to obtain the asymptotic distribution of $T_{D^{(k)}_{\phi}}$ under $f^{(P_{n})}_{X,Y}$.
To proceed, we define
\begin{align}
I^{(P_{n})}_{D^{(k)}_{\phi}}
= D^{(k)}_{\phi}(f^{(P_{n})}_{X,Y}, f^{(P_{n})}_{X}f^{(P_{n})}_{Y}),
\end{align}
where $f^{(P_{n})}_{X}$ and $f^{(P_{n})}_{Y}$ denote the marginal densities corresponding to the joint density $f^{(P_{n})}_{X,Y}$, as understood from the context.
The following lemma establishes an asymptotic expression for $I^{(P_{n})}_{D^{(k)}_{\phi}}$.

\begin{Lemma}
	\label{ch6:MI under contaminated contiguous}
	Under the Assumption \descref{(A1)} it holds that  
	\begin{align}
		I^{(P_{n})}_{D^{(k)}_{\phi}}
		&=\frac{k^{2}}{2nh^{1/2}_{n}}
		\sum_{x=0}^{1} \int (f_{x}f_{y})^{2k}
		\Bigg\{
		d\Bigg(
		\frac{\Delta_{x}}{f_{x}}
		+ \frac{\Delta_{y}}{f_{y}}
        -\frac{\Delta_{x,y}}{f_{x}f_{y}}
		\Bigg)
		+
		\epsilon
		\Bigg(
		\frac{\Delta'_{x}}{f_{x}}
		+\frac{\Delta'_{y}}{f_{y}}
        -\frac{\Delta'_{x,y}}{f_{x}f_{y}}
		\Bigg)\Bigg\}^{2}\phi''\Big(f^{k}_{x}f^{k}_{y}\Big)dy
		\nonumber\\
		&+o\Bigg(\frac{(d+\epsilon)^{2}}{nh^{1/2}_{n}}\Bigg).
	\end{align}
\end{Lemma}	

Now, we present the asymptotic distribution of $I^{(P_{n})}_{D^{(k)}_{\phi}}$ under $f^{(P_{n})}_{X,Y}$.

\begin{Theorem}
	\label{theorem: asymptotic dist under contaminated contiguous}
	Suppose the Assumptions \descref{(A1)}-\descref{(A5)} are true and $0 \le \max\{d, \epsilon\} \le C^{*} \sup_{x,y}|f_{x,y}-f_{x}f_{y}|$ for some $C^{*}>0$. Then it follows that 
	\begin{align}
		T_{D^{(k)}_{\phi}} 
		-
		\frac{k^{2}}{2\sigma_{\phi}}
		\sum_{x=0}^{1} \int (f_{x}f_{y})^{2k}
		\Bigg\{
		d\Bigg(		
		\frac{\Delta_{x}}{f_{x}}
		  + \frac{\Delta_{y}}{f_{y}}
        -\frac{\Delta_{x,y}}{f_{x}f_{y}}
		\Bigg)
		+
		\epsilon
		\Bigg(
		\frac{\Delta'_{x}}{f_{x}}
		+ \frac{\Delta'_{y}}{f_{y}}
        -\frac{\Delta'_{x,y}}{f_{x}f_{y}}
		\Bigg)\Bigg\}^{2}\phi''\Big(f^{k}_{x}f^{k}_{y}\Big)dy
		\overset{\mathcal{L}}{\longrightarrow} 
		\mathcal{N}\Big( 0,1\Big)
	\end{align}
	under $f^{(P_{n})}_{X,Y}$ as $n \to \infty$.	
\end{Theorem}

\begin{Remark}
The distribution of the test statistic under $f^{(P_{n})}_{X,Y}$ coincides with that given in Theorem \ref{Theorem: Asymptotic normality of MI under contiguous alternative} when $\epsilon=0$.
\end{Remark}

We are now in a position to derive the expressions for $\mathcal{LIF}$ and $\mathcal{PIF}$. The exact values of the level and power under the $\epsilon$-contaminated versions of the true null and contiguous distributions are generally difficult to compute. However, their limiting values can be obtained under suitable regularity conditions. To achieve this, it is necessary to interchange the order of the limit and differentiation, which is valid only under specific assumptions. These results are formally stated in the next theorem. To compute the power influence function, we take
\begin{align}
\Delta'_{x,y}
=\delta{x_{1}}(x)\delta_{y_{1}}(y)-f^{(P_{n})}_{x,y}.
\end{align}

\begin{Theorem}
	\label{PIF and LIF theorem}
	Suppose that the sequence $\Big\{ \frac{\partial}{\partial \epsilon } \pi(f^{(P_{n})}_{X,Y}, t_{1})\Big\}$ converges uniformly for all $\epsilon \in [0,1]$ at fixed $t_{1}=(x_{1}, y_{1})$, and the conditions of Theorem \ref{theorem: asymptotic dist under contaminated contiguous} are true. Then we have the following results.
	\begin{itemize}
		\item[(a)] The power influence function is given by
		\begin{align}
			\mathcal{PIF}( \pi, t_{1})
			&=\frac{d}{\sigma_{\phi}}
			\sum_{x=0}^{1} \int k^{2}(f_{x}f_{y})^{2k}
			\Bigg(
			\frac{\Delta^{*}_{x}}{f_{x}}
			+ \frac{\Delta^{*}_{y}}{f_{y}}
            -\frac{\Delta^{*}_{x,y}}{f_{x}f_{y}}
			\Bigg)
			\Bigg(
            \frac{\Delta_{x}}{f_{x}}
			+ \frac{\Delta_{y}}{f_{y}}
            -\frac{\Delta_{x,y}}{f_{x}f_{y}}
			\Bigg)\phi''\Big(f^{k}_{x}f^{k}_{y}\Big)dy \nonumber \\
			&\times 
			\phi_{1}
			\Bigg( \tau_{c}- \frac{d^{2}}{2\sigma_{\phi}}
			\mathcal{IF}_{2}(I_{D^{(k)}_{\phi}}, f_{X}f_{Y}, t_{0}) \Bigg)
		\end{align}	
		where  $\Delta^{*}_{x,y}=(\delta_{x_{1}}(x)\delta_{y_{1}}(y)-f_{x}f_{y})$, $\Delta^{*}_{x}=(\delta_{x_{1}}(x)-f_{x})$ and $\Delta^{*}_{y}=(\delta_{y_{1}}(y)-f_{y})$.  
		
		\item[(b)] When $t_{0}=t_{1}$, we obtain
		\begin{align}
			\mathcal{PIF}( \pi, t_{1})
			= \frac{d}{\sigma_{\phi}}
			\mathcal{IF}_{2}(I_{D^{(k)}_{\phi}}, f_{X}f_{Y}, t_{1}) \cdot
			\phi_{1}
			\Bigg( \tau_{c}- \frac{d^{2}}{2\sigma_{\phi}}
			\mathcal{IF}_{2}(I_{D^{(k)}_{\phi}}, f_{X}f_{Y}, t_{1}) \Bigg).
		\end{align}	
		\item[(c)] The level influence function is given by
		\begin{align}
			\mathcal{LIF}( \pi, t_{0})\equiv 0. 	
		\end{align}	
	\end{itemize}	
\end{Theorem}

\begin{Remark}	
	The stability of the power influence function is proportional to the second-order influence function of $I_{D^{(k)}_{\phi}}$ under the null distribution. However, $\mathcal{LIF}$ is zero identically. Like the influence function, the level influence function also fails to reveal the robustness of the type-I error, which is not the case in reality. In simulation studies, we observe that the robustness of the levels of this class of tests depends on the choice of the $\phi$-function as well.      
\end{Remark}

\subsection{Asymptotic Breakdown Point Analysis of \texorpdfstring{$I_{D^{(k)}_{\phi}}$}{}}
\label{Asymptotic Breakdown Point Analysis}

Earlier, we studied the robustness of $I_{D^{(k)}_{\phi}}$ through influence function analysis. The influence function, being defined as a directional derivative in the direction of a point mass, describes the stability of the functional for infinitesimal contamination at that point. Although a useful concept, it sometimes fails to reveal the desired robustness of a functional that may otherwise be apparent. Since the influence function is entirely a local concept, it should be complemented by a global measure of robustness. One such global measure is the breakdown point, which quantifies the maximum proportion of outlying observations in a data set that a statistical functional can tolerate before it yields erratic values. The notion of outlying observations in this nonparametric setup has already been discussed in Remark \ref{Remark: outliers in nonparametric setup}.

Sometimes, it is intractable to compute the actual breakdown point using the definitions given in Hampel (1986) \cite*{hampel1986robust}. However, the asymptotic breakdown point is often easier to calculate in many situations.

Before presenting our working definition, we introduce some useful notations.
Let $k_{XY,m}(x,y):=k_{xy,m}$ be a sequence of contaminating densities, and let $k_{x,m}=\int_{f_{y}>0} k_{xy,m}dy$, $k_{y,m}=\sum_{x=0}^{1}k_{xy,m}$ be its marginals. The true densities $f_{x,y}$, $f_{x}$, and $f_{y}$ are respectively contaminated as
\begin{align}
f^{m}_{x,y}
&=(1-\epsilon)f_{x,y}+\epsilon k_{xy,m},
\nonumber \\
f^{m}_{x}
&=(1-\epsilon)f_{x}+\epsilon k_{x,m},
\nonumber \\
f^{m}_{y}
&=(1-\epsilon)f_{y}+\epsilon k_{y,m}
\end{align}
with $0\le \epsilon \le 1$. Often, we express $f^{m}_{x,y}=f^{m}_{x}f^{m}_{y|x}$ for all $x,y$, where $f^{m}_{y|x}$ is the conditional density of $y$ given $x$, obtained from $f^{m}_{x,y}$. Similar notations are used for the other densities as well. Recall that $I_{D^{(k)}_{\phi}}=D^{(k)}_{\phi}(f_{X,Y}, f_{X}f_{Y})$, and further define
\begin{gather}
\label{ch6: contaminated MI}
I_{m}
=D^{(k)}_{\phi}(f^{m}_{X,Y},
f^{m}_{X} f^{m}_{Y})
\mbox{ and }
I'_{m}
=D^{(k)}_{\phi}(f^{m}_{X,Y}, f_{X}f_{Y}).
\end{gather}

The definition of the asymptotic breakdown point used so far in a parametric setup does not directly fit here because of the different nature of the problem. Here, the goal is to compute the mutual information for a given joint density. Naturally, kernel density estimates are used in practice to approximate the mutual information. When robust to a particular sequence of contaminating densities, the generalized mutual information should not vary erratically as we move along that sequence. From this perspective, the contaminated mutual information $I_{m}$ in (\ref{ch6: contaminated MI}) should not break down at a contamination proportion $\epsilon$ if $\lim_{m \to \infty} I_{m}$ exists.

With this background, we now provide the working definition of the breakdown point in the present context.
\begin{Definition}
\label{ch6:def1: breakdown point}
Given the true density $f_{X, Y}$, the generalized mutual information functional $I_{D^{(k)}_{\phi}}$ is said to break down asymptotically at $\epsilon$ if its contaminated version $I_{m}$ does not converge when $I_{D^{(k)}_{\psi}} < \infty$. Then $\epsilon^{*}$ is called the asymptotic breakdown point of $I_{D^{(k)}_{\phi}}$ if
\begin{align}
\label{ch6: breakdown point definition 2}
\epsilon^{*}=
\inf \Big\{\epsilon : I_{m}
\mbox{ does not converge but } I_{D^{(k)}_{\psi}} < \infty \Big\}.
\end{align}
\end{Definition}

From Definition \ref{ch6:def1: breakdown point}, it follows that $I_{D^{(k)}_{\phi}}$ may break down at $\epsilon$ only when it is bounded. The exclusion of divergences for which $I_{D^{(k)}_{\phi}}=\infty$ may somewhat limit the generality of this definition, but it does not significantly affect the analysis, as it only omits some special cases that are not statistically meaningful.

Following Park and Basu (2004) \cite*{park2004minimum} and Roy et al. (2023) \cite*{roy2026asymptotic}, we adopt a similar approach to establish the asymptotic breakdown point of $I_{D^{(k)}_{\phi}}$ using Definition \ref{ch6:def1: breakdown point}. Let us make the following assumptions:
\begin{description}
\descitem{(BP1)} $\int \min \{f_{y|x}, k_{y|x,m}\}dy \longrightarrow 0$ as $m \to \infty$ for $x=0, 1$,

\descitem{(BP2)} $\int \min\{ f_{y|x}, f^{m}_{y}\}dy \longrightarrow 0$ as $m \to \infty$ for $x=0,1$,  

\descitem{(BP3)} $\int \min \{f_{y}, k_{y|x,m}\}dy \longrightarrow 0$ as $m \to \infty$ for $x=0,1$,  

\descitem{(BP4)} $\phi(0),\phi'(0)$ are finite, and $Y$ has a bounded support,  

\descitem{(BP5)} $I_{m} \le I'_{m}$ for all $m \ge 1$,  

\descitem{(BP6)} there exists $\tilde{\epsilon} \in [0, \frac{1}{2}]$ such that for all $\epsilon < \tilde{\epsilon}$,  
\begin{align}
	\underset{m \to \infty}{\liminf}
	D^{(k)}_{\phi}(\epsilon k_{XY,m}, f^{m}_{X}f^{m}_{Y}) 
	&> \underset{m \to \infty}{\limsup}
	\sum \int \Bigg[\phi( (\epsilon k_{xy,m})^{k})
	-(\epsilon k_{xy,m})^{k}\phi'(0)\Bigg] dy
	\nonumber \\
	&+\sum \int \Bigg[ ((1-\epsilon)f_{xy})^{k}\phi'(0)
	-\phi((f_{x}f_{y})^{k}) \nonumber \\
	&-\Big\{ ((1-\epsilon)f_{xy})^{k}-(f_{x}f_{y})^{k} \Big\}\phi'((f_{x}f_{y})^{k})\Bigg] dy.
\end{align}

\end{description}

Assumption \descref{(BP1)} ensures that the conditional contaminating densities $k_{y|x,m}$ are asymptotically singular with respect to the conditional densities $f_{y|x}$ for $x=0,1$. Assumption \descref{(BP2)} ensures that the contaminating densities $f^{m}_{y}$ are also asymptotically singular to $f_{y|x}$ for all $x$. Similarly, Assumption \descref{(BP3)} ensures that the conditional contaminating densities $k_{y|x,m}$ are asymptotically singular to the marginal density $f_{y}$ for all $x$. 

These three conditions together represent worst-case scenarios that may cause $I_{m}$ to diverge for sufficiently large $m$. Assumption \descref{(BP4)} imposes certain technical conditions on the $\phi$-function, which are generally satisfied in most cases. It also assumes that the support of $Y$ is bounded, ensuring that certain integrals remain finite even in degenerate cases when a density is replaced by zero in the divergence measure.

Assumption \descref{(BP5)} is a technical requirement for which a formal proof is not available. It states that the $\mathscr{B}$-MI between the contaminated joint density and the uncontaminated marginals should be at least as large as that between the contaminated joint and its own marginals. If \descref{(BP5)} were violated, a contradiction would arise when $X$ and $Y$ are independent under the contaminated joint densities $f^{m}_{x,y}$.

Finally, Assumption \descref{(BP6)} specifies the extremal level of contamination for which the subsequent result remains valid. Note that if $X$ and $Y$ are independent, Assumptions \descref{(BP1)} and \descref{(BP3)} become equivalent.

\begin{Theorem}
\label{ch6:first BP theorem}
Suppose that Assumptions \descref{(BP1)}–\descref{(BP6)} hold with $I_{D^{(k)}_{\phi}}< \infty$. Then the asymptotic breakdown point of the generalized mutual information $I_{D^{(k)}_{\phi}}$ is at least $ \min\Big\{\frac{1}{2}, \tilde{\epsilon}\Big\}$, where $\tilde{\epsilon}$ is defined in Assumption \descref{(BP6)}.
\end{Theorem}

Theorem \ref{ch6:first BP theorem} depends heavily on Assumption \descref{(BP6)}, and the asymptotic breakdown point can be found as an implicit solution of $\epsilon$ that satisfies Assumption \descref{(BP6)}. This is not easy in practice. However, we can do better. We can find a closed-form expression for the asymptotic breakdown point, albeit with further conditions. Next, we shall state a set of sufficient conditions for Assumption \descref{(BP6)} in the spirit of Roy et al. (2023) \cite*{roy2026asymptotic}. To do this, firstly, we define $M_{g,f}=\sum \int g^{k}\phi'(f^{k})$, and start with the following lemma that provides a lower (or upper) bound for the $\phi$-generated extended Bregman divergence. This lemma will be further used to provide a second version of the asymptotic breakdown point of $I_{D^{(k)}_{\phi}}$. 

\begin{Lemma}
	\label{ch6:lemma1: Breakdown point}
	Assume that $ M_{f, f} \gtreqless M_{g, f}$, where $g=f_{X,Y}$ and $f=f_{X}f_{Y}$. Then $D^{(k)}_{\phi}(\epsilon g, f ) \gtreqless D^{(k)}_{\phi}(\epsilon g, g)$ for $\epsilon^{k}  \lesseqgtr 1+\frac{\sum \int [\phi(g^{k})-\phi(f^{k})]}
	{M_{f,f}-M_{g,g}}$ when $g \ne f$. 
\end{Lemma}

When the divergence is defined as a limit for some tuning parameters, the same limit is taken over the appropriate assumptions, e.g., $M_{f,f} \le M_{g,f}$ and the bound of $\epsilon$ as well. Now, we state some additional sufficient conditions for Assumption \descref{(BP6)}. 
\begin{description}
	\descitem{(BP7)}
	Let the densities $g=f_{X,Y}$, $f=f_{X}f_{Y}$, $f^{m}=f^{m}_{X}f^{m}_{Y}$ and $k_{m}=K_{XY, m}$ satisfy
	\begin{gather}
		M_{f,f}  \le  M_{g,f}  \mbox{ and }
		\underset{m \to \infty}{\liminf} M_{f^{m}, f^{m}} \ge  
		\underset{m \to \infty}{\limsup} M_{k_{m}, f^{m}}.
	\end{gather}
	Also, $X, Y$ are not independent for any joint distribution under consideration. 
	
	\descitem{(BP8)} For all $\epsilon < \tilde{\epsilon}$, it is true that
	\begin{align}
		\underset{m \to \infty}{\liminf}D^{(k)}_{\phi}(\epsilon k_{m}, k_{m}) 
		&>  \underset{m \to \infty}{\limsup} D^{(k)}_{\phi}(\epsilon k_{m}, 0) \nonumber\\
		&+D^{(k)}_{\phi}((1-\epsilon)f_{X,Y}, f_{X,Y})
		-D^{(k)}_{\phi}((1-\epsilon)f_{X,Y}, 0).
	\end{align} 
\end{description}
In the next result, we shall see that the Assumptions \descref{(BP7)} and \descref{(BP8)} will together imply Assumption \descref{(BP6)}. Further, the bounds of the asymptotic breakdown point will be derived using Lemma \ref{ch6:lemma1: Breakdown point}. Let us define the following quantities:
\begin{align}
	\label{ch6:epsilon1}
	\epsilon_{1}
	&=\Bigg[
	1+\underset{m \to \infty}{\limsup}\frac{\sum\int \Big\{\phi[(k_{xy,m})^{k}]
		-\phi[ (f^{m}_{x}f^{m}_{y})^{k}]\Big\}}
	{(M_{f^{m},f^{m}} - M_{k_{m},k_{m}})} \Bigg]^{1/k},
	\\
	\label{ch6:epsilon2}  
	\epsilon_{2}
	&=1- \Bigg[ 1+ \frac{\sum \int \Big[\phi(f^{k}_{x,y})-\phi(f^{k}_{x}f^{k}_{y}) \Big]}
	{ M_{f_{X}f_{Y}, f_{X}f_{Y}} -M_{f_{X,Y}, f_{X,Y}}} \Bigg]^{1/k}. 
\end{align}

\begin{Theorem}
	\label{ch6:second BP theorem}
	Suppose the Assumptions \descref{(BP1)}-\descref{(BP5)} and \descref{(BP7)}-\descref{(BP8)} are true. Then the asymptotic breakdown point of $I_{D^{(k)}_{\phi}}$ is atleast $\min\big\{\epsilon_{1}, \epsilon_{2}, \frac{1}{2}\big\}$.
\end{Theorem}

Theorem \ref{ch6:first BP theorem} can be easily translated to the GSB divergence. Since it heavily depends on Assumption \descref{(BP6)}, we give a set of sufficient conditions that imply Assumption \descref{(BP6)}. In the view of Lemma \ref{ch6:lemma1: Breakdown point} we find that $D_{*}(\epsilon g, f) \ge D_{*}(\epsilon g, g)$ for 
\begin{align}
	\label{ch7:gsb:epsilon bound}
	\epsilon^{A} \le  1+ 
	\frac{ \sum \int \Big[ AB( e^{\beta g^{A}}-e^{\beta f^{A}})
		+A(g^{1+\alpha}-f^{1+\alpha}) \Big] }
	{\sum \int \Big[AB \{(\beta f^{A})e^{\beta f^{A}} -(\beta g^{A})e^{\beta g^{A}} \}+ (1+\alpha)(f^{1+\alpha}- g^{1+\alpha})\Big] },
\end{align}
provided it is true that
\begin{align}
	\label{ch7:gsb:epsilon bound condition}
	\sum \int \Big[AB\beta e^{\beta f^{A}}( f^{A}- g^{A})
	+(1+\alpha)(f^{A}-g^{A})f^{B} \Big]
	\ge 0 \mbox{ with } B>0, 
\end{align}
where $g=f_{X,Y}, f=f_{X}f_{Y}$, and $X,Y$ are not independent. When the divergence is defined as a limit, all the conditions as in (\ref{ch7:gsb:epsilon bound}) and (\ref{ch7:gsb:epsilon bound condition}) will be similarly restated using the appropriate limit(s). 
If $A=0$, the expression in (\ref{ch7:gsb:epsilon bound}) will become independent of $\epsilon$, hence may be discarded. So, without loss of generality, we will implicitly assume that $A > 0$ without bothering about the degenerate case. The other degenerate case occurs when $B \downarrow 0+$. Further, if $B \downarrow 0+$ but $\alpha > -1$, then $A>0$ which implies that $\underset{B \downarrow 0+}{\lim}D_{*}(\epsilon g, f) \ge \underset{B \downarrow 0+}{\lim}D_{*}(\epsilon g, g)$ for $\epsilon=0$, when $\sum \int (f^{1+\alpha}-g^{1+\alpha}) \ge 0$ and $X,Y$ are not independent. In this case, the tuning parameters belong to a sufficiently small neighbourhood containing $\mathds{S}_{5}=\Big\{(\alpha, \frac{\alpha}{1-\alpha}, \beta): \alpha > -1, \beta \in \mathds{R}\Big\}$. Similarly,  $\alpha=-1$ and $B \downarrow 0+$ imply that $A \downarrow 0+$. Consequently, the expression in (\ref{ch7:gsb:epsilon bound}) will become independent of $\epsilon$, hence will be discarded.  

\begin{Corollary}
	When $\beta=0, \alpha \ge 0$, the GSB divergence becomes the S-divergence. In that case $D_{*}(\epsilon g, f) \ge D_{*}(\epsilon g, g)$ for $\epsilon^{A} \le \frac{B}{1+\alpha}$, provided
	\begin{align}
		\label{S-div:epsilon bound condition}
		\sum \int \Big\{f^{1+\alpha}-g^{A}f^{B}\Big\} 
		\ge  0 \mbox{ with } B>0,
	\end{align}
	where $g=f_{X,Y}, f=f_{X}f_{Y}$ and $X, Y$ are not independent. If we assume that $\sum \int f^{1+\alpha} \ge \sum \int g^{1+\alpha}$, then (\ref{S-div:epsilon bound condition}) is implied by a simple application of H\H{o}lder's inequality. The degenerate case of $B \downarrow 0+$ for the S-divergence may be similarly characterized as before.  
	
\end{Corollary}
\begin{Corollary}
	For the density power divergence, $A=1, B=\alpha$. So $D_{*}(\epsilon g, f) \ge D_{*}(\epsilon g, g)$ for $\epsilon \le \frac{\alpha}{1+\alpha}$ provided 
	\begin{align}
		\label{DPD:epsilon bound condition}
		\sum \int \Big\{f^{1+\alpha}-g^{1+\alpha}\Big\} 
		\ge  0, 
	\end{align}
	where $g=f_{X,Y}, f=f_{X}f_{Y}$ and $X, Y$ are not independent. 
\end{Corollary}
\begin{Corollary}
	When $A=1+\lambda, B=-\lambda$, it becomes the power divergence family. Since $A >0$, i.e., $\lambda > -1$. Combining the degenerate case $\lambda=-1$, we get  $D_{*}(\epsilon g, f) \ge D_{*}(\epsilon g, g)$ when 
	\begin{align}
		\epsilon \le (-\lambda)^{\frac{1}{1+\lambda}}
		\mbox{ for } -1 \le \lambda \le 0,
	\end{align}
	such that $X, Y$ are not independent. Since the conditions as in Lemma \ref{ch6:lemma1: Breakdown point} are trivially true, we do not need any further restrictions. 
\end{Corollary}
\begin{Corollary}
	When $\alpha=-1, \lambda=0$, it becomes a scaled BED family with tuning parameter $\beta$. Here $A=1, B=-1$. Suppose $\beta \ne 0$ then $D_{*}(\epsilon g, f) \ge D_{*}(\epsilon f, f)$ such that   
	\begin{align}
		\label{bed:epsilon bound}
		\epsilon \le  1+ 
		\frac{ \sum \int \Big( e^{\beta g}-e^{\beta f} \Big) }
		{\beta \sum \int \Big( fe^{\beta f} -g e^{\beta g}\Big) }
		\mbox{ for }
		\beta \sum \int \Big( e^{\beta f}( f- g) \Big)
		\ge 0,
	\end{align}
	and $X,Y$ are not independent. When $\beta=0$, Lemma \ref{ch6:lemma1: Breakdown point} holds in limit as $\beta \to 0$ for $\epsilon \le \frac{1}{2}$ such that $\sum \int f^{2} > \sum \int g^{2}$ and $X, Y$ are not independent. 
\end{Corollary}

In Theorem \ref{ch6:second BP theorem} we find that the asymptotic breakdown point of $I_{D_{*}}$ is at least $\min\{\epsilon_{1}, \epsilon_{2}, \frac{1}{2}\}$  where    
\begin{align}
	\epsilon_{1} 
	&= \Bigg( 1+ \underset{m \to \infty}{\limsup}
	\frac{ \sum \int \Big[ AB\{ e^{\beta k_{xy,m}^{A}}-e^{\beta (f^{m}_{x}f^{m}_{y})^{A}}\}+A(k_{xy,m}^{1+\alpha}-(f^{m}_{x}f^{m}_{y})^{1+\alpha}) \Big] }
	{\sum \int \Big[AB\beta \{ (f^{m}_{x}f^{m}_{y})^{A}e^{\beta (f^{m}_{x}f^{m}_{y})^{A}} -( k_{xy,m}^{A})e^{\beta k_{xy,m}^{A}}\}+ (1+\alpha)((f^{m}_{x}f^{m}_{y})^{1+\alpha}- k_{xy,m}^{1+\alpha})\Big] }\Bigg)^{1/A},   \\
	\epsilon_{2} 
	&=  1-
	\Bigg(1+
	\frac{ \sum \int \Big[ AB\{e^{\beta f_{x,y}^{A}}-e^{\beta (f_{x}f_{y})^{A}}\}+A(f_{x,y}^{1+\alpha}-(f_{x}f_{y})^{1+\alpha}) \Big] }
	{\sum \int \Big[AB\{\beta (f_{x}f_{y})^{A}e^{\beta (f_{x}f_{y})^{A}} -(\beta f_{x,y}^{A})e^{\beta f_{x,y}^{A}}\}+ (1+\alpha)((f_{x}f_{y})^{1+\alpha}- f_{x,y}^{1+\alpha})\Big] }\Bigg)^{1/A}.
\end{align}
As it turns out, the result of the asymptotic breakdown point of the GSB divergence family is valid for $A >0$ and $B >0$, i.e., $-\frac{1}{1-\alpha} \le \lambda \le \frac{1}{1-\alpha}$ and $\alpha \ge -1$. 

\begin{Corollary}
	For the S-divergence family $\beta=0$, and  
	\begin{align}
		\epsilon_{1}
		= \Bigg( \frac{B}{1+\alpha}\Bigg)^{1/A}
		\mbox{ and }
		\epsilon_{2}
		=  1- 
		\Bigg(\frac{B}{1+\alpha}\Bigg)^{1/A}.
	\end{align}
	So the asymptotic breakdown point of $I_{D_{*}}$ will be 
	\begin{align}
		\min\Bigg\{ \Bigg( \frac{B}{1+\alpha}\Bigg)^{1/A}, 
		1- \Bigg(\frac{B}{1+\alpha}\Bigg)^{1/A}, \frac{1}{2} \Bigg\}
		\mbox{ with } A> 0, B >0,
	\end{align}
	under the assumptions of Theorem \ref{ch6:second BP theorem}. 
\end{Corollary}

\begin{Corollary}
	Next, we consider the power divergence family, i.e., $\alpha=\beta=0$. Then under the assumptions of Theorem \ref{ch6:second BP theorem}, the asymptotic breakdown point of $I_{D_{*}}$ will be 
	\begin{align}
		\min\Bigg\{ \big( -\lambda\big)^{1/(1+\lambda)}, 1- \big( -\lambda\big)^{1/(1+\lambda)}, \frac{1}{2} \Bigg\}
		\mbox{ for } -1 \le \lambda \le 0.
	\end{align}
\end{Corollary}

\begin{Corollary}
	When $\lambda=\beta=0$, it becomes the DPD. Then under the assumptions of Theorem \ref{ch6:second BP theorem} the asymptotic breakdown point of $I_{D_{*}}$ becomes  
	\begin{align}
		\min\Bigg\{ \frac{\alpha}{1+\alpha}, \frac{1}{1+\alpha}, \frac{1}{2} \Bigg\}. 
	\end{align}
\end{Corollary}
It is evident that as $\alpha$ increases, the asymptotic breakdown point of the generalized MI based on the DPD increases considerably, while increasing $\lambda$ should have the opposite effect of reducing the asymptotic breakdown point. Figure 1 of Roy et al. (2023) \cite*{roy2026asymptotic} depicts a wide range of $(\alpha, \lambda )$ where the resulting minimum divergence estimates based on the S-divergence would be highly robust. Observe that if $\beta \ne 0$, the asymptotic breakdown point of $I_{D_{*}}$ will depend on both the densities of $Y$ and its contaminating sequences. Hence, it is hard to characterize the tuning parameters producing highly asymptotic breakdown points, in general situations, unless specific examples are considered. The important thing to note here is that the asymptotic breakdown points do not depend on the dimension of $Y$.   

\clearpage
\newpage

\section{Numeric Studies}
\label{Numeric Studies}

\subsection{Simulation Results}
\label{Simulation Results}

\begin{description}
    
\descitem{(A)} \textbf{Uncontaminated Null:}  In this section, we report the simulation results for the following sets of models. 

\begin{description}
    \descitem{Model 0}: (Null Model)
    $Y_{0} \sim \mathcal{N}(0, 1)$,

    \descitem{Model 1} : 
    $Y_{1} \sim \mathcal{N}(0, (1+a)^{2})$ with $a=0.75$,
    
    \descitem{Model 2}: 
    $Y_{1} \sim (1-a)\mathcal{N}(-1, 1)+a \mathcal{N}(1, 1)$ with $a=0.6$.    
\end{description}

 \descitem{(B)} \textbf{Contaminated Null:} To study the robustness of type-I error, the first sample is contaminated as $(1-\epsilon)\mathcal{N}(0,1)+\epsilon \mathcal{N}(5,2)$ with $\epsilon=0.05, 0.10, 0.12, 0.15$, when the second sample comes from $\mathcal{N}(0,1)$. 

\end{description}

Samples of sizes $n_{0}=100$ and $n_{1}=100$ are respectively drawn from both the distributions over $500$ replications. The tests are conducted at $5\%$ nominal level of significance with the tuning parameters chosen as $\alpha=(0, 0.1, 0.2, 0.3, 0.4, 0.5, 0.6, 0.7, 0.8, 0.9, 1)^{T}$, $\beta=(0, -0.05)^{T}$ and $\lambda=(-0.5, -0.3, -0.2, -0.1, 0, 0.25, 0.50, 1)$.

Since the distributions in this setup have unbounded supports, we cannot use the asymptotic null distribution to simulate powers for the power divergence family. In that case, we need to use permutation tests using the algorithm of Guha et al. (2021) \cite{guha2021nonparametric}. In these cases, we consider $500$ permutations. However, the unbounded support of the continuous distribution should not cause a problem in using the asymptotic null distribution for other members of the generalized S-Bregman divergence family except for $\alpha=0$. Using the asymptotic null distribution saves a lot of computational burden. Throughout the numerical studies, the kernels and the bandwidth sequence are chosen as  
\begin{align}
\label{kernel and bandwidth}
    K(u)=\frac{3}{4}(1-u^{2})\mathds{1}\Big\{ |u| \le 1 \Big\}
    \mbox{ and }
    h_{n} =1.06 \times sd(Y)n^{-1/5},
\end{align}
where $n=n_{0}+n_{1}$ being the combined sample size. This kernel is called the Epanechnikov kernel, and it satisfies Assumption \descref{(A3)}. As it is well-known that kernel density estimates are generally robust to the choice of kernels, it is sometimes useful to use bounded and symmetric kernels. The optimum choice of bandwidth sequence for kernel density estimates depends on the problem at hand. However, a {\it rule-of-thumb} optimal bandwidth sequence $h_{n}$ considered here is best suited when the original distribution is Gaussian or symmetric, and fairly works well even if the distribution is not heavily skewed. See Silverman (2018) \cite{silverman2018density} for further details.

In Table \ref{table: unbounded model0 with beta=0} and Table \ref{table: unbounded model0 with beta=-0.05}, the observed levels of the tests under pure models are reported. We see that when $\beta=-0.05$, sometimes the tests become conservative in the sense of producing very low observed levels. When the null model is contaminated, the observed levels increase along with the amount of contamination. However, all members of the GSB divergence family except $\alpha=0$ exhibit better robustness. These values are reported in Table \ref{table: 0.05 contam unbounded model0 with beta=0} to Table \ref{table: 0.15 contam unbounded model0 with beta=-0.05}. The null hypothesis $\mathds{H}$ is rejected if the first sample comes from \descref{Model 0} but the second sample comes from \descref{Model 1} or \descref{Model 2}. The proportions of rejections (i.e., observed powers) in such cases are reported in Table \ref{table: unbounded model1 with beta=0} to Table \ref{table: unbounded model2 with beta=-0.05}. Sometimes we see that the observed powers are comparatively low for $\beta < 0$, which may be due to the slow rate of convergence of distributions. However, as we increase the sample sizes, observed powers improve.

\clearpage 
\newpage



\begin{table}
 \caption{ Proportion of Rejections under \descref{Model 0} with $\beta=0$}
\begin{tabular}
{ |p{1cm}|p{1cm}|p{1cm}|p{1cm}|p{1cm}|p{1cm}| p{1cm}|p{1cm}|p{1cm}|p{1cm}|p{1cm}|p{1cm}|}
 \hline
  \backslashbox{$\lambda$}{$\alpha$}  & $0$ & $0.1$ & $0.2$ & $0.3$ & $0.4$ & $0.5$ & $0.6$ &  $0.7$ & $0.8$ & $0.9$ & $1$ \\
   \hline

   $-0.5$  & $0.058$ & $0.054$ & $0.040$ & $0.034$ & $0.034$ & $0.036$ & $0.032$ & $0.026$ & $0.026$ & $0.022$ & $0.024$   \\
   \hline 
   
   $-0.3$  & $0.058$ & $0.040$ & $0.034$ & $0.034$ & $0.032$ &  $0.034$ & $0.030$ & $0.026$ & $0.026$ & $0.022$ & $0.024$  \\
   \hline 

   $-0.2$  & $0.058$ & $0.036$ & $0.030$ & $0.032$ & $0.032$ &  $0.034$ & $0.030$ & $0.026$ & $0.026$ & $0.022$ & $0.024$   \\
   \hline 
   
   $-0.1$  & $0.058$ & $0.034$ & $0.030$ & $0.032$ & $0.032$ &  $0.034$ & $0.030$ & $0.026$ & $0.026$ & $0.022$ & $0.024$   \\
   \hline 
   
   $0.0$  & $0.040$ & $0.036$ & $0.030$ & $0.030$ & $0.032$ &  $0.032$ & $0.030$ & $0.026$ & $0.024$ & $0.022$ & $0.024$   \\
   \hline 
   
   $0.25$  & $0.042$ & $0.098$ & $0.096$ & $0.028$ & $0.032$ &  $0.030$ & $0.028$ & $0.026$ & $0.024$ & $0.022$ & $0.024$  \\
   \hline
   
   $0.50$  & $0.042$ & $0.092$ & $0.084$ & $0.078$ & $0.032$ &  $0.026$ & $0.028$ & $0.026$ & $0.022$ & $0.022$ & $0.024$   \\
   \hline
   
   $1.0$  & $0.042$ & $0.092$ & $0.082$ & $0.076$ & $0.076$ &  $0.098$ & $0.024$ & $0.026$ & $0.022$ & $0.022$ & $0.024$   \\
   \hline
  
 \end{tabular}
\label{table: unbounded model0 with beta=0} 
\end{table}


\begin{table}
 \caption{ Proportion of Rejections under \descref{Model 0} with $\beta=-0.05$}
\begin{tabular}
{ |p{1cm}|p{1cm}|p{1cm}|p{1cm}|p{1cm}|p{1cm}| p{1cm}|p{1cm}|p{1cm}|p{1cm}|p{1cm}|p{1cm}|}
 \hline
  \backslashbox{$\lambda$}{$\alpha$}  & $0$ & $0.1$ & $0.2$ & $0.3$ & $0.4$ & $0.5$ & $0.6$ &  $0.7$ & $0.8$ & $0.9$ & $1$ \\
   \hline

   $-0.5$  & $0.028$ & $0.054$ & $0.040$ & $0.034$ & $0.034$ & $0.036$ & $0.032$ & $0.026$ & $0.026$ & $0.022$ & $0.024$   \\
   \hline 
   
   $-0.3$  & $0.018$ & $0.012$ & $0.008$ & $0.008$ & $0.006$ &  $0.004$ & $0.004$ & $0.004$ & $0.004$ & $0.004$ & $0.024$  \\
   \hline 

   $-0.2$  & $0.012$ & $0.008$ & $0.008$ & $0.006$ & $0.004$ &  $0.004$ & $0.004$ & $0.004$ & $0.004$ & $0.004$ & $0.024$   \\
   \hline 
   
   $-0.1$  & $0.012$ & $0.008$ & $0.006$ & $0.006$ & $0.004$ &  $0.004$ & $0.004$ & $0.004$ & $0.004$ & $0.004$ & $0.024$   \\
   \hline 
   
   $0.0$  & $0.036$ & $0.010$ & $0.006$ & $0.006$ & $0.004$ &  $0.004$ & $0.004$ & $0.004$ & $0.004$ & $0.004$ & $0.024$   \\
   \hline 
   
   $0.25$  & $0.034$ & $0.036$ & $0.042$ & $0.006$ & $0.004$ &  $0.004$ & $0.004$ & $0.004$ & $0.004$ & $0.004$ & $0.024$  \\
   \hline
   
   $0.50$  & $0.034$ & $0.036$ & $0.036$ & $0.040$ & $0.004$ &  $0.004$ & $0.004$ & $0.004$ & $0.004$ & $0.004$ & $0.024$   \\
   \hline
   
   $1.0$  & $0.054$ & $0.040$ & $0.036$ & $0.036$ & $0.036$ &  $0.052$ & $0.004$ & $0.004$ & $0.004$ & $0.004$ & $0.024$   \\
   \hline
  
 \end{tabular}
\label{table: unbounded model0 with beta=-0.05} 
\end{table}


\begin{table}
 \caption{ Proportion of Rejections under \descref{Model 0} with $\beta=0$ when $5\%$ obs. of the first sample come from $\mathcal{N}(5,2)$}
\begin{tabular}
{ |p{1cm}|p{1cm}|p{1cm}|p{1cm}|p{1cm}|p{1cm}| p{1cm}|p{1cm}|p{1cm}|p{1cm}|p{1cm}|p{1cm}|}
 \hline
  \backslashbox{$\lambda$}{$\alpha$}  & $0$ & $0.1$ & $0.2$ & $0.3$ & $0.4$ & $0.5$ & $0.6$ &  $0.7$ & $0.8$ & $0.9$ & $1$ \\
   \hline

   $-0.5$  & $1$ & $0.148$ & $0.094$ & $0.078$ & $0.056$ & $0.048$ & $0.044$ & $0.040$ & $0.038$ & $0.036$ & $0.038$   \\
   \hline 
   
   $-0.3$  & $1$ & $0.094$ & $0.076$ & $0.058$ & $0.052$ &  $0.046$ & $0.040$ & $0.040$ & $0.038$ & $0.036$ & $0.038$  \\
   \hline 

   $-0.2$  & $1$ & $0.084$ & $0.072$ & $0.056$ & $0.048$ &  $0.046$ & $0.040$ & $0.040$ & $0.038$ & $0.036$ & $0.038$   \\
   \hline 
   
   $-0.1$  & $1$ & $0.084$ & $0.070$ & $0.056$ & $0.048$ &  $0.046$ & $0.040$ & $0.040$ & $0.038$ & $0.036$ & $0.038$   \\
   \hline 
   
   $0.0$  & $0.996$ & $0.124$ & $0.064$ & $0.054$ & $0.046$ &  $0.044$ & $0.040$ & $0.036$ & $0.036$ & $0.036$ & $0.038$   \\
   \hline 
   
   $0.25$  & $0.998$ & $0.226$ & $0.222$ & $0.056$ & $0.046$ &  $0.044$ & $0.038$ & $0.036$ & $0.036$ & $0.036$ & $0.038$  \\
   \hline
   
   $0.50$  & $0.998$ & $0.208$ & $0.184$ & $0.164$ & $0.048$ &  $0.044$ & $0.038$ & $0.036$ & $0.036$ & $0.036$ & $0.038$   \\
   \hline
   
   $1.0$  & $0.998$ & $0.260$ & $0.188$ & $0.162$ & $0.146$ &  $0.258$ & $0.038$ & $0.036$ & $0.036$ & $0.036$ & $0.038$   \\
   \hline
  
 \end{tabular}
\label{table: 0.05 contam unbounded model0 with beta=0} 
\end{table}


\begin{table}
 \caption{ Proportion of Rejections under \descref{Model 0} with $\beta=-0.05$ when $5\%$ obs. of the first sample come from $\mathcal{N}(5,2)$}
\begin{tabular}
{ |p{1cm}|p{1cm}|p{1cm}|p{1cm}|p{1cm}|p{1cm}| p{1cm}|p{1cm}|p{1cm}|p{1cm}|p{1cm}|p{1cm}|}
 \hline
  \backslashbox{$\lambda$}{$\alpha$}  & $0$ & $0.1$ & $0.2$ & $0.3$ & $0.4$ & $0.5$ & $0.6$ &  $0.7$ & $0.8$ & $0.9$ & $1$ \\
   \hline

   $-0.5$  & $1$ & $0.148$ & $0.094$ & $0.078$ & $0.056$ & $0.048$ & $0.044$ & $0.040$ & $0.038$ & $0.036$ & $0.038$   \\
   \hline 
   
   $-0.3$  & $1$ & $0.024$ & $0.016$ & $0.014$ & $0.014$ &  $0.008$ & $0.006$ & $0.004$ & $0.004$ & $0.004$ & $0.038$  \\
   \hline 

   $-0.2$  & $1$ & $0.020$ & $0.016$ & $0.014$ & $0.010$ &  $0.008$ & $0.006$ & $0.004$ & $0.004$ & $0.004$ & $0.038$   \\
   \hline 
   
   $-0.1$  & $1$ & $0.018$ & $0.014$ & $0.012$ & $0.010$ &  $0.008$ & $0.006$ & $0.004$ & $0.004$ & $0.004$ & $0.038$   \\
   \hline 
   
   $0.0$  & $0.996$ & $0.024$ & $0.012$ & $0.012$ & $0.010$ &  $0.008$ & $0.006$ & $0.004$ & $0.004$ & $0.004$ & $0.038$   \\
   \hline 
   
   $0.25$  & $0.998$ & $0.074$ & $0.102$ & $0.012$ & $0.010$ &  $0.008$ & $0.004$ & $0.004$ & $0.004$ & $0.004$ & $0.038$  \\
   \hline
   
   $0.50$  & $0.998$ & $0.070$ & $0.062$ & $0.060$ & $0.010$ &  $0.008$ & $0.004$ & $0.004$ & $0.004$ & $0.004$ & $0.038$   \\
   \hline
   
   $1.0$  & $0.998$ & $0.086$ & $0.060$ & $0.058$ & $0.056$ &  $0.140$ & $0.004$ & $0.004$ & $0.004$ & $0.004$ & $0.038$   \\
   \hline
  
 \end{tabular}
\label{table: 0.05 contam unbounded model0 with beta=-0.05} 
\end{table}

\clearpage 
\newpage


\begin{table}
 \caption{ Proportion of Rejections under \descref{Model 0} with $\beta=0$ when $10\%$ obs. of the first sample come from $\mathcal{N}(5,2)$}
\begin{tabular}
{ |p{1cm}|p{1cm}|p{1cm}|p{1cm}|p{1cm}|p{1cm}| p{1cm}|p{1cm}|p{1cm}|p{1cm}|p{1cm}|p{1cm}|}
 \hline
  \backslashbox{$\lambda$}{$\alpha$}  & $0$ & $0.1$ & $0.2$ & $0.3$ & $0.4$ & $0.5$ & $0.6$ &  $0.7$ & $0.8$ & $0.9$ & $1$ \\
   \hline

   $-0.5$  & $1$ & $0.770$ & $0.510$ & $0.308$ & $0.214$ & $0.148$ & $0.122$ & $0.106$ & $0.100$ & $0.092$ & $0.088$   \\
   \hline 
   
   $-0.3$  & $1$ & $0.578$ & $0.390$ & $0.250$ & $0.180$ &  $0.136$ & $0.120$ & $0.106$ & $0.100$ & $0.092$ & $0.088$  \\
   \hline 

   $-0.2$  & $1$ & $0.522$ & $0.338$ & $0.236$ & $0.168$ &  $0.134$ & $0.118$ & $0.106$ & $0.100$ & $0.092$ & $0.088$   \\
   \hline 
   
   $-0.1$  & $1$ & $0.488$ & $0.310$ & $0.220$ & $0.162$ &  $0.132$ & $0.116$ & $0.106$ & $0.100$ & $0.092$ & $0.088$   \\
   \hline 
   
   $0.0$  & $1$ & $0.522$ & $0.300$ & $0.212$ & $0.146$ &  $0.130$ & $0.116$ & $0.104$ & $0.098$ & $0.092$ & $0.088$   \\
   \hline 
   
   $0.25$  & $1$ & $0.664$ & $0.402$ & $0.220$ & $0.146$ &  $0.122$ & $0.114$ & $0.104$ & $0.098$ & $0.092$ & $0.088$  \\
   \hline
   
   $0.50$  & $1$ & $0.632$ & $0.520$ & $0.428$ & $0.158$ &  $0.120$ & $0.110$ & $0.104$ & $0.096$ & $0.092$ & $0.088$   \\
   \hline
   
   $1.0$  & $1$ & $0.668$ & $0.522$ & $0.424$ & $0.336$ &  $0.502$ & $0.108$ & $0.100$ & $0.096$ & $0.090$ & $0.088$   \\
   \hline
  
 \end{tabular}
\label{table: 0.10 contam unbounded model0 with beta=0} 
\end{table}


\begin{table}
 \caption{ Proportion of Rejections under \descref{Model 0} with $\beta=-0.05$ when $10\%$ obs. of the first sample come from $\mathcal{N}(5,2)$}
\begin{tabular}
{ |p{1cm}|p{1cm}|p{1cm}|p{1cm}|p{1cm}|p{1cm}| p{1cm}|p{1cm}|p{1cm}|p{1cm}|p{1cm}|p{1cm}|}
 \hline
  \backslashbox{$\lambda$}{$\alpha$}  & $0$ & $0.1$ & $0.2$ & $0.3$ & $0.4$ & $0.5$ & $0.6$ &  $0.7$ & $0.8$ & $0.9$ & $1$ \\
   \hline

   $-0.5$  & $1$ & $0.770$ & $0.510$ & $0.308$ & $0.214$ & $0.148$ & $0.122$ & $0.106$ & $0.100$ & $0.092$ & $0.088$   \\
   \hline 
   
   $-0.3$  & $1$ & $0.214$ & $0.130$ & $0.090$ & $0.064$ &  $0.054$ & $0.042$ & $0.036$ & $0.036$ & $0.032$ & $0.088$  \\
   \hline 

   $-0.2$  & $1$ & $0.182$ & $0.114$ & $0.084$ & $0.060$ &  $0.052$ & $0.040$ & $0.036$ & $0.036$ & $0.032$ & $0.088$   \\
   \hline 
   
   $-0.1$  & $ 1$ & $0.172$ & $0.100$ & $0.076$ & $0.058$ &  $0.048$ & $0.040$ & $0.036$ & $0.036$ & $0.032$ & $0.088$   \\
   \hline 
   
   $0.0$  & $1$ & $0.198$ & $0.096$ & $0.066$ & $0.058$ &  $0.048$ & $0.040$ & $0.036$ & $0.036$ & $0.032$ & $0.088$   \\
   \hline 
   
   $0.25$  & $1$ & $0.356$ & $0.246$ & $0.070$ & $0.054$ &  $0.046$ & $0.040$ & $0.036$ & $0.036$ & $0.032$ & $0.088$  \\
   \hline
   
   $0.50$  & $1$ & $0.316$ & $0.248$ & $0.204$ & $0.054$ &  $0.040$ & $0.038$ & $0.036$ & $0.034$ & $0.032$ & $0.088$   \\
   \hline
   
   $1.0$  & $1$ & $0.354$ & $0.254$ & $0.196$ & $0.156$ &  $0.308$ & $0.036$ & $0.034$ & $0.034$ & $0.032$ & $0.088$   \\
   \hline
  
 \end{tabular}
\label{table: 0.10 contam unbounded model0 with beta=-0.05} 
\end{table}


\begin{table}
 \caption{ Proportion of Rejections under \descref{Model 0} with $\beta=0$ when $12\%$ obs. of the first sample come from $\mathcal{N}(5,2)$}
\begin{tabular}
{ |p{1cm}|p{1cm}|p{1cm}|p{1cm}|p{1cm}|p{1cm}| p{1cm}|p{1cm}|p{1cm}|p{1cm}|p{1cm}|p{1cm}|}
 \hline
  \backslashbox{$\lambda$}{$\alpha$}  & $0$ & $0.1$ & $0.2$ & $0.3$ & $0.4$ & $0.5$ & $0.6$ &  $0.7$ & $0.8$ & $0.9$ & $1$ \\
   \hline

   $-0.5$  & $1$ & $0.932$ & $0.792$ & $0.540$ & $0.360$ & $0.260$ & $0.206$ & $0.168$ & $0.148$ & $0.134$ & $0.128$   \\
   \hline
   
   $-0.3$  & $1$ & $0.848$ & $0.648$ & $0.434$ & $0.300$ &  $0.236$ & $0.194$ & $0.164$ & $0.146$ & $0.134$ & $0.128$  \\
   \hline 

   $-0.2$  & $1$ & $0.792$ & $0.588$ & $0.408$ & $0.292$ &  $0.230$ & $0.192$ & $0.164$ & $0.142$ & $0.134$ & $0.128$   \\
   \hline 
   
   $-0.1$  & $1$ & $0.750$ & $0.540$ & $0.392$ & $0.280$ &  $0.228$ & $0.184$ & $0.164$ & $0.142$ & $0.134$ & $0.128$   \\
   \hline 
   
   $0.0$  & $1$ & $0.754$ & $0.524$ & $0.364$ & $0.268$ &  $0.226$ & $0.184$ & $0.162$ & $0.142$ & $0.134$ & $0.128$   \\
   \hline 
   
   $0.25$  & $1$ & $0.882$ & $0.516$ & $0.350$ & $0.246$ &  $0.210$ & $0.182$ & $0.160$ & $0.140$ & $0.134$ & $0.128$  \\
   \hline
   
   $0.50$  & $1$ & $0.858$ & $0.756$ & $0.650$ & $0.260$ &  $0.200$ & $0.176$ & $0.160$ & $0.140$ & $0.134$ & $0.128$   \\
   \hline
   
   $1.0$  & $1$ & $0.844$ & $0.748$ & $0.618$ & $0.506$ &  $0.606$ & $0.170$ & $0.154$ & $0.140$ & $0.134$ & $0.128$   \\
   \hline
  
 \end{tabular}
\label{table: 0.12 contam unbounded model0 with beta=0} 
\end{table}


\begin{table}
 \caption{ Proportion of Rejections under \descref{Model 0} with $\beta=-0.05$ when $12\%$ obs. of the first sample come from $\mathcal{N}(5,2)$}
\begin{tabular}
{ |p{1cm}|p{1cm}|p{1cm}|p{1cm}|p{1cm}|p{1cm}| p{1cm}|p{1cm}|p{1cm}|p{1cm}|p{1cm}|p{1cm}|}
 \hline
  \backslashbox{$\lambda$}{$\alpha$}  & $0$ & $0.1$ & $0.2$ & $0.3$ & $0.4$ & $0.5$ & $0.6$ &  $0.7$ & $0.8$ & $0.9$ & $1$ \\
   \hline

   $-0.5$  & $1$ & $0.932$ & $0.792$ & $0.540$ & $0.360$ & $0.260$ & $0.206$ & $0.168$ & $0.148$ & $0.134$ & $0.128$   \\
   \hline
   
   $-0.3$  & $1$ & $0.492$ & $0.290$ & $0.168$ & $0.124$ &  $0.094$ & $0.076$ & $0.070$ & $0.068$ & $0.064$ & $0.128$  \\
   \hline 

   $-0.2$  & $1$ & $0.414$ & $0.236$ & $0.156$ & $0.118$ &  $0.094$ & $0.074$ & $0.070$ & $0.066$ & $0.064$ & $0.128$   \\
   \hline 
   
   $-0.1$  & $1$ & $0.364$ & $0.214$ & $0.150$ & $0.114$ &  $0.092$ & $0.072$ & $0.070$ & $0.066$ & $0.064$ & $0.128$   \\
   \hline 
   
   $0.0$  & $1$ & $0.392$ & $0.200$ & $0.140$ & $0.104$ &  $0.090$ & $0.072$ & $0.070$ & $0.066$ & $0.064$ & $0.128$   \\
   \hline 
   
   $0.25$  & $1$ & $0.610$ & $0.318$ & $0.130$ & $0.096$ &  $0.088$ & $0.070$ & $0.068$ & $0.066$ & $0.062$ & $0.128$  \\
   \hline
   
   $0.50$  & $1$ & $0.546$ & $0.428$ & $0.332$ & $0.104$ &  $0.076$ & $0.066$ & $0.066$ & $0.066$ & $0.062$ & $0.128$   \\
   \hline
   
   $1.0$  & $1$ & $0.582$ & $0.430$ & $0.318$ & $0.262$ &  $0.446$ & $0.068$ & $0.066$ & $0.066$ & $0.062$ & $0.128$   \\
   \hline
  
 \end{tabular}
\label{table: 0.12 contam unbounded model0 with beta=-0.05} 
\end{table}

\clearpage 
\newpage


\begin{table}
 \caption{ Proportion of Rejections under \descref{Model 0} with $\beta=0$ when $15\%$ obs. of the first sample come from $\mathcal{N}(5,2)$}
\begin{tabular}
{ |p{1cm}|p{1cm}|p{1cm}|p{1cm}|p{1cm}|p{1cm}| p{1cm}|p{1cm}|p{1cm}|p{1cm}|p{1cm}|p{1cm}|}
 \hline
  \backslashbox{$\lambda$}{$\alpha$}  & $0$ & $0.1$ & $0.2$ & $0.3$ & $0.4$ & $0.5$ & $0.6$ &  $0.7$ & $0.8$ & $0.9$ & $1$ \\
   \hline

   $-0.5$  & $1$ & $0.994$ & $0.972$ & $0.892$ & $0.710$ & $0.530$ & $0.394$ & $0.300$ & $0.254$ & $0.234$ & $0.232$   \\
   \hline
   
   $-0.3$  & $1$ & $0.978$ & $0.934$ & $0.802$ & $0.626$ &  $0.472$ & $0.350$ & $0.290$ & $0.252$ & $0.234$ & $0.232$  \\
   \hline 

   $-0.2$  & $1$ & $0.972$ & $0.906$ & $0.778$ & $0.600$ &  $0.446$ & $0.344$ & $0.284$ & $0.252$ & $0.234$ & $0.232$   \\
   \hline 
   
   $-0.1$  & $1$ & $0.962$ & $0.880$ & $0.740$ & $0.590$ &  $0.436$ & $0.334$ & $0.280$ & $0.250$ & $0.234$ & $0.232$   \\
   \hline 
   
   $0.0$  & $1$ & $0.954$ & $0.854$ & $0.708$ & $0.560$ &  $0.424$ & $0.334$ & $0.276$ & $0.248$ & $0.234$ & $0.232$  \\
   \hline
   
   $0.25$  & $1$ & $0.980$ & $0.652$ & $0.654$ & $0.522$ &  $0.404$ & $0.314$ & $0.272$ & $0.244$ & $0.234$ & $0.232$   \\
   \hline 
   
   $0.50$  & $1$ & $0.974$ & $0.954$ & $0.906$ & $0.498$ &  $0.384$ & $0.306$ & $0.268$ & $0.240$ & $0.234$ & $0.232$   \\
   \hline
   
   $1.0$  & $1$ & $0.974$ & $0.952$ & $0.892$ & $0.796$ &  $0.764$ & $0.296$ & $0.266$ & $0.240$ & $0.234$ & $0.232$   \\
   \hline
  
 \end{tabular}
\label{table: 0.15 contam unbounded model0 with beta=0} 
\end{table}


\begin{table}
 \caption{ Proportion of Rejections under \descref{Model 0} with $\beta=-0.05$ when $15\%$ obs. of the first sample come from $\mathcal{N}(5,2)$}
\begin{tabular}
{ |p{1cm}|p{1cm}|p{1cm}|p{1cm}|p{1cm}|p{1cm}| p{1cm}|p{1cm}|p{1cm}|p{1cm}|p{1cm}|p{1cm}|}
 \hline
  \backslashbox{$\lambda$}{$\alpha$}  & $0$ & $0.1$ & $0.2$ & $0.3$ & $0.4$ & $0.5$ & $0.6$ &  $0.7$ & $0.8$ & $0.9$ & $1$ \\
   \hline

   $-0.5$  & $1$ & $0.994$ & $0.972$ & $0.892$ & $0.710$ & $0.530$ & $0.394$ & $0.300$ & $0.254$ & $0.234$ & $0.232$   \\
   \hline
   
   $-0.3$  & $1$ & $0.872$ & $0.694$ & $0.456$ & $0.288$ &  $0.216$ & $0.184$ & $0.158$ & $0.142$ & $0.132$ & $0.232$  \\
   \hline 

   $-0.2$  & $1$ & $0.810$ & $0.604$ & $0.388$ & $0.262$ &  $0.208$ & $0.182$ & $0.158$ & $0.142$ & $0.132$ & $0.232$   \\
   \hline 
   
   $-0.1$  & $1$ & $0.760$ & $0.554$ & $0.352$ & $0.248$ &  $0.200$ & $0.180$ & $0.154$ & $0.142$ & $0.132$ & $0.232$   \\
   \hline 
   
   $0.0$  & $1$ & $0.742$ & $0.506$ & $0.332$ & $0.232$ &  $0.194$ & $0.174$ & $0.152$ & $0.142$ & $0.132$ & $0.232$  \\
   \hline
   
   $0.25$  & $1$ & $0.910$ & $0.472$ & $0.290$ & $0.220$ &  $0.186$ & $0.172$ & $0.148$ & $0.140$ & $0.132$ & $0.232$   \\
   \hline 
   
   $0.50$  & $1$ & $0.886$ & $0.776$ & $0.662$ & $0.216$ &  $0.180$ & $0.166$ & $0.148$ & $0.138$ & $0.132$ & $0.232$   \\
   \hline
   
   $1.0$  & $1$ & $0.858$ & $0.758$ & $0.618$ & $0.538$ &  $0.608$ & $0.166$ & $0.146$ & $0.134$ & $0.132$ & $0.232$   \\
   \hline
  
 \end{tabular}
\label{table: 0.15 contam unbounded model0 with beta=-0.05} 
\end{table}


\begin{table}
 \caption{ Proportion of Rejections under \descref{Model 1} with $\beta=0$}
\begin{tabular}
{ |p{1cm}|p{1cm}|p{1cm}|p{1cm}|p{1cm}|p{1cm}| p{1cm}|p{1cm}|p{1cm}|p{1cm}|p{1cm}|p{1cm}|}
 \hline
  \backslashbox{$\lambda$}{$\alpha$}  & $0$ & $0.1$ & $0.2$ & $0.3$ & $0.4$ & $0.5$ & $0.6$ &  $0.7$ & $0.8$ & $0.9$ & $1$ \\
   \hline

   $-0.5$  & $1$ & $0.992$ & $0.980$ & $0.958$ & $0.936$ & $0.914$ & $0.886$ & $0.862$ & $0.846$ & $0.822$ & $0.806$   \\
   \hline 
   
   $-0.3$  & $1$ & $0.986$ & $0.966$ & $0.946$ & $0.928$ &  $0.904$ & $0.886$ & $0.858$ & $0.838$ & $0.822$ & $0.806$  \\
   \hline 

   $-0.2$  & $1$ & $0.980$ & $0.960$ & $0.942$ & $0.924$ &  $0.898$ & $0.878$ & $0.858$ & $0.838$ & $0.822$ & $0.806$   \\
   \hline 
   
   $-0.1$  & $1$ & $0.978$ & $0.954$ & $0.938$ & $0.924$ &  $0.896$ & $0.874$ & $0.858$ & $0.838$ & $0.822$ & $0.806$   \\
   \hline 
   
   $0.0$  & $1$ & $0.984$ & $0.948$ & $0.932$ & $0.918$ &  $0.894$ & $0.874$ & $0.854$ & $0.838$ & $0.822$ & $0.806$   \\
   \hline 
   
   $0.25$  & $1$ & $0.986$ & $0.964$ & $0.932$ & $0.912$ &  $0.888$ & $0.866$ & $0.848$ & $0.836$ & $0.822$ & $0.806$  \\
   \hline
   
   $0.50$  & $1$ & $0.982$ & $0.974$ & $0.964$ & $0.910$ &  $0.884$ & $0.864$ & $0.846$ & $0.832$ & $0.822$ & $0.806$   \\
   \hline
   
   $1.0$  & $1$ & $0.986$ & $0.968$ & $0.960$ & $0.942$ &  $0.964$ & $0.860$ & $0.842$ & $0.830$ & $0.816$ & $0.806$   \\
   \hline
  
 \end{tabular}
\label{table: unbounded model1 with beta=0} 
\end{table}


\begin{table}
 \caption{ Proportion of Rejections under \descref{Model 1} with $\beta=-0.05$}
\begin{tabular}
{ |p{1cm}|p{1cm}|p{1cm}|p{1cm}|p{1cm}|p{1cm}| p{1cm}|p{1cm}|p{1cm}|p{1cm}|p{1cm}|p{1cm}|}
 \hline
  \backslashbox{$\lambda$}{$\alpha$}  & $0$ & $0.1$ & $0.2$ & $0.3$ & $0.4$ & $0.5$ & $0.6$ &  $0.7$ & $0.8$ & $0.9$ & $1$ \\
   \hline

   $-0.5$  & $1$ & $0.992$ & $0.980$ & $0.958$ & $0.936$ & $0.914$ & $0.886$ & $0.862$ & $0.846$ & $0.822$ & $0.806$   \\
   \hline 
   
   $-0.3$  & $1$ & $0.952$ & $0.924$ & $0.902$ & $0.866$ &  $0.840$ & $0.792$ & $0.768$ & $0.736$ & $0.708$ & $0.806$  \\
   \hline 

    $-0.2$  & $1$ & $0.944$ & $0.922$ & $0.892$ & $0.860$ &  $0.832$ & $0.792$ & $0.764$ & $0.736$ & $0.708$ & $0.806$   \\
   \hline 

   $-0.1$  & $1$ & $0.940$ & $0.912$ & $0.888$ & $0.856$ &  $0.826$ & $0.792$ & $0.764$ & $0.734$ & $0.708$ & $0.806$   \\
   \hline 
   
   $0.0$  & $1$ & $0.948$ & $0.906$ & $0.880$ & $0.854$ &  $0.818$ & $0.790$ & $0.762$ & $0.734$ & $0.708$ & $0.806$   \\
   \hline 
   
   $0.25$  & $1$ & $0.958$ & $0.938$ & $0.870$ & $0.840$ &  $0.802$ & $0.786$ & $0.750$ & $0.732$ & $0.708$ & $0.806$  \\
   \hline
   
   $0.50$  & $1$ & $0.950$ & $0.942$ & $0.926$ & $0.840$ &  $0.792$ & $0.772$ & $0.744$ & $0.732$ & $0.706$ & $0.806$   \\
   \hline
   
   $1.0$  & $1$ & $0.950$ & $0.940$ & $0.924$ & $0.910$ &  $0.942$ & $0.762$ & $0.740$ & $0.724$ & $0.702$ & $0.806$   \\
   \hline
  
 \end{tabular}
\label{table: unbounded model1 with beta=-0.05} 
\end{table}

\clearpage 
\newpage


\begin{table}
 \caption{ Proportion of Rejections under \descref{Model 2} with $\beta=0$}
\begin{tabular}
{ |p{1cm}|p{1cm}|p{1cm}|p{1cm}|p{1cm}|p{1cm}| p{1cm}|p{1cm}|p{1cm}|p{1cm}|p{1cm}|p{1cm}|}
 \hline
  \backslashbox{$\lambda$}{$\alpha$}  & $0$ & $0.1$ & $0.2$ & $0.3$ & $0.4$ & $0.5$ & $0.6$ &  $0.7$ & $0.8$ & $0.9$ & $1$ \\
   \hline

   $-0.5$  & $1$ & $0.868$ & $0.822$ & $0.780$ & $0.720$ & $0.674$ & $0.634$ & $0.604$ & $0.586$ & $0.568$ & $0.544$   \\
   \hline 
   
   $-0.3$  & $1$ & $0.830$ & $0.796$ & $0.754$ & $0.694$ &  $0.668$ & $0.628$ & $0.600$ & $0.586$ & $0.562$ & $0.544$  \\
   \hline 

   $-0.2$  & $1$ & $0.820$ & $0.786$ & $0.738$ & $0.690$ &  $0.664$ & $0.628$ & $0.600$ & $0.586$ & $0.562$ & $0.544$   \\
   \hline 
   
   $-0.1$  & $1$ & $0.816$ & $0.774$ & $0.724$ & $0.680$ &  $0.652$ & $0.622$ & $0.596$ & $0.586$ & $0.562$ & $0.544$   \\
   \hline 
   
   $0.0$  & $0.962$ & $0.882$ & $0.766$ & $0.718$ & $0.672$ &  $0.650$ & $0.618$ & $0.592$ & $0.584$ & $0.562$ & $0.544$   \\
   \hline 
   
   $0.25$  & $0.980$ & $0.914$ & $0.898$ & $0.710$ & $0.666$ &  $0.638$ & $0.608$ & $0.590$ & $0.580$ & $0.562$ & $0.544$  \\
   \hline
   
   $0.50$  & $0.980$ & $0.910$ & $0.896$ & $0.886$ & $0.674$ &  $0.630$ & $0.604$ & $0.584$ & $0.578$ & $0.562$ & $0.544$   \\
   \hline
   
   $1.0$  & $0.980$ & $0.914$ & $0.896$ & $0.880$ & $0.864$ &  $0.900$ & $0.594$ & $0.582$ & $0.572$ & $0.560$ & $0.544$   \\
   \hline
  
 \end{tabular}
\label{table: unbounded model2 with beta=0} 
\end{table}


\begin{table}
 \caption{ Proportion of Rejections under \descref{Model 2} with $\beta=-0.05$}
\begin{tabular}
{ |p{1cm}|p{1cm}|p{1cm}|p{1cm}|p{1cm}|p{1cm}| p{1cm}|p{1cm}|p{1cm}|p{1cm}|p{1cm}|p{1cm}|}
 \hline
  \backslashbox{$\lambda$}{$\alpha$}  & $0$ & $0.1$ & $0.2$ & $0.3$ & $0.4$ & $0.5$ & $0.6$ &  $0.7$ & $0.8$ & $0.9$ & $1$ \\
   \hline

   $-0.5$  & $1$ & $0.868$ & $0.822$ & $0.780$ & $0.720$ & $0.674$ & $0.634$ & $0.604$ & $0.586$ & $0.568$ & $0.544$   \\
   \hline 
   
    $-0.3$  & $1$ & $0.614$ & $0.552$ & $0.500$ & $0.454$ &  $0.420$ & $0.400$ & $0.382$ & $0.372$ & $0.360$ & $0.544$   \\
   \hline 
   
   $-0.2$  & $1$ & $0.598$ & $0.532$ & $0.488$ & $0.442$ &  $0.414$ & $0.398$ & $0.380$ & $0.372$ & $0.358$ & $0.544$  \\
   \hline

   $-0.1$  & $1$ & $0.586$ & $0.516$ & $0.476$ & $0.432$ &  $0.414$ & $0.398$ & $0.376$ & $0.372$ & $0.356$ & $0.544$   \\
   \hline 
   
   $0.0$  & $0.962$ & $0.594$ & $0.506$ & $0.462$ & $0.426$ &  $0.410$ & $0.396$ & $0.376$ & $0.370$ & $0.356$ & $0.544$   \\
   \hline 
   
   $0.25$  & $0.980$ & $0.796$ & $0.756$ & $0.450$ & $0.416$ &  $0.394$ & $0.382$ & $0.374$ & $0.366$ & $0.354$ & $0.544$  \\
   \hline
   
   $0.50$  & $0.980$ & $0.780$ & $0.760$ & $0.724$ & $0.422$ &  $0.392$ & $0.374$ & $0.368$ & $0.366$ & $0.354$ & $0.544$   \\
   \hline
   
   $1.0$  & $0.980$ & $0.778$ & $0.746$ & $0.718$ & $0.694$ &  $0.764$ & $0.374$ & $0.366$ & $0.364$ & $0.350$ & $0.544$   \\
   \hline
  
 \end{tabular}
\label{table: unbounded model2 with beta=-0.05} 
\end{table}

\subsection{Tuning Parameter Selection}
\label{Tuning Parameter Selection}

In simulation studies, we have seen that the choice of tuning parameters plays an important role in the level and power of the tests. It is therefore necessary to have a data-driven approach for optimal tuning parameter selection. We start with the test function, which is given by 
\begin{align}
    \psi(Y)=\begin{cases}
    1 & \mbox{ if } \widehat{T}_{D^{(k)}_{\phi}} > \tau_{c}, \\
    0 & \mbox{ otherwise },
    \end{cases}
\end{align}
where $Y=(Y_{0}, Y_{1})$ being the combined sample. The $0-1$ loss function for the {\it decision rule} $\psi(\cdot)$ is defined as 
\begin{align}
    L(I_{D^{(k)}_{\phi}}, \psi(Y))
    =\begin{cases}
    1 &\mbox{ if true } 
    I_{D^{(k)}_{\phi}}=0  \mbox{ and } \widehat{T}_{D^{(k)}_{\phi}} > \tau_{c},
    \\
    1 &\mbox{ if true } 
    I_{D^{(k)}_{\phi}} >0  \mbox{ and } \widehat{T}_{D^{(k)}_{\phi}} \le \tau_{c}, \\
    0 &\mbox{ otherwise. }
    \end{cases}
\end{align}
The risk function associated with the loss function is given by
\begin{align}
\label{risk function}
   R(I_{D^{(k)}_{\phi}}, \psi)
   =\mathds{E}_{Y}\Big[ L(I_{D^{(k)}_{\phi}}, \psi(Y))\Big]
  =\begin{cases}
  \mathds{P}\Big\{ \widehat{T}_{D^{(k)}_{\phi}} > \tau_{c} \Big\}
  &\mbox{ when true } I_{D^{(k)}_{\phi}}=0, \\ \\
  1-\mathds{P}\Big\{ \widehat{T}_{D^{(k)}_{\phi}} > \tau_{c} \Big\}
  &\mbox{ when true } I_{D^{(k)}_{\phi}} > 0.
  \end{cases}
\end{align}
The risk function $R(I_{D^{(k)}_{\phi}}, \psi)$ plays a similar role in the hypothesis testing problem, as the {\it mean squared error} (MSE) in the context of estimation. Consider the case of the GSB divergence. In the view of decision theory, an optimum ({\it admissible}) set of tuning parameters would be $(\alpha_{*}, \lambda_{*}, \beta_{*})$, if \begin{align}
    (\alpha_{*}, \lambda_{*}, \beta_{*})
    :=\arg\min_{\alpha, \lambda, \beta} R(I_{D^{(k)}_{\phi}}, \psi)
    \mbox{ for all true } I_{D^{(k)}_{\phi}}.
\end{align}
True $I_{D^{(k)}_{\phi}}$ is unknown to us. Note that optimum tuning parameters may not be unique. For multiple optimizers, all the tests are called {\it risk equivalent}. As a function of tuning parameters $(\alpha, \lambda, \beta)$, the risk function is defined as a map such that $ R(I_{D^{(k)}_{\phi}}, \psi): [-1,1] \times \mathds{R}^{2} \mapsto [0,1] $, whose minimizer may not always exist. Given a data set, we need a `{\it suitable}' estimate of the risk function defined in (\ref{risk function}). The probability is estimated using resamples from the combined data $Y$. The acceptance or rejection of the null hypothesis may be decided by p-value using {\it true tuning parameters}, which are unknown to us. To do that, we need to start with some {\it robust pilot tuning parameters}. Taking a cue from the simulation studies, we know that a robust pilot $(\alpha, \lambda, \beta)$ may lead to reasonably low type I and type II errors, even if the data set is pure or contaminated. The algorithm of tuning parameter selection is stated as follows. 

\begin{Algorithm} \hfill
\label{algorithm}
\begin{description}
   
    \descitem {Step 1} The first sample $Y_{0}$ of size $n_{0}$ is combined with the second sample $Y_{1}$ of size $n_{1}$ as $Y=(Y_{0}, Y_{1})$. Define a $0-1$ valued dummy variable $X$ as in (\ref{X vector}). 
    
    \descitem {Step 2} Take a robust pilot $(\alpha_{1}, \lambda_{1}, \beta_{1})$, and compute the p-value ($p_{1}$) using Theorem \ref{Asymptotic Normality under Independence}. The pilot is chosen such that the p-value should be consistent with the data.
    
    \descitem {Step 3} Reject the null hypothesis $\mathds{H}$ at $100c\%$ nominal level if $p_{1} \le c$, or accept otherwise.
    
    \descitem{Step 4} If $p_{1} \le c$, draw independent random resamples $Y_{0b}$ and $Y_{1b}$ respectively from $Y_{0}$ and $Y_{1}$, otherwise draw both $Y_{0b}$ and $Y_{1b}$ from $Y:=(Y_{0}, Y_{1})$ where $b=1,2, \ldots,B$. Using the resamples, the test statistics are computed as $T^{(b)}_{D^{(k)}_{\phi}}, b=1,\ldots,B$. 
    
    \descitem{Step 5} Calculate 
    \begin{equation}
    \label{empirical level/power}
        \hat{P}=\frac{1}{B}\sum_{b=1}^{B} \mathds{1}
        \Big\{ T^{(b)}_{D^{(k)}_{\phi}} > \tau_{c}\Big\},
    \end{equation}
   where $\tau_{c}$ being the $100(1-c)\%$ point of $\mathcal{N}(0,1)$. 
   
   \descitem{Step 6} Define 
   \begin{align}
   \label{estimate of risk function}
       \widehat{R}(I_{D^{(k)}_{\phi}}, \psi)
       =\begin{cases}
       \hat{P}    &\mbox{ if } p_{1} >  c, \\
       1-\hat{P}  &\mbox{ if } p_{1} \le c .
       \end{cases}
   \end{align}
   
    \descitem{Step 7} Compute the optimum tuning parameters as $(\alpha_{2}, \lambda_{2}, \beta_{2}):=\arg\min_{\alpha, \lambda, \beta} \widehat{R}(I_{D^{(k)}_{\phi}}, \psi)$.

\end{description}
\end{Algorithm}
It may be easy to see that when the combined sample $Y$ is fixed 
\begin{align}
    \widehat{R}(I_{D^{(k)}_{\phi}}, \psi)
    \overset{\mathds{P}}{\longrightarrow} R(I_{D^{(k)}_{\phi}}, \psi)
    \mbox{ as } B \to \infty
    \mbox{ at a true pilot } (\alpha_{1}, \lambda_{1}, \beta_{1}).
\end{align}
So $\widehat{R}(I_{D^{(k)}_{\phi}}, \psi)$ is a good proxy for $R(I_{D^{(k)}_{\phi}}, \psi)$. An empirical version of the risk function $R(I_{D^{(k)}_{\phi}}, \psi)$ is minimized over a fine grid of tuning parameters as a function of suitable robust pilot tuning parameters. Although the Algorithm \ref{algorithm} depends heavily on the choice of robust pilot tuning parameters, nevertheless, it will not cause a serious problem if they lead to a reasonable optimum set of parameters, consistent with the data set.

\subsection{Real Data Examples}
\label{Real Data Examples}

Here we take up two real data examples and consider the problem of choosing ‘optimal’ tuning parameters using Algorithm \ref{algorithm}. The primary aim here is to show that Algorithm \ref{algorithm} leads to a set of tuning parameters which are at least as good as the power divergence family, if not better. The number of resamples $B$ is chosen to be $200$. In the following examples, the continuous random variables cannot be unbounded from practical considerations. So it is safe to use the asymptotic null distributions for all combinations of the tuning parameters.  

\begin{Example}
\label{ds salaries}
(Data science salaries in $2023$)
This data set contains salaries (in USD) of the employees in the `data science' profession in the year $2023$ across different companies and countries. We are interested in seeing if the distributions of salaries among the `Research scientists' differ significantly from those of the `Data engineers'. We plot this data set in the right panel of Figure \ref{plots of data sets}. There are $1040$ data engineers and $82$ research scientists in the data set. We start with the pilot tuning parameters $(\alpha, \lambda, \beta)=(0.1, 0.5, 0)$ which gives the p-value as $0.01739 < 0.05$. So both distributions are different at $5\%$ level of significance, and this conclusion is apparently consistent with Figure \ref{plots of data sets}. In Table \ref{table: ds salaries}, the lowest risks are highlighted in green, and the tuning parameters corresponding to them are color-coded in blue. Here, the set of optimum tuning parameters does not belong to the PD family, but outside of it.

\begin{figure}
\begin{center}
 \includegraphics[width=0.8\textwidth]{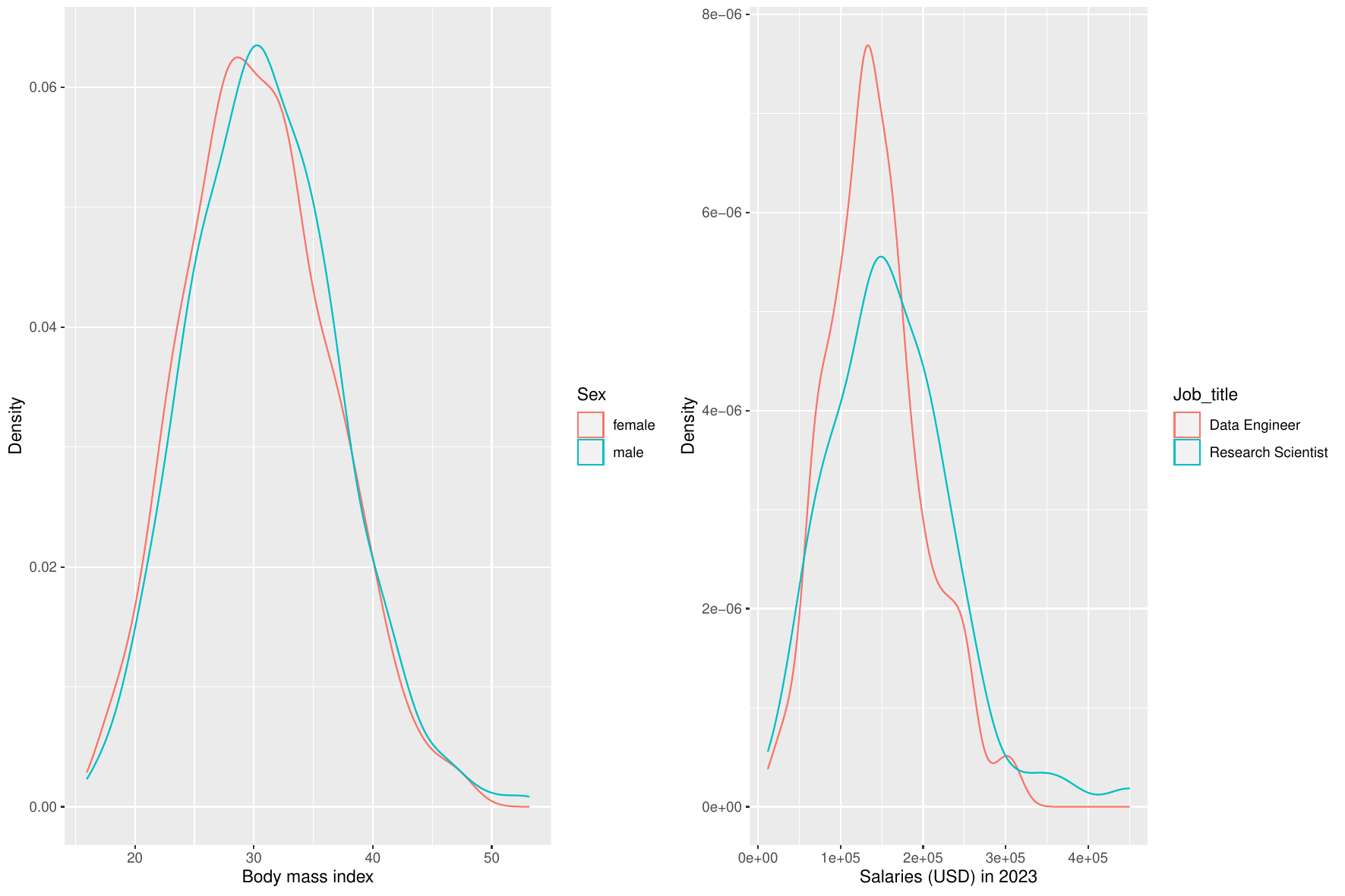}
 \end{center}
   \caption{Plots of data sets}
    \label{plots of data sets}
\end{figure}

\begin{table}[!htb]
 \caption{ Comparison of risks for different methods for the data set in Example \ref{ds salaries}}
\begin{tabular}
{ |p{2cm}p{5cm}p{5cm}p{2.5cm}|}
 \hline
       Method & $(\alpha, \lambda, \beta)$  & Risk & Power   \\
   \hline 
  PD family & $(0, -0.5, 0 )$  & $0.640$ & $0.360$ \\

             & $(0, -0.3, 0 )$  & $0.690$ & $0.310$  \\

             & $(0, -0.2, 0 )$  & $0.665$ & $0.335$ \\

             & $(0, -0.1, 0 )$  & $0.580$ & $0.420$ \\

            & $(0, 0, 0 )$  & $0.425$ & $0.575$ \\

           & $(0, 0.25, 0 )$  & $0.380$ & $0.620$  \\

           & $(0, 0.5, 0 )$  & $0.285$ & $0.715$ \\
           
           & $(0, 1, 0 )$  & $0.105$ & $0.895$ \\
   
   \hline 
   GSB family   & \textcolor{blue}{$(-0.1718, 1.0348, -0.0204)$}  & \textcolor{green}{$0.040$} & $0.960$ \\

   \hline 
 \end{tabular}
 \label{table: ds salaries}
\end{table}

\begin{table}[!htb]
 \caption{ Comparison of risks for different methods for the data set in Example \ref{BMI over sex}}
\begin{tabular}
{ |p{2cm}p{5cm}p{4cm}p{3.5cm}|}
 \hline
       Method & $(\alpha, \lambda, \beta)$  & Risk & Pr. truly accept $\mathds{H}$\\
   \hline 
  PD family & $(0, -0.5, 0 )$  & $0.020$ & $0.080 $ \\

             & $(0, -0.3, 0 )$  & $0.010$ & $0.090$  \\

             & $(0, -0.2, 0 )$  & $0.010$ & $0.090$\\

             & $(0, -0.1, 0 )$  & $0.015$ & $0.985$\\

            & $(0, 0, 0 )$  & $0.020$ & $0.980$\\

           & $(0, 0.25, 0 )$ & $0.020$ &  $0.980$\\

           & $(0, 0.5, 0 )$  & $0.020$ & $0.980$\\
           
           & $(0, 1, 0 )$   & $0.050$ & $0.950$\\
   
   \hline 
   GSB family  & \textcolor{blue}{$(0.5968, 1.2306, -0.5153)$} 
   & \textcolor{green}{$0.005$} & $0.995$ \\
               & \textcolor{blue}{$(0.1248, 0.0802, -0.4720 )$}  
               & \textcolor{green}{$0.005$} & $0.995$\\

   \hline 
 \end{tabular}
 \label{table: bmi sex}
\end{table}

\end{Example}

\begin{Example}(USA health insurance data set) 
\label{BMI over sex}
This data set contains information about insurance charges based on different categories, such as age, sex, and BMI, among many others. An important question is to test the equality of BMI between males and females who buy some insurance policies. Distributions of BMI are plotted for males and females in the left panel of Figure \ref{plots of data sets}. Here the Pilot is taken as $(\alpha, \lambda, \beta)=(0, 0.8, 0)$, which gives p-value as $0.86274 > 0.05$. In Table \ref{table: bmi sex}, the lowest risks are highlighted in green, and the tuning parameters corresponding to them are color-coded in blue. Here, the set of optimum tuning parameters does not belong to the PD family, but outside of it.
\end{Example}

\section{Concluding Remarks}
\label{Concluding Remarks}

From the discussions of this paper, we see that all the test statistics based on the generalized definition of mutual information produce consistent tests for the equality between two completely unstructured absolutely continuous distributions. However, contamination in one data set may push the observed level (Type I error) unacceptably high. In the face of contamination, the members of the GSB divergence family generally exhibit much better robustness when chosen outside the power divergence family. Also, see that observed powers at different choices of the tuning parameters do not vary much. Finally, an algorithm for choosing optimum tuning parameters for a given dataset is provided. This may be quite helpful to the applied scientists.

\section*{Acknowledgement}

This work is part of the author's PhD thesis submitted to the Indian Statistical Institute, Kolkata, India. The author gratefully acknowledges the guidance and critical input received from his supervisors, Prof. Ayanendranath Basu and Prof. Abhik Ghosh. The author also acknowledges the supportive environment he enjoys at Novartis AG.  

\section*{Disclosure statement}
No potential conflict of interest is reported by the author.

\section*{Data availability statement}
The data sets that support the findings of this study are openly available in the Kaggle repository at \url{https://www.kaggle.com/datasets/arnabchaki/data-science-salaries-2023?resource=download} and \url{https://www.kaggle.com/code/teertha/us-health-insurance-eda/data}.

\section{Appendix}
\label{appendix}
This section includes the necessary proofs of the results. 

\subsection*{Proposition \ref{proposition: mathematical properties of MI}}

\descref{(P1)} Suppose the random variables are independent. Then it follows from Definition \ref{extended bregman MI} that $I_{D^{(k)}_{\phi}}=0$. The converse follows from the strict convexity of $\phi$.   
	
	\descref{(P2)} Consider any permutation $\pi$ of the random vector $(X_{1}, \ldots , X_{m})$ as a map $\pi(X_{1}, \ldots, X_{m})=(X_{i_{1}}, \ldots , X_{i_{m}})$. Since $\pi(X_{1}, \ldots , X_{m})=\pi(Y_{1}, \ldots , Y_{m})$ implies that $X_{i_{j}}=Y_{i_{j}}$ for $j=1, 2, \ldots ,m$, the map $\pi$ is one-to-one. Also, $\pi$ is an onto map. Thus $\pi$ is bijective, therefore $\pi^{-1}$ exists. Thus we get
	\begin{align}
		f_{\pi(X_{1}, \ldots , X_{m})}(\pi(x_{1}, \ldots, x_{m}))
		&=f_{X_{1}, \ldots , X_{m}}(\pi^{-1}\pi(x_{1}, \ldots, x_{m}))
		|\pi| \nonumber \\
		&=f_{X_{1}, \ldots , X_{m}}(x_{1}, \ldots, x_{m})
		|\pi|,
	\end{align}
	where $|\pi|$ denotes the Jacobian of this map, which turns out to be $1$. Therefore we have 
	\begin{align}
		I_{D^{(k)}_{\phi}}(X_{i_{1}}, \ldots , X_{i_{m}})
		&=I_{D^{(k)}_{\phi}}(\pi(X_{1}, \ldots , X_{m}),
	\end{align}
	for any permutation $\pi$. 
	
	\descref{(P3)} Since $(X_{1}, \ldots, X_{m}) \mapsto (T_{1}(X_{1}), \ldots, T_{m}(X_{m}))$ is a one-one and onto transformation, it is a permutation. Thus, we can use the same argument as before to get the result. 
	
	\descref{(P4)}  Choose a finite constant $c \phi'(t) \ne k^{2}t^{2k-1} \phi''(t)$ for all $t$. Define         
	\begin{align}
		R(s,t)=c\phi\Big(\frac{s}{t}\Big)t
		-\Big\{\phi(s^{k})-\phi(t^{k})-(s^{k}-t^{k})\phi'(t^{k})\Big\}
		\mbox{ for } s, t >0. 
	\end{align}
	Note that $R(s,s)=0$. See that 
	\begin{align}
		\frac{\partial}{\partial s}
		R(s,t)
		&=c\phi'(s/t)-ks^{k-1}\Big\{\phi'(s^{k})-\phi'(t^{k})\Big\}, \\
		\frac{\partial^{2}}{\partial s^{2}} R(s,t)
		&=c\phi''(s/t) \frac{1}{t}-k(k-1)s^{k-2}
		\Big\{\phi'(s^{k})-\phi'(t^{k})\Big\}-(ks^{k-1})^{2}\phi''(s^{k})
		\ne 0 \mbox{ at } s\ne t. 
	\end{align}
	When $t$ is fixed, $\frac{\partial}{\partial s}R(s,t)=0$ gives a stationary point at $s=t$. See that $R(s, t)$ attains minimum or maximum at $s=t$ according to  
	\begin{gather}
		\Big[\frac{\partial^{2}}{\partial s^{2}}R(s,t)\Big]_{s=t} \gtrless 0 
		\iff 
		c \gtrless \frac{k^{2}t^{2k-1}\phi''(t^{k})}{\phi''(1)}.
	\end{gather}
	Since $s=t$ is a stationary point, $R(s,t) \gtreqless R(t,t)=0$ for all $s,t$. Substituting the functions $g(x)$ and $f(x)$ respectively for $s,t$, we get that $R(g(x), f(x)) \gtreqless 0$ when 
	\begin{gather}
		cf(x) \gtrless \frac{ k^{2}f(x)^{2k}\phi''(f(x)^{k})}{\phi''(1)}=
		\tilde{c}_{L}(x) 
		\mbox{ (say)}
	\end{gather}
	for each point $x$ such that $f(x),g(x)>0$. Since $\phi$ is twice continuously differentiable and strictly convex, $\phi''(t) >0$ for all $t > 0$. Therefore $\tilde{c}_{L}(x) >0$ on the common support of $g,f$. Define $c_{L}:=\int_{f>0}\tilde{c}_{L}(x)dx$. It is easy to see that $c \lessgtr c_{L}$ when $c \lessgtr \tilde{c}_{L}(x)$ for all $x$ such that $f,g>0$. Further, integrating $R(g(x), f(x))$ over the common support of $f,g$, we obtain  
	\begin{align}
		\label{ch6:phi and bregman ineq}
		c_{1}d_{\phi}(g, f) \le D^{(k)}_{\phi}(g,f) \le c_{2} d_{\phi}(g, f)
		\mbox{ for all } 0 < c_{1} < c_{L} < c_{2}.
	\end{align}
	Take $g=f_{X_{1},X_{2}, \ldots, X_{m}}$ and $f=f_{X_{1}}\cdots f_{X_{m}}$. Note that $c_{L}$ may be $\infty$. 
	
	Assuming $\phi_{2}=\infty$, Vajda (\cite*{vajda1972f}) shows that $d_{\phi}(g, f)=\infty$ when $f,g$ are singular. Thus (\ref{ch6:phi and bregman ineq}) implies that $D^{(k)}_{\phi}(g,f)=\infty$.

	\descref{(P5)} Assume $\phi_{2},c_{L} < \infty$ and $\phi \notin C_{B\phi}$. Then $d_{\phi}(g,f)=\phi_{2}$ (Vajda, \cite*{vajda1972f}) also $D^{(k)}_{\phi}(g, f) < \infty$. At fixed finite constants $c_{1}, c_{2}$, choose two convergent sequences $\{c_{1n}\}$ and $\{c_{2n}\}$ such that 
	\begin{align}
		0< c_{1} \le c_{1n} < c_{1(n+1)} < \cdots < c_{L}
		< \cdots < c_{2(n+1)} < c_{2n} \le c_{2}.
	\end{align}
	It is easy to see that (\ref{ch6:phi and bregman ineq}) holds for these two sequences. Define a sequence of closed and bounded intervals $C_{n}:=\Big[D^{(k)}_{\phi}/c_{2n}, D^{(k)}_{\phi}/c_{1n}\Big]$ where $d_{\phi}:=d_{\phi}(g,f)$ and $D^{(k)}_{\phi}:=D^{(k)}_{\phi}(g,f)$. Clearly 
	\begin{align}
		d_{\phi}, \frac{D^{(k)}_{\phi}}{c_{L}} \in C_{n}
		\mbox{ and }
		C_{n+1} \subset C_{n}
		\mbox{ for all } n .
	\end{align}
	So $C:=\cap_{n \ge 1} C_{n}$ will be a non-empty singleton set. Since $d_{\phi}, \frac{D^{(k)}_{\phi}}{c_{L}} \in C$, they must be the same. Hence we have $D^{(k)}_{\phi}(g,f)=c_{L}d_{\phi}(g,f)$. In this condition, Vajda (\cite*{vajda1972f}) shows that
	\begin{align}
		f \perp g \iff 
		d_{\phi}(g, f)=\phi_{2} \iff D^{(k)}_{\phi}(g, f) =c_{L}\phi_{2}.
	\end{align}
	Hence, the result follows. 

\subsection*{Lemma \ref{Lemma: Asymptotic representation of MI under independence}}

	Define $g_{x,y}(\lambda)=(f_{x}+\lambda h_{x})(f_{y}+\lambda h_{y})$ as a function of $\lambda \in [0,1]$. See that 
	\begin{align}
		g_{x,y}(0)=f_{x}f_{y},
		\mbox{ and }
		g_{x,y}(1)=\widehat{f}_{x}\widehat{f}_{y}.
	\end{align}
	Consider the $\phi$-generated extended Bregman divergence with the following arguments  
	\begin{align}
		\Lambda(\lambda)
		&=\int_{f_{y}>0} \sum_{x=0}^{1} 
		\Bigg\{\phi \Big((f_{x,y}+\lambda h_{x,y})^{k}\Big)-\phi \Big(g_{x,y}(\lambda)^{k}\Big)  
		-\Big((f_{x,y}+\lambda h_{x,y})^{k}-g_{x,y}(\lambda)^{k}\Big)\phi'\Big(g_{x,y}(\lambda)^{k}\Big)  \Bigg\}dy.
	\end{align}
	See that 
	\begin{align}
		\Lambda(1)
		&=D_{\phi}^{(k)}(\widehat{f}_{XY}, \widehat{f}_{x}\widehat{f}_{y})=\widehat{I}_{D^{(k)}_{\phi}}(X, Y), \\
		\Lambda(0)
		&=D_{\phi}^{(k)}(f_{XY}, f_{X}f_{Y})=I_{D^{(k)}_{\phi}}(X, Y).
	\end{align}
	Since $\phi$ is assumed to be four times differentiable, we can expand $\Lambda(1)$ around $\lambda=0$ up to the second-order term with Lagrange's form of the remainder as
	\begin{align}
		\label{ch6:second-order taylor expansion}
		\Lambda(1)
		&=\Lambda(0)
		+ \frac{\partial \Lambda(0)}{\partial \lambda}
		+\frac{1}{2} \cdot 
		\frac{\partial^{2} \Lambda(0)}{\partial \lambda^{2}}
		+\underbrace{ \frac{1}{6} \cdot \frac{\partial^{3} \Lambda(\lambda^{*})}{\partial \lambda^{3}}}_{R_{n}}
		\mbox{ where } \lambda^{*} \in (0, 1).
	\end{align}
	It is implicitly assumed that the partial derivatives with respect to $\lambda$ are interchangeable with integration over $y$. It is easy to verify that 
	\begin{align}
		\frac{\partial g_{x,y}(0)}{\partial \lambda}
		&=f_{x}h_{y}+f_{y}h_{x}, 
		\\
		\frac{\partial^{2}g_{x,y}(0)}{\partial \lambda^{2}}
		&=2h_{x}h_{y}.
	\end{align}
	When $\mathds{H}$ is true $f_{x,y}=f_{x}f_{y}$ for all $(x,y)$. Then we know that $I_{D^{(k)}_{\phi}}=0$, 
	\begin{align}
		\frac{\partial \Lambda(\lambda)}{\partial \lambda}
		&=\int_{f_{y}>0} \sum_{x=0}^{1}\Bigg[k(f_{x,y}+\lambda h_{x,y})^{k-1}h_{x,y}
		\phi'\Big((f_{x,y}+\lambda h_{x,y})^{k}\Big)- k g_{x,y}(\lambda)^{k-1}
		\frac{\partial g_{x,y}(\lambda)}{\partial \lambda}
		\phi'\Big(g_{x,y}(\lambda)^{k}\Big) \nonumber \\
		&  
		-\Big((f_{x,y}+\lambda h_{x,y})^{k}-(g_{x,y}(\lambda))^{k}
		\Big)
		k g_{x,y}(\lambda)^{k-1}
		\frac{\partial g_{x,y}(\lambda)}{\partial \lambda}
		\phi''\Big(g_{x,y}(\lambda)^{k}\Big) \nonumber \\
		&-\Big(k(f_{x,y}+\lambda h_{x,y})^{k-1}h_{x,y}- k(g_{x,y}(\lambda))^{k-1}\frac{\partial g_{x,y}(\lambda)}{\partial \lambda}
		\Big)\phi'\Big(g_{x,y}(\lambda)^{k}\Big)
		\Bigg] dy=0
		\mbox{ at } \lambda=0, 
	\end{align}
	
	\begin{align}
		\frac{\partial^{2}\Lambda(0)}{\partial \lambda^{2}}
		&=\int_{f_{y}>0}\sum_{x=0}^{1}\Bigg[
		k(k-1)f_{x,y}^{k-2}h_{x,y}^{2}
		\phi'(f_{x,y}^{k})+k^{2}f_{x,y}^{2k-2}h^{2}_{x,y}\phi''(f_{x,y}^{k}) \nonumber \\
		&-k(k-1)(f_{x}f_{y})^{k-2}(f_{x}h_{y}+f_{y}h_{x})^{2}\phi'\Big(f^{k}_{x}f^{k}_{y}\Big)
		-k(f_{x}f_{y})^{k-1}(2h_{x}h_{y})\phi'\Big(f^{k}_{x}f^{k}_{y}\Big) \nonumber \\
		&-k^{2}(f_{x}f_{y})^{2k-2}(f_{x}h_{y}+f_{y}h_{x})^{2}\phi''\Big(f^{k}_{x}f^{k}_{y}\Big) \nonumber \\
		&-\Bigg\{
		k f_{x,y}^{k-1}h_{x,y}-k(f_{x}f_{y})^{k-1}(f_{x}h_{y}+f_{y}h_{x})\Bigg\}k(f_{x}f_{y})^{k-1}
		(f_{x}h_{y}+f_{y}h_{x})\phi''\Big(f^{k}_{x}f^{k}_{y}\Big)
		\nonumber \\
		&-\Bigg\{ kf_{x,y}^{k-1}h_{x,y}-k(f_{x}f_{y})^{k-1} (f_{x}h_{y}+f_{y}h_{x})\Bigg\} k(f_{x}f_{y})^{k-1}(f_{x}h_{y}+f_{y}h_{x})\phi''\Big(f^{k}_{x}f^{k}_{y}\Big) \nonumber \\
		&-\Bigg\{k(k-1)f_{x,y}^{k-2}h^{2}_{x,y}
		-k(f_{x}f_{y})^{k-1}(2h_{x}h_{y})-k(k-1)(f_{x}f_{y})^{k-2}
		(f_{x}h_{y}+f_{y}h_{x})^{2}\Bigg\} \phi'\Big(f^{k}_{x}f^{k}_{y}\Big)\Bigg]dy.
	\end{align}
	Upon algebraic simplification, we obtain
	\begin{align}
		\frac{\partial^{2}\Lambda(0)}{\partial \lambda^{2}}  
		&=\int_{f_{y}>0} \sum_{x=0}^{1}k^{2}(f_{x}f_{y})^{2k}
		\phi''(f^{k}_{x}f^{k}_{y})
		\Bigg(\frac{h_{x}}{f_{x}}+\frac{h_{y}}{f_{y}}-\frac{h_{x,y}}{f_{x}f_{y}}\Bigg)^{2} dy.
	\end{align}
	The expression of the remainder $R_{n}$ may be obtained from (\ref{eq:zeta3}) with $\Delta_{x,y}, \Delta_{x}, \Delta_{y}, \theta, t$ in that expression being replaced by $h_{x,y}, h_{x}, h_{y}, \lambda, 1$ respectively. Since $f_{x}$ is a probability mass function, $|h_{x}|$ is always bounded in $x$. When $\sup_{y}f_{y} < \infty$, we get  
	\begin{align*}
		|g'_{x,y}(\lambda)|
		&< \sup_{y}|h_{y}|(1+2 \sup_{x}|h_{x}|)+\sup_{y}(f_{y}) \times \sup_{x}|h_{x}|
		=\mathcal{O}_{\mathds{P}}(\sup_{y}|h_{y}|), \\
		|g''_{x,y}(\lambda)|
		&=\mathcal{O}_{\mathds{P}}(\sup_{y}|h_{y}|^{2})
	\end{align*}
	for all $0 < \lambda < 1$. Later we shall see that $\sup_{x}|h_{x}|$, $\sup_{y}|h_{y}|$ and $\sup_{x,y}|h_{x,y}|$ are of the same stochastic order. Since $f_{x,y}, f_{y}$ are assumed to have bounded first derivatives, they are Lipschitz functions. Hence, they are uniformly continuous. In the view of (\ref{eq:zeta3}) and Assumption \descref{(A1)}, we see that $|R_{n}|$ can be bounded by the product of $\mathcal{O}_{\mathds{P}}(\sup_{y}|h_{y}|^{3})$ and an integral of finite integrand. Hence we get $R_{n}=\mathcal{O}_{\mathds{P}}(\sup_{y}|h_{y}|^{3})$. 
	
	From Fernandes and Ne\'ri (2006) \cite*{fernandes2009nonparametric} we know that $\sup_{y}|h_{y}|^{3}=o_{\mathds{P}}\Big(\frac{1}{nh^{1/2}_{n}}\Big)$ and $\sup_{x,y}|h_{x,y}|^{3}=o_{\mathds{P}}\Big(\frac{1}{nh^{1/2}_{n}}\Big)$. We already know that  $h_{x}=n^{-1}\sum_{i=1}^{n}Z_{ix}$. Using Markov's inequality, we get
	\begin{align}
		\mathds{P}\Big\{ (nh^{1/2}_{n})^{1/3}|h_{x}| > \epsilon \Big\}
		\le \frac{1}{\epsilon^{4}}
		\mathds{E}\Big\{(nh^{1/2}_{n})^{1/3}|h_{x}|\Big\}^{4}
		=\frac{1}{\epsilon^{4}}\frac{h^{2/3}_{n}}{n^{5/3}}\mathds{E}(Z_{1x})^{4}
		\longrightarrow 0 
		\mbox{ as } n \to \infty 
	\end{align}
	for fix $\epsilon>0$. This gives $|h_{x}|^{3}=o_{\mathds{P}}\Big( \frac{1}{nh^{1/2}_{n}} \Big)$ for all $x$. Since $x$ takes only $0-1$ values $\sup_{x}|h_{x}|=\max\{|h_{0}|, |h_{1}|\}=o_{\mathds{P}}\Big( \frac{1}{nh^{1/2}_{n}} \Big)$. This implies that $R_{n}=o_{\mathds{P}}\Big( \frac{1}{nh^{1/2}_{n}} \Big)$. Hence, the proof is complete.  

\subsection*{Lemma \ref{Lemma: consistency of MI}}

Recall that the true joint density at a point $(x,y)$ is given by $f_{X, Y}(x, y):=f_{x,y}$. As in the proof of Lemma \ref{Lemma: Asymptotic representation of MI under independence}, we expand $\Lambda(1)$ around $\lambda=0$ up to first-order term as 
	\begin{align}
		\Lambda(1)=\Lambda(0)+\Lambda'(0)+
		\underbrace{\frac{1}{2}\Lambda''(\lambda^{*})}_{R_{n}}
		\mbox{ for } \lambda^{*} \in (0, 1) 
	\end{align}
	with $R_{n}$ being the remainder term. We know that $\Lambda(1)=\widehat{I}_{D^{(k)}_{\phi}}$, $\Lambda(0)=I_{D^{(k)}_{\phi}}$ and
	\begin{align}
		\frac{\partial}{\partial \lambda}\Lambda(0)
		&=\int_{f_{y}>0} \sum_{x=0}^{1}\Bigg[k(f_{x,y})^{k}
		\Big\{\phi'\Big(f^{k}_{x,y}\Big)-\phi'\Big((f_{x}f_{y})^{k} \Big\} 
		\frac{h_{x,y}}{f_{x,y}}
		\nonumber \\
		&-\Big(f_{x,y}^{k}-(f_{x}f_{y})^{k} \Big)
		k (f_{x}f_{y})^{k} \Bigg(\frac{h_{x}}{f_{x}}+\frac{h_{y}}{f_{y}}\Bigg)
		\phi''\Big((f_{x}f_{y})^{k}\Big)
		\Bigg] dy. 
	\end{align}
	Write $\widehat{I}_{D^{(k)}_{\phi}}- I_{D^{(k)}_{\phi}}=\frac{1}{n}\sum_{i=1}^{n} V_{i}+R_{n}=\overline{V}_{n} +R_{n}$ where  
	\begin{align}
		V_{i}
		&=\frac{1}{2}\int_{f_{y}>0}\sum_{x=0}^{1}
		\Bigg[k (f_{x,y})^{k} 
		\Big\{\phi'(f^{k}_{x,y})-\phi'\Big( (f_{x}f_{y})^{k} \Big) \Big\}
		\Bigg( \frac{K_{h_{n}i}(y)\mathds{1}_{ix}}{h_{n}f_{x,y}} -1 \Bigg) \nonumber \\
		&-\Big(f_{x,y}^{k}-(f_{x}f_{y})^{k} \Big)
		k (f_{x}f_{y})^{k}
		\phi''\Big((f_{x}f_{y})^{k}\Big)
		\Bigg(\frac{\mathds{1}_{ix}}{f_{x}}+\frac{K_{h_{n}i}(y)}{h_{n}f_{y}}-2\Bigg)
		\Bigg] dy,.
	\end{align}
	See that 
	\begin{align}
		\mathds{E}_{f_{X,Y}}
		\Bigg(\frac{K_{h_{n}i}(y)\mathds{1}_{ix}}{h_{n}f_{x,y}} \Bigg)
		&=\frac{1}{h_{n}f_{x,y}}\sum_{t=0}^{1}\int_{w}
		K\Big(\frac{w-y}{h_{n}}\Big)
		\mathds{1}(t=x)
		f_{X,Y}(t, w)dw
		\nonumber \\
		&=\frac{1}{h_{n}f_{x,y}}\int_{u} 
		K(u)
		f_{X,Y}(x, y+uh_{n})h_{n}du
		\nonumber \\
		&=\frac{1}{f_{x,y}}\int K(u)\big[f_{x,y}+\mathcal{O}(h_{n})\big]du
		\mbox{ \Big(Assumption \descref{(A2)}\Big) }
		\nonumber \\
		&=1+\mathcal{O}(h_{n})
		\mbox{ \Big(Assumption \descref{(A3)}\Big) }.  
	\end{align}
	Similarly, we have 
	\begin{align}
		\mathds{E}_{f_{X,Y}}\Bigg(\frac{K_{h_{n}i}(y)}{h_{n}f_{y}} \Bigg)
		=1+\mathcal{O}(h_{n})
	\end{align}
	which yields $\mathds{E}_{f_{X,Y}}(V_{i})=\mathcal{O}(h_{n})$. See that $V_{i}$s are independent because the pairs $(Y_{i}, X_{i})$s are independent. So the weak laws of large numbers imply that $\overline{V}_{n} \overset{\mathds{P}}{\longrightarrow}  0$. The remainder term $R_{n}$ involves the terms which are cross products of $h_{x}, h_{y}, h_{x,y}$ of order $2$, multiplied as factors, with $\phi'', \phi'''$ inside an integration. This implies that $R_{n}=o_{\mathds{P}}\Big(\frac{1}{(nh^{1/2}_{n})^{2/3}}\Big)$, and finally 
	\begin{align}
		\widehat{I}_{D^{(k)}_{\phi}}=I_{D^{(k)}_{\phi}}
		+o_{\mathds{P}}(1)+o_{\mathds{P}}\Bigg(\frac{1}{(nh^{1/2}_{n})^{2/3}} \Bigg)
		\longrightarrow I_{D^{(k)}_{\phi}}
		\mbox{ as } n \to \infty.
	\end{align}
	This completes the proof. 

\subsection*{Lemma \ref{conditions of hall and heyde}}
    
    The null hypothesis $\mathds{H}$ is implicitly assumed throughout this proof. We see that $\mathds{E}(S_{n})=0$ for all $n$, and $\mathds{E}\Big( S_{n}| \mathcal{F}_{m}\Big)=S_{m}$ a.s.-$\mathcal{F}_{m}$ for all $m < n$. So $\{S_{n}, \mathcal{F}_{n}: n \ge 1\}$ is a zero-mean, square-integrable martingale and $\{T_{i}: i \ge 1\}$ being the martingale difference. The conditional variance $v^{2}_{n}=\sum_{i=1}^{n}\mathds{E}(T^{2}_{i}|\mathcal{F}_{i-1})$ is closely tied to the behavior of $S_{n}$. In the first result, we establish that the conditional variance $v^{2}_{n}$ may be well approximated by the squared variation $u^{2}_{n}:=\sum_{i=1}^{n}T^{2}_{i}$. The second result implies that the tails of $S_{n}$ are asymptotically negligible. 
	
	\descref{(a)} 
	First we shall show that $\mathds{E}\Big(\sum_{i=1}^{n}T^{2}_{i} \Big) \to 4 \sigma^{2}_{\phi}$. See that 
	\begin{align}
		\sum_{i=1}^{n}\mathds{E}(T^{2}_{i})
		=\frac{4}{n^{2}h^{3}_{n}}\Bigg[
		\sum_{i=1}^{n} \sum_{j < i}\mathds{E}(V^{2}_{ij})
		+ 2\sum_{i=1}^{n}\sum_{j_{1}< j_{2} < i}\mathds{E}(V_{ij_{1}}V_{ij_{2}})\Bigg]
		=\frac{4}{n^{2}h^{3}_{n}}
		\sum_{i=1}^{n} \sum_{j < i}\mathds{E}(V^{2}_{ij}),
	\end{align}
	as the second term is $0$ by independence under $\mathds{H}$. Therefore
	\begin{align}
		\label{exp of summed T squared}	
		\sum_{i=1}^{n}\mathds{E}(T^{2}_{i})
		=\frac{4}{n^{2}h^{3}_{n}}
		\sum_{i=1}^{n} \sum_{j < i}\mathds{E}(V^{2}_{ij})
		=\frac{4}{n^{2}h^{3}_{n}}\binom{n}{2}\mathds{E}(V^{2}_{ij})
		=\frac{\mathds{E}(C^{2}_{n})}{n^{2}h^{3}_{n}}
		=4\sigma^{2}_{\phi}+o(1),
	\end{align}
	where $C_{n}=nh^{3/2}_{n}S_{n}$ as in Theorem \ref{Theorem: Asymptotic null distribution}. Next we show that $Var(\sum_{i=1}^{n}T^{2}_{i}) \longrightarrow 0$. Observe that
	\begin{align}
		Var(\sum_{i=1}^{n}T^{2}_{i})
		=\sum_{i=1}^{n}Var(T^{2}_{i})+2\sum_{j < k}Cov(T^{2}_{j}, T^{2}_{k}),
	\end{align}
	where 
	\begin{align}
		Cov(T^{2}_{j}, T^{2}_{k})
		&=\frac{2^{4}}{n^{4}h^{6}_{n}}
		Cov\Big(\sum_{l < j} V^{2}_{lj}, \sum_{m < k}V^{2}_{mk}  \Big)
		\nonumber \\
		&=\frac{2^{4}}{n^{4}h^{6}_{n}}
		\Bigg[4 \sum_{l_{1}< l_{2}< j} \sum_{m_{1}< m_{2}< k}
		Cov(V_{l_{1}j} V_{l_{2}j}, V_{m_{1}k}V_{m_{2}k})
		+
		\sum_{l < j}\sum_{m < k}Cov(V^{2}_{lj}, V^{2}_{mk})
		\Bigg].
	\end{align}
	From (\ref{ch6:after variable transformation}) we will know that 
	\begin{align}
		\mathds{E}V^{2}_{ij}=\mathds{E}\Bigg[\sum_{x=0}^{1}
		\int_{f_{y}>0}
		k^{2}(f_{x}f_{y})^{2k-2}\phi''(f^{k}_{x}f^{k}_{y})
		Z_{ix}Z_{jx}
		K_{h_{n}i}(y)K_{h_{n}j}(y)dy\Bigg]^{2}=\mathcal{O}(h^{3}_{n}),
	\end{align}
	and 
	\begin{gather}
		\mathds{E}(V_{lj}V_{mk})^{2}
		\nonumber \\
		=\mathds{E}\Bigg[\sum_{x}\int C_{x,y}Z_{lx}Z_{jx}K_{h_{n}l}(y)K_{h_{n}j}(y)dy
		\times
		\sum_{x}\int C_{x,y}Z_{mx}Z_{kx}K_{h_{n}m}(y)K_{h_{n}k}(y)dy
		\Bigg]^{2}
		\nonumber \\
		=\mathds{E}\Bigg[\sum_{x_{1},x_{2}}\int 
		C_{x_{1},y_{1}}C_{x_{2},y_{2}}(Z_{lx_{1}}Z_{jx_{1}}Z_{kx_{2}}Z_{mx_{2}})
		K_{h_{n}l}(y_{1})K_{h_{n}j}(y_{1})K_{h_{n}m}(y_{2})K_{h_{n}k}(y_{2})
		dy_{1}dy_{2}
		\Bigg]^{2}
		\nonumber \\
		=\mathds{E}\Bigg[\sum_{x_{1},x_{2},x_{3},x_{4}}\int 
		C_{x_{1},y_{1}}C_{x_{2},y_{2}}C_{x_{3},y_{3}}C_{x_{4},y_{4}}
		(Z_{lx_{1}}Z_{lx_{3}})
		(Z_{jx_{1}}Z_{jx_{3}})
		(Z_{kx_{2}}Z_{kx_{4}})
		(Z_{mx_{2}}Z_{mx_{4}})
		\nonumber \\  
		\times 
		( K_{h_{n}l}(y_{1})K_{h_{n}l}(y_{3}))
		( K_{h_{n}j}(y_{1})K_{h_{n}j}(y_{3}))
		( K_{h_{n}m}(y_{2})K_{h_{n}m}(y_{4}))
		( K_{h_{n}k}(y_{2})K_{h_{n}k}(y_{4}))
		dy_{1}dy_{2}dy_{3}dy_{4}
		\Bigg]
		\nonumber \\
		=\Bigg[\sum_{x_{1},x_{2},x_{3},x_{4}}\int 
		C_{x_{1},y_{1}}C_{x_{2},y_{2}}C_{x_{3},y_{3}}C_{x_{4},y_{4}}
		\mathds{E}^{2}(Z_{lx_{1}}Z_{lx_{3}})
		\mathds{E}^{2}(Z_{kx_{2}}Z_{kx_{4}})
		\nonumber \\
		\times 
		\mathds{E}^{2}\big( K_{h_{n}l}(y_{1})K_{h_{n}l}(y_{3})\big)
		\mathds{E}^{2}\big( K_{h_{n}m}(y_{2})K_{h_{n}m}(y_{4})\big)
		dy_{1}dy_{2}dy_{3}dy_{4}
		\Bigg],
	\end{gather}
	where $C_{x,y}=k^{2}(f_{x}f_{y})^{2k-2}\phi''(f^{k}_{x}f^{k}_{y})$. See that 
	\begin{align}
		dy_{3}\mathds{E}^{2}\Big(K_{h_{n}l}(y_{1}) K_{h_{n}l}(y_{3}) \Big) 
		&=dy_{3}\Bigg[\int  K\Big(\frac{w-y_{1}}{h_{n}} \Big)
		K\Big(\frac{w-y_{3}}{h_{n}} \Big)f_{Y}(w)dw \Bigg]^{2} \nonumber \\ 
		&=-h_{n}dz\Bigg[\int  K(u)
		K(u+z)f_{Y}(y_{1}+uh_{n})h_{n}du \Bigg]^{2} 
		\nonumber \\
		&=-h^{3}_{n}dz\Bigg[\int  K(u)
		K(u+z)f_{Y}(y_{1}+uh_{n})du \Bigg]^{2}
		\nonumber \\
		&=-f^{2}_{y_{1}}h^{3}_{n}dz\Bigg[\int 
		K(u) K(u+z)du \Bigg]^{2} +o(h^{3}_{n})
		\mbox{ \Big\{Assumptions \descref{(A2)} and \descref{(A3)}\Big\} }
	\end{align}
	by transforming $u=\frac{w-y_{1}}{h_{n}}, u+z=\frac{w-y_{3}}{h_{n}}$,
	$y_{3}=y_{1}-zh_{n}$. Similarly, 
	\begin{align}
		dy_{2}\Bigg[\int  K\Big(\frac{w-y_{2}}{h_{n}} \Big)
		K\Big(\frac{w-y_{4}}{h_{n}} \Big)f_{Y}(w)dw \Bigg]^{2}  
		&=-f^{2}_{y_{4}}h^{3}_{n}dz\Bigg[\int 
		K(u) K(u+z)du  \Bigg]^{2}+o(h^{3}_{n}).
	\end{align}
	Also, see that 
	\begin{align}
		\mathds{E}(V_{lj}V_{mk})^{2}
		&=\sum_{x_{1},x_{2}, x_{3}x_{4}}
		\int C_{x_{1},y_{1}}C_{x_{2}y_{4}+zh_{n}}C_{x_{3}y_{1}-zh_{n}}C_{x_{4},y_{4}}
		\mathds{E}^{2}(Z_{lx_{1}}Z_{lx_{3}})
		\mathds{E}^{2}(Z_{kx_{2}}Z_{kx_{4}})
		\nonumber \\
		&\times f^{2}_{y_{1}}f^{2}_{y_{4}}h^{6}_{n}
		\Bigg(\int \Big[\int K(u)K(u+z)du \Big]^{2}dz\Bigg)^{2} dy_{1}dy_{4}
		\nonumber \\
		&=h^{6}_{n}
		\sum_{x_{1},x_{2}, x_{3}x_{4}}
		\int C_{x_{1},y_{1}}C_{x_{2}y_{4}}C_{x_{3}y_{1}}C_{x_{4},y_{4}}
		f^{2}_{y_{1}}f^{2}_{y_{4}}dy_{1}dy_{4}
		\nonumber \\
		& \times 
		\mathds{E}^{2}(Z_{lx_{1}}Z_{lx_{3}})
		\mathds{E}^{2}(Z_{kx_{2}}Z_{kx_{4}})
		\Bigg(\int \Big[\int K(u)K(u+z)du \Big]^{2}dz\Bigg)^{2} 
		+\mathcal{O}(h^{8}_{n}),
	\end{align}
	by expanding $C_{x_{2},y_{4}+zh_{n}}, C_{x_{3},y_{1}-zh_{n}}$ respectively around $y_{4}$ and $y_{1}$. Thus 
	\begin{align}
		\frac{1}{n^{4}h^{6}_{n}}\sum_{j < k}\sum_{l < j}\sum_{m < k}Cov(V^{2}_{lj}, V^{2}_{mk})
		&=\frac{1}{n^{4}h^{6}_{n}}\binom{n}{3}Cov(V^{2}_{lj}, V^{2}_{mk})
		\nonumber \\
		&=\frac{1}{n^{4}h^{6}_{n}}\binom{n}{3}
		\big[ \mathds{E}(V_{lj}V_{mk})^{2}-\mathds{E}^{2}(V^{2}_{lj}) \big]
		\nonumber \\
		&=\frac{1}{nh^{6}_{n}}\Big(\mathcal{O}(h^{6}_{n})
		+ \mathcal{O}(h^{6}_{n})\Big)
		\longrightarrow 0.
	\end{align}
	The order of $Cov(V_{l_{1}j} V_{l_{2}j}, V_{m_{1}k}V_{m_{2}k})$ depends on the order of $\mathds{E}(V_{l_{1}j}V_{l_{2}j} V_{m_{1}k}V_{m_{2}k})$ which depends on the following term  
	\begin{align}
		&=dy_{1}dy_{2}dy_{3}dy_{4}
		\mathds{E}\Bigg[
		K_{h_{n}l_{1}}(y_{1})K_{h_{n}j}(y_{1})
		K_{h_{n}l_{2}}(y_{2})K_{h_{n}j}(y_{2})
		K_{h_{n}m_{1}}(y_{3})K_{h_{n}k}(y_{3})
		K_{h_{n}m_{2}}(y_{4})K_{h_{n}k}(y_{4})\Bigg]
		\nonumber \\
		&=dy_{1}\mathds{E}\Bigg[K_{h_{n}j}(y_{1})K_{h_{n}j}(y_{2})\Bigg]
		\times 
		dy_{3}\mathds{E}\Bigg[K_{h_{n}k}(y_{3})K_{h_{n}k}(y_{4})\Bigg]
		\nonumber \\
		&\times dy_{2}dy_{4} \mathds{E}\Big[K_{h_{n}l_{1}}(y_{1})\Big] \mathds{E}\Big[K_{h_{n}l_{2}}(y_{2})\Big]
		\mathds{E}\Big[K_{h_{n}m_{1}}(y_{3})\Big]
		\mathds{E}\Big[K_{h_{n}m_{2}}(y_{4})\Big]
		\Bigg]
		\nonumber \\
		&=h^{4}_{n}f_{y_{1}}f_{y_{3}} \Big[\iint K(u)K(u+z)du dz\Big]^{2}
		\times f_{y_{1}}f_{y_{2}}f_{y_{3}}f_{y_{4}}h^{4}_{n}
		\nonumber \\
		&=h^{8}_{n}(f_{y_{2}+zh_{n}} f_{y_{4}+zh_{n}})^{2}f_{y_{2}}f_{y_{4}}
		\Big[\iint K(u)K(u+z)du dz\Big]^{2}
		dy_{2}dy_{4}
		\nonumber \\
		&=h^{8}_{n}(f_{y_{2}}f_{y_{4}})^{3}
		\Big[\iint K(u)K(u+z)du dz\Big]^{2}
		dy_{2}dy_{4}+\mathcal{O}(h^{10}_{n}). 
	\end{align}
	Therefore   
	\begin{align}
		\frac{2^{4}}{n^{4}h^{6}_{n}}
		\sum_{l_{1}< l_{2}< j} \sum_{m_{1}< m_{2}< k}
		Cov(V_{l_{1}j} V_{l_{2}j}, V_{m_{1}k}V_{m_{2}k})    
		&=\frac{2^{4}}{n^{4}h^{6}}Cov(V_{l_{1}j} V_{l_{2}j}, V_{m_{1}k}V_{m_{2}k})
		\sum_{l_{1}< l_{2}< j} \sum_{m_{1}< m_{2}< k} 1
		\nonumber \\
		&=\frac{2^{4}}{n^{4}h^{6}} \binom{n}{4}
		Cov(V_{l_{1}j} V_{l_{2}j}, V_{m_{1}k}V_{m_{2}k})
		\nonumber \\
		&=\mathcal{O}(h^{2}_{n}).
	\end{align}
	So we have $\sum_{j < k} Cov(T^{2}_{j}, T^{2}_{k}) \to 0$ when $n \to \infty$. For $j < i$, observe that  
	\begin{gather}
		\mathds{E}(V^{4}_{ij})
		=\mathds{E}
		\Bigg[\sum_{x}\int C_{x,y}Z_{ix}Z_{jx}K_{h_{n}i}(y)K_{h_{n}j}dy \Bigg]^{4}
		\nonumber \\
		=\mathds{E}\Bigg[
		\sum_{x_{1},x_{2},x_{3},x_{4}}\int 
		C_{x_{1},y_{1}}C_{x_{2},y_{2}}C_{x_{3},y_{3}}C_{x_{4},y_{4}}
		(Z_{ix_{1}}Z_{jx_{1}})
		(Z_{ix_{2}}Z_{jx_{2}})
		(Z_{ix_{3}}Z_{jx_{3}})
		(Z_{ix_{4}}Z_{jx_{4}})
		\nonumber \\
		\times 
		\big(K_{h_{n}i}(y_{1})K_{h_{n}j}(y_{1})\big)
		\big(K_{h_{n}i}(y_{2})K_{h_{n}j}(y_{2})\big)
		\big(K_{h_{n}i}(y_{3})K_{h_{n}j}(y_{3})\big)
		\big(K_{h_{n}i}(y_{4})K_{h_{n}j}(y_{4})\big)
		\Bigg]dy_{1}dy_{2}dy_{3}dy_{4}
		\nonumber \\
		=\sum_{x_{1},x_{2},x_{3},x_{4}}\int 
		C_{x_{1},y_{1}}C_{x_{2},y_{2}}C_{x_{3},y_{3}}C_{x_{4},y_{4}}
		\mathds{E}^{2}\big(Z_{ix_{1}}Z_{ix_{2}}Z_{ix_{3}}Z_{ix_{4}}\big)
		\nonumber \\
		\times 
		\mathds{E}^{2}\big[K_{h_{n}i}(y_{1})K_{h_{n}i}(y_{2})K_{h_{n}i}(y_{3})
		K_{h_{n}i}(y_{4})
		\big]dy_{1}dy_{2}dy_{3}dy_{4},
	\end{gather}
	where 
	\begin{align}
		&dy_{1}dy_{2}dy_{3}dy_{4} 
		\mathds{E}^{2}\big[ K_{h_{n}i}(y_{1})K_{h_{n}i}(y_{2})K_{h_{n}i}(y_{3})
		K_{h_{n}i}(y_{4}) \big]
		\nonumber \\
		&=dy_{1}dy_{2}dy_{3}dy_{4}
		\Bigg[\int
		K\Big(\frac{y_{1}-w}{h_{n}}\Big)
		K\Big(\frac{y_{2}-w}{h_{n}}\Big)
		K\Big(\frac{y_{3}-w}{h_{n}}\Big)
		K\Big(\frac{y_{4}-w}{h_{n}}\Big)f_{Y}(w)dw
		\Bigg]^{2}.
	\end{align}
	Transforming $u_{1}=\frac{y_{1}-w}{h_{n}}, u_{2}=\frac{y_{2}-w}{h_{n}}, u_{3}=\frac{y_{3}-w}{h_{n}}$ and $u_{4}=\frac{y_{4}-w}{h_{n}}$, the above expression simplifies to 
	\begin{align}
		&h^{3}_{n}dy_{1}du_{2}du_{3}du_{4}
		\Bigg[\int K(u_{1})K(u_{2})K(u_{3})K(u_{4})f_{Y}(y_{1}-u_{1}h_{n})
		h_{n}du_{1} \Bigg]^{2}
		\nonumber \\
		&=h^{5}_{n} 
		\Bigg[\int K(u_{1})K(u_{2})K(u_{3})K(u_{4})(f_{y_{1}}+\mathcal{O}(h_{n}))
		du_{1} \Bigg]^{2}du_{2}du_{3}dy_{4}dy_{1}
		\nonumber \\
		&=h^{5}_{n}
		\Bigg[\int K(u_{1})K(u_{2})K(u_{3})K(u_{4})du_{1} \Bigg]^{2}du_{2}du_{2}dy_{3} \times f^{2}_{y_{1}}dy_{1}
		+\mathcal{O}(h^{6}_{n}).
	\end{align}
	So, we get $\mathds{E}(V^{4}_{ij})=\mathcal{O}(h^{5}_{n})$ for all $j < i$. Similarly when $j_{1}< j_{2}< i$, we get 
	\begin{align}
		&\mathds{E}\big(V^{3}_{ij_{1}}V_{ij_{2}} \big)  
		\nonumber \\
		&=\mathds{E}\Bigg[\sum_{x}\int C_{x,y}Z_{ix}Z_{j_{1}x}
		K_{h_{n}i}(y) K_{h_{n}j_{1}}(y) \Bigg]^{3}
		\Bigg[\sum_{x}\int C_{x,y}Z_{ix}Z_{j_{2}x}
		K_{h_{n}i}(y) K_{h_{n}j_{2}}(y) \Bigg]
		\nonumber \\
		&=\mathds{E}\Bigg[\sum_{x_{1},x_{2},x_{3}}\int 
		C_{x_{1},y_{1}}C_{x_{2},y_{2}}C_{x_{3},y_{3}}
		(Z_{ix_{1}}Z_{j_{1}x_{1}})
		(Z_{ix_{2}}Z_{j_{1}x_{2}})
		(Z_{ix_{3}}Z_{j_{1}x_{3}})
		\nonumber \\
		&\times 
		\Big(K_{h_{n}i}(y_{1})K_{h_{n}j_{1}}(y_{1}) \Big)
		\Big(K_{h_{n}i}(y_{2})K_{h_{n}j_{1}}(y_{2}) \Big)
		\Big(K_{h_{n}i}(y_{3})K_{h_{n}j_{1}}(y_{3}) \Big)
		\Bigg]
		\nonumber \\
		& \times 
		\Bigg[\sum_{x}\int C_{x,y}Z_{ix}Z_{j_{2}x}
		K_{h_{n}i}(y) K_{h_{n}j_{2}}(y) \Bigg]
		\nonumber \\
		&=\mathds{E}\Bigg[\sum_{x_{1},x_{2},x_{3},x_{4}}\int 
		C_{x_{1},y_{1}}C_{x_{2},y_{2}}C_{x_{3},y_{3}}C_{x_{4}y_{4}}
		(Z_{ix_{1}}Z_{j_{1}x_{1}})
		(Z_{ix_{2}}Z_{j_{1}x_{2}})
		(Z_{ix_{3}}Z_{j_{1}x_{3}})
		(Z_{ix_{4}}Z_{j_{2}x_{4}})
		\nonumber \\
		&\times 
		\Big(K_{h_{n}i}(y_{1})K_{h_{n}j_{1}}(y_{1}) \Big)
		\Big(K_{h_{n}i}(y_{2})K_{h_{n}j_{1}}(y_{2}) \Big)
		\Big(K_{h_{n}i}(y_{3})K_{h_{n}j_{1}}(y_{3}) \Big)
		\Big( K_{h_{n}i}(y_{4}) K_{h_{n}j_{2}}(y_{4}) \Big) \Bigg].
	\end{align}
	This becomes 
	\begin{align}
		&\sum_{x_{1},x_{2},x_{3},x_{4}}\int 
		C_{x_{1},y_{1}}C_{x_{2},y_{2}}C_{x_{3},y_{3}}C_{x_{4}y_{4}}
		\mathds{E}(Z_{ix_{1}}Z_{ix_{2}} Z_{ix_{3}}Z_{ix_{4}}    )
		\mathds{E}(Z_{j_{1}x_{1}} Z_{j_{1}x_{2}} Z_{j_{1}x_{3}} )
		\underbrace{\mathds{E}(Z_{j_{2}x_{4}})}_{=0}
		\nonumber \\
		&\times 
		\mathds{E}\Big(K_{h_{n}i}(y_{1}) K_{h_{n}i}(y_{2})
		K_{h_{n}i}(y_{3}) K_{h_{n}i}(y_{4}) \Big)
		\mathds{E} \Big( K_{h_{n}j_{1}}(y_{1})K_{h_{n}j_{1}}(y_{2}) 
		K_{h_{n}j_{1}}(y_{3}) \Big)
		\mathds{E}\Big( K_{h_{n}j_{2}}(y_{4}) \Big) \Bigg] 
		=0.
	\end{align}
	The remaining cross-product terms will be $0$ similarly. This yields  
	\begin{align}
		\sum_{i=1}^{n}\mathds{E}(T^{4}_{i})=
		\frac{2^{4}}{n^{4}h^{6}_{n}}\sum_{i=1}^{n}
		\sum_{j < i}\mathds{E}(V^{4}_{ij})
		=\frac{2^{4}}{n^{4}h^{6}_{n}} \times \binom{n}{2} \mathcal{O}(h^{5}_{n})
		=\mathcal{O}\Big(\frac{1}{n^{2}h_{n}}\Big)=o(1),
	\end{align}
	as $nh^{1/2}_{n} \to \infty$. Therefore, we get $\sum_{i=1}^{n}Var(T^{2}_{i})\le \sum_{i=1}^{n}\mathds{E}(T^{4}_{i}) \to 0$ as $n \to \infty$. Finally, applying Chebyshev's inequality, we get 
	\begin{align}
		\mathds{P}\Bigg\{ \big|\sum_{i=1}^{n} T^{2}_{i} - 4 \sigma^{2}_{\phi}\big| > \epsilon  \Bigg\}
		\le \frac{Var\Big(\sum_{i=1}^{n}T^{2}_{i}\Big)}{\epsilon^{2}}
		\longrightarrow 0 
		\mbox{ for any } \epsilon >0.
	\end{align}
	
	\descref{(b)} For any $\epsilon > 0$,   
	\begin{align}
		\mathds{P}\Big\{ \max_{1 \le i \le n}|T_{i}|> \epsilon \Big\}
		\le 
		\frac{ \mathds{E} \big(\max_{1 \le i \le n}|T_{i}|^{4} \big)}{\epsilon^{4}}
		\le 
		\frac{\sum_{i=1}^{n}\mathds{E}(T^{4}_{i})}{\epsilon^{4}}
		=\mathcal{O}\Big(\frac{1}{\epsilon^{4}n^{2}h_{n}}\Big)
		\longrightarrow 0,
	\end{align}
	as $nh^{1/2}_{n} \to \infty$ for $n \to \infty$.

	\descref{(c)} We already know that $\sum_{i=1}^{n}T^{2}_{i} \overset{\mathds{P}}{\longrightarrow} 4\sigma^{2}_{\phi}$ and $\underset{1 \le i \le n}{\max}|T_{i}| \overset{\mathds{P}}{\longrightarrow} 0$. To prove the third part, first, we shall prove that $Z_{n}=\prod_{j=1}^{n}(1+itT_{j})$ is uniformly integrable.
	
	Fix any $M$ such that $0< 4\sigma^{2}_{\phi}< M$, and let $t_{n}:=\sum_{i=1}^{n}T^{2}_{i}$. Then there exists a subsequence $\{t_{k_{n}}: k_{n} \ge n\}$ such that $t_{k_{n}} \overset{a.s.}{\longrightarrow} 4\sigma^{2}_{\phi} < M$.  Define $A_{n}=\Big\{\omega: t_{n} \le M\Big\}$. Note that $A^{c}_{n} \subseteq A^{c}_{n+1}$ as $t_{n} \le t_{n+1}$ for all $n \ge 1$. Notice that 
	\begin{align*}
		\mathds{P}(A^{c}_{n})
		\le \mathds{P}\Big\{ t_{n} > M \mbox{ infinitely often }\Big\} 
		\le \mathds{P}\Big\{ t_{k_{n}} > M \mbox{ infinitely often }\Big\}=0
	\end{align*}
	for all $n \ge 1$. Using the Cauchy-Schwarz inequality, we obtain  
	\begin{align}
		R_{n}:=\mathds{E}\Bigg( \prod_{j=1}^{n}(1+t^{2}T^{2}_{j}) \mathds{1}(A^{c}_{n}) \Bigg)
		\le \sqrt{\mathds{E}\Bigg( \prod_{j=1}^{n}(1+t^{2}T^{2}_{j})^{2} \Bigg)\mathds{P}(A^{c}_{n})}=0.
	\end{align}
	Now, see that 
	\begin{align}
		\mathds{E} |Z_{n}|^{2}
		&=\mathds{E}\Bigg( \prod_{j=1}^{n}(1+t^{2}T^{2}_{j}) \mathds{1}(A_{n}) \Bigg)
		+\underbrace{\mathds{E}\Bigg( \prod_{j=1}^{n}(1+t^{2}T^{2}_{j}) \mathds{1}(A^{c}_{n}) \Bigg)}_{R_{n}}
		\nonumber \\
		&\le \mathds{E}\Big[\Big\{e^{t^{2}\sum_{j=1}^{n-1}T^{2}_{j}} \Big\}\Big(1+t^{2}T^{2}_{n}\Big)\mathds{1}(A_{n})\Big]
		\mbox{ ( as } e^{x} \ge 1+x)
		\nonumber \\
		& \le \Big\{e^{t^{2}M} \Big\}(1+t^{2}\mathds{E}T^{2}_{n})
		\nonumber \\
		&< \Big\{e^{t^{2}M} \Big\}\Big(1+t^{2}\sum_{n=1}^{n}\mathds{E}T^{2}_{n} \Big)
		\nonumber \\
		&=\Big\{e^{t^{2}M} \Big\}\Big(1+t^{2}4\sigma^{2}_{\phi}\Big)+o(1)
		< \infty
		\mbox{ uniformly in } n 
	\end{align}
	by (\ref{exp of summed T squared}). So $Z_{n}$ is uniformly integrable (UI). Next we shall prove that $\mathds{E}\Big(Z_{n}\mathds{1}(E)\Big) \longrightarrow \mathds{P}(E)$ for any $E \in \mathcal{F}$. Define 
	\begin{align}
		J_{n}=\begin{cases}
			\min\Big\{m \le n : \sum_{i=1}^{m}T^{2}_{i} > 2M\Big\}
			&\mbox{ if } \sum_{i=1}^{n}T^{2}_{i} > 2M ,
			\nonumber \\
			n &\mbox{ Otherwise}. 
		\end{cases}
	\end{align}
	See that $J_{n} \le n$. Recall that $\mathds{E}(S_{n}| \mathcal{F}_{m})=S_{m}$ for all $m \le n$, so 
	\begin{align*}
		\mathds{E}(T_{j}| \mathcal{F}_{j-1})
		=\mathds{E}(S_{j}- S_{j-1}| \mathcal{F}_{j-1})  \
		=\mathds{E}(S_{j}| \mathcal{F}_{j-1})-S_{j-1}
		=S_{j-1}-S_{j-1}=0. 
	\end{align*}
	See that
	\begin{align}
		\mathds{E}\Big[Z_{n}\mathds{1}(E)\Big]
		&=\mathds{E}\Bigg\{\mathds{1}(E)\prod_{j=1}^{J_{n}}(1+itT_{j})
		\prod_{j=J_{n}+1}^{n}\Big(1+it \mathds{E}\Big(T_{j}| \mathcal{F}_{j-1}\Big)\Big) \Bigg\} \nonumber \\
		&=\mathds{E}\Bigg\{\mathds{1}(E)\prod_{j=1}^{J_{n}}(1+itT_{j}) \Bigg\} \nonumber \\
		&=\mathds{P}(E)+R'_{n},
	\end{align}
	where the remainder term $R'_{n}$ consists of at most $(2^{J_{n}}-1)$ terms of the following form
	\begin{align}
		\mathds{E}
		\Big[ \mathds{1}(E) (it)^{r}T_{i_{1}}T_{i_{2}} \cdots T_{i_{r}}\Big]
	\end{align}
	such that $1 \le r \le J_{n}$ and $1 \le i_{1} \le i_{2} \le \cdots \le i_{r} \le J_{n}$. Since 
	\begin{align}
		\Big|T^{2}_{i_{1}}T^{2}_{i_{2}} \cdots T^{2}_{i_{r-1}}\Big|^{\frac{1}{r-1}}
		&\le \frac{1}{r-1}\sum_{t=1}^{r-1}T^{2}_{i_{t}} 
		\le \sum_{t=1}^{r-1}T^{2}_{i_{t}}
		\le \sum_{t=1}^{J_{n}-1}T^{2}_{i_{t}} 
		\nonumber \\
		\implies 
		\Big|T^{2}_{i_{1}}T^{2}_{i_{2}} \cdots T^{2}_{i_{r-1}}\Big|
		&\le \Bigg(\sum_{t=1}^{J_{n}-1}T^{2}_{i_{t}} \Bigg)^{r-1}
		\nonumber \\
		\implies 
		\Big|T^{2}_{i_{1}}T^{2}_{i_{2}} \cdots T^{2}_{i_{r}}\Big|
		&\le \Bigg(\sum_{t=1}^{J_{n}-1}T^{2}_{i_{t}} \Bigg)^{r-1} 
		\Bigg(\max_{1 \le i \le n}T^{2}_{i} \Bigg)
		\le  (2M)^{r-1} \Bigg( \max_{1\le i \le n}T^{2}_{i} \Bigg) 
		\nonumber \\
		\implies 
		\mathds{E}\Big|T_{i_{1}}T_{i_{2}} \cdots T_{i_{r}}\Big|
		&\le (2M)^{\frac{r-1}{2}} \mathds{E}
		\Bigg(\max_{1 \le i\le n} |T_{i}|\Bigg),
	\end{align}
	it follows that  
	\begin{align}
		|R'_{n}| \le (2^{J_{n}}-1)(2M)^{\frac{J_{n}}{2}} 
		\mathds{E}\Big(\max_{1 \le i \le n}|T_{i}|\Big)
	\end{align}
	as the remainder term contains at most $(2^{J_{n}}-1)$ terms. But, for any $\epsilon > 0$, 
	\begin{align}
		\mathds{E}\Bigg(\max_{1 \le i \le n}|T_{i}|\Bigg)
		&\le \epsilon +
		\mathds{E}\Bigg[\max_{1 \le i \le n}|T_{i}| 
		\mathds{1}( |T_{i}|> \epsilon \Big)\Bigg]\nonumber \\
		&\le  
		\epsilon + \Bigg[ \mathds{E}\Bigg(\max_{1 \le i \le n}T^{2}_{i}\Bigg) \mathds{P}\Bigg(\max_{1\le i \le n}|T_{i}|>\epsilon \Bigg)\Bigg]^{\frac{1}{2}}
		\mbox{ (Cauchy-schwarz inequality) } \nonumber \\
		&\longrightarrow \epsilon
		\mbox{ when } n \to \infty,
	\end{align}
	as \descref{(b)} holds and $\mathds{E}(T^{2}_{i})$ is finite for all $i$. Since $\epsilon>0$ is arbitrary, $\mathds{E}\Big(\max_{1 \le i \le n}|T_{i}|\Big) \to 0$. So we get $|R'_{n}| \le (2^{J_{n}}-1)(2M)^{\frac{J_{n}}{2}}\epsilon $ with $J_{n} \le n$ which yields $R'_{n} \to 0$ as $n \to \infty$. Hence we get $\mathds{E}(Z_{n}\mathds{1}(E)) \longrightarrow \mathds{P}(E)$ for any $E \in \mathcal{F}$, so $Z_{n} \longrightarrow 1$ weakly in $L_{1}$.

\subsection*{Theorem \ref{Theorem: Asymptotic null distribution}} 

Recall the notations $K_{h_{n}i}(y)$, $Z_{ix}$ and $\mathds{1}_{ix}$. Let us define 
	\begin{align}
		I_{n}
		&=\frac{1}{2}\sum_{x=0}^{1}\int_{f_{y}>0}k^{2}(f_{x}f_{y})^{2k}
		\phi''(f_{x}^{k}f^{k}_{y})\Bigg(\frac{h_{x}}{f_{x}}+
		\frac{h_{y}}{f_{y}}-\frac{h_{x,y}}{f_{x}f_{y}}\Bigg)^{2}dy, \\
		\widetilde{I}_{n} 
		&=\frac{1}{2}\sum_{x=0}^{1}\int_{f_{y}>0}k^{2}
		(f_{x}f_{y})^{2k}
		\phi''(f_{x}^{k}f^{k}_{y})\Bigg(\frac{h_{x,y}}{f_{x}f_{y}}
		-\frac{h_{y}}{f_{y}}\Bigg)^{2}dy.
	\end{align}
	From Lemma \ref{Lemma: Asymptotic representation of MI under independence} we know that  
	\begin{align}
		\widehat{I}_{D^{(k)}_{\phi}}=I_{n}
		+ o_{\mathds{P}}\Big(\frac{1}{nh^{1/2}_{n}}\Big)
		\mbox{ under } \mathds{H}. 
	\end{align}
	First we shall show that $nh^{1/2}_{n}|I_{n}-\widetilde{I}_{n}| \overset{\mathds{P}}{\longrightarrow} 0$ which implies that $I_{n}$ and $\widetilde{I}_{n}$ have the same asymptotic distribution under the null hypothesis $\mathds{H}$. See that 
	\begin{align}
		|I_{n}-\widetilde{I}_{n}|
		&\le \frac{1}{2}\sum_{x=0}^{1}\int_{f_{y}>0}
		k^{2}(f_{x}f_{y})^{2k}
		\big|\phi''(f^{k}_{x}f^{k}_{y})\big|\Bigg|\Bigg(\frac{h_{x,y}}{f_{x}f_{y}}-\frac{h_{x}}{f_{x}} -\frac{h_{y}}{f_{y}}\Bigg)^{2} 
		-\Bigg(\frac{h_{x,y}}{f_{x}f_{y}}   -\frac{h_{y}}{f_{y}}\Bigg)^{2}
		\Bigg| dy \nonumber \\
		&\le\sum_{x=0}^{1}\int_{f_{y}>0}k^{2}(f_{x}f_{y})^{2k-1}
		\big|\phi''(f^{k}_{x}f^{k}_{y})\big|
		\frac{|h_{x}|}{f_{x}}
		\Big| 2h_{x,y}-h_{x}f_{y} -2h_{y}f_{x} \Big|dy
		\nonumber \\
		& \le \frac{\sup_{x}\big|h_{x}\big|}{\min_{x} f_{x}}
		\sup_{x,y} \Big| 2h_{x,y}-h_{x}f_{y} - 2h_{y}f_{x} \Big|
		\times 
		\sum_{x=0}^{1}\int_{f_{y}>0}k^{2}(f_{x}f_{y})^{2k-1}
		\big|\phi''(f^{k}_{x}f^{k}_{y})\big|dy
		\nonumber \\
		& \le 
		\frac{2\sup_{x}\big|h_{x}\big|}{\min_{x} f_{x}} \times \Bigg\{
		\sup_{x,y} \Big| h_{x,y}| + \sup_{x}|h_{x}|\sup_{y}|f_{y}|
		+ \sup_{y}|h_{y}| \sup_{x}|f_{x} \Big|
		\Bigg\}
		\nonumber \\
		& \times 
		\sum_{x=0}^{1}\int_{f_{y}>0}k^{2}(f_{x}f_{y})^{2k-1}
		\big|\phi''(f^{k}_{x}f^{k}_{y})\big|
		dy.
	\end{align}
	We know that $\sup_{y}|h_{y}|^{3}=o_{\mathds{P}}\Big(\frac{1}{nh^{1/2}_{n}}\Big)$ and $\sup_{y}|h_{x,y}|^{3}=o_{\mathds{P}}\Big(\frac{1}{nh^{1/2}_{n}}\Big)$ uniformly. Further, by Assumption \descref{(A1)} , $\sum_{x=0}^{1}k^{2}(f_{x}f_{y})^{2k-1} \phi''(f^{k}_{x}f^{k}_{y})$ is uniformly bounded by integrable function. Also, see that  
	\begin{align}
		\mathds{P}\Big\{(nh^{1/2}_{n})^{2/3}|h_{x}| \ge \epsilon \Big\}
		\le \frac{1}{\epsilon^{4}}
		\mathds{E}\Big((nh^{1/2}_{n})^{2/3}h_{x}\Big)^{4}
		&=\frac{1}{\epsilon^{4}} \cdot \frac{h^{4/3}_{n}}{n^{1/3}}
		\cdot \mathds{E}[Z_{1x}]^{4}
		\to 0
		\mbox{ as }
		n \to \infty,
	\end{align}
	for any fixed $\epsilon>0$. Thus we get $(nh^{1/2}_{n})^{2/3}|h_{x}|=o_{\mathds{P}}(1)$ and $nh^{1/2}_{n}|I_{n}-\widetilde{I}_{n}| \overset{\mathds{P}}{\longrightarrow} 0$. 
	
	Next, we shall find the asymptotic distribution of $\widetilde{I}_{n}$. See that 
	\begin{align}
		\widetilde{I}_{n}
		&=\frac{1}{2}\sum_{x=0}^{1}\int_{f_{y}>0}k^{2}(f_{x}f_{y})^{2k}\phi''(f^{k}_{x}f^{k}_{y})\Bigg[\frac{1}{nh_{n}}
		\sum_{i=1}^{n}
		\Bigg\{\frac{K_{h_{n}i}(y)1_{ix}}{f_{x}f_{y}}-\frac{K_{h_{n}i}(y)}{f_{y}} \Bigg\} \Bigg]^{2}dy \nonumber \\
		&=\frac{1}{2(nh_{n})^{2}}\sum_{x=0}^{1}\int_{f_{y}>0}k^{2}
		(f_{x}f_{y})^{2k-2}\phi''(f^{k}_{x}f^{k}_{y})
		\Bigg[\sum_{i=1}^{n}Z^{2}_{ix}K^{2}_{h_{n}i}(y)
		+2 \sum_{i=1}^{n}\sum_{j < i}
		Z_{ix}Z_{jx}K_{h_{n}i}(y)K_{h_{n}j}(y)
		\Bigg]dy \nonumber \\
		&=\frac{B_{n}+C_{n}}{2(nh_{n})^{2}},
	\end{align}
	where 
	\begin{align}
		B_{n}
		&=\sum_{i=1}^{n} \sum_{x=0}^{1}\int_{f_{y}>0}k^{2}(f_{x}f_{y})^{2k-2}
		\phi''(f^{k}_{x}f^{k}_{y})
		Z^{2}_{ix}
		K^{2}_{h_{n}i}(y) dy, \\
		C_{n}
		&=2\sum_{i=1}^{n}\sum_{j < i}\sum_{x=0}^{1}
		\int_{f_{y}>0}
		k^{2}(f_{x}f_{y})^{2k-2}
		\phi''(f^{k}_{x}f^{k}_{y})
		Z_{ix}Z_{jx}K_{h_{n}i}(y)K_{h_{n}j}(y)dy.
	\end{align}
	Notice that under the null hypothesis $\mathds{E}(Z^{2}_{ix})=f_{x}(1-f_{x})$ and  
	\begin{align}
		\mathds{E}(K^{2}_{h_{n}i}(y))
		&=\int_{w}K^{2}\Big(\frac{w-y}{h_{n}}\Big)f_{Y}(w)dw \nonumber \\
		&=h_{n}\int_{u}K^{2}(u)f_{Y}(y+uh_{n})du \nonumber \\
		&=h_{n}\Big\{f_{y}\int_{u}K^{2}(u)du+\mathcal{O}(h_{n})\Big\} \mbox{ \Big\{ Assumptions \descref{(A2)} and \descref{(A3)} \Big\} } \nonumber \\
		&=h_{n}f_{y}\int_{u}K^{2}(u)du+\mathcal{O}(h^{2}_{n}). 
	\end{align}
	Thus, we find that  
	\begin{align}
		\mathds{E}(B_{n})
		&=nh_{n}\int_{u}K^{2}(u)du\int_{f_{y}>0}\sum_{x=0}^{1}k^{2}
		(f_{x}f_{y})^{2k-1}\phi''(f^{k}_{x}f^{k}_{y})(1-f_{x})dy
		+\mathcal{O}(nh^{2}_{n})  \\
		&=2nh_{n}\mu_{\phi}+o(nh_{n}).
	\end{align}
	Similarly, under independence, 
	\begin{align}
		Var(B_{n})
		&=\sum_{i=1}^{n} Var\Bigg[\sum_{x=0}^{1}\int_{f_{y}>0}
		\underbrace{k^{2}(f_{x}f_{y})^{2k-2}\phi''(f^{k}_{x}f^{k}_{y})}_{C_{x,y}}
		Z^{2}_{ix}K^{2}_{h_{n}i}(y)dy    
		\Bigg]
		\nonumber \\
		&\le \sum_{i=1}^{n} \mathds{E}\Bigg[\sum_{x=0}^{1}
		\int
		C_{x,y}
		Z^{2}_{ix}K^{2}_{h_{n}i}(y) dy   
		\Bigg]^{2}
		\nonumber \\
		&=n\mathds{E}\Bigg[\sum_{x=0}^{1}\int
		C_{x,y}
		Z^{2}_{ix}K^{2}_{h_{n}i}(y)  dy  
		\Bigg]^{2}
		\nonumber \\
		&=n\mathds{E}\Bigg[\sum_{x_{1},x_{2}}\iint 
		C_{x_{1},y_{1}}C_{x_{2}, y_{2}}
		Z^{2}_{ix_{1}}K^{2}_{h_{n}i}(y_{1})
		Z^{2}_{ix_{2}}K^{2}_{h_{n}i}(y_{2}) dy_{1}dy_{2}
		\Bigg]
		\nonumber \\
		&=n\sum_{x_{1}, x_{2}}\iint 
		C_{x_{1},y_{1}}C_{x_{2}, y_{2}}
		\mathds{E}\Big(Z^{2}_{ix_{1}}Z^{2}_{ix_{2}}\Big)
		\mathds{E}\Big(K^{2}_{h_{n}i}(y_{1}) K^{2}_{h_{n}i}(y_{2}) \Big)dy_{1}dy_{2}.
	\end{align}
	Also, it holds that  
	\begin{align}
		dy_{1}dy_{2}\mathds{E}\Big(K^{2}_{h_{n}i}(y_{1})K^{2}_{h_{n}y_{2}} \Big)
		&=dy_{1}dy_{2}\int
		\Bigg[K^{2}\Big(\frac{w-y_{1}}{h_{n}}\Big)
		K^{2}\Big( \frac{y_{2}-w}{h_{n}}\Big)f_{Y}(w)dw
		\Bigg]
		\nonumber \\
		&=h_{n}dzdy_{1} \Big[\int K^{2}(u)K^{2}(u+z)f_{Y}(y_{1}+uh_{n})h_{n}du \Big]
		\nonumber \\
		&=h^{2}_{n}(f_{y_{1}}dy_{1})
		\times 
		\Big(\int  K^{2}(u)K^{2}(u+z)du \Big) dz+\mathcal{O}(h^{3}_{n})
	\end{align}
	by Assumption \descref{(A3)}. Thus we have  
	\begin{gather}
		Var(B_{n})
		\le nh^{2}_{n}
		\sum_{x_{1},x_{2}}
		\mathds{E}(Z^{2}_{ix_{1}}Z^{2}_{ix_{2}})
		\int 
		C_{x_{1},y_{1}}
		C_{x_{2},y_{1}+h_{n}(u+z) }
		f_{y_{1}}dy_{1}
		\nonumber \\
		\times 
		\iint K^{2}(u)K^{2}(u+z)dudz 
		=\mathcal{O}(nh^{2}_{n}).
	\end{gather}
	Next see that $\mathds{E}(C_{n})=0$ and 
	\begin{gather}
		\mathds{E}(C^{2}_{n})=
		4\binom{n}{2}\mathds{E}\Bigg[\sum_{x=0}^{1}\int k^{2}
		(f_{x}f_{y})^{2k-2} \phi''(f^{k}_{x}f^{k}_{y})Z_{ix}Z_{jx}
		K_{h_{n}i}(y) K_{h_{n}j}(y) dy \Bigg]^{2} \nonumber \\
		=4\binom{n}{2}k^{4}\iint \Bigg[\sum_{x=0}^{1}(f_{x}f_{y_{1}})^{2k-2}
		(f_{x}f_{y_{2}})^{2k-2}\phi''(f^{k}_{x}f^{k}_{y_{1}})
		\phi''(f^{k}_{x}f^{k}_{y_{2}})
		\mathds{E}^{2}(Z^{2}_{ix})
		\nonumber \\
		\label{ch6:var of C_n eq0}
		+2(f_{x_{0}}f_{y_{1}})^{2k-2}
		(f_{x_{1}}f_{y_{2}})^{2k-2}\phi''(f^{k}_{x_{0}}f^{k}_{y_{1}})
		\phi''(f^{k}_{x_{1}}f^{k}_{y_{2}})
		\mathds{E}^{2}(Z_{ix_{0}}Z_{ix_{1}})\Bigg]
		\mathds{E}^{2}\Big(K_{h_{n}i}(y_{1})K_{h_{n}i}(y_{2})\Big)
		dy_{1}dy_{2}.
	\end{gather}
	Observe that $Z_{ix_{1}}=-Z_{ix_{2}}$ when $x_{1} \ne x_{2}$, and  
	\begin{align}
		\mathds{E}\Big(Z_{ix_{1}}Z_{ix_{2}}\Big)
		=\begin{cases}
			-f_{x_{1}}(1-f_{x_{1}}) &\mbox{ if } x_{1}\ne x_{2},\\
			f_{x_{1}}(1-f_{x_{1}})  &\mbox{ if } x_{1}=x_{2}.
		\end{cases}
	\end{align}
	Thus $\mathds{E}^{2}\Big[Z_{ix_{1}}Z_{ix_{2}}\Big]=
	f^{2}_{x_{1}}(1-f_{x_{1}})^{2}$. See that 
	\begin{align}
		\label{ch6:variable transformation}
		dy_{2}\mathds{E}^{2}\Big( K_{h_{n}i}(y_{1})K_{h_{n}i}(y_{2})\Big)
		=dy_{2}\Bigg[\int_{w}K\Big(\frac{y_{1}-w}{h_{n}} \Big)
		K\Big(\frac{y_{2}-w}{h_{n}} \Big)f_{Y}(w)dw\Bigg]^{2}. 
	\end{align}
	Substituting $\frac{y_{1}-w}{h_{n}}=z, \frac{y_{2}-w}{h_{n}}=z+u$ and $y_{2}=y_{1}+uh_{n}$ in (\ref{ch6:variable transformation}), we obtain   
	\begin{align}
		&h_{n}du
		\Bigg[h_{n}\int_{w}K(z)K(z+u)f_{Y}(y_{1}-zh_{n})dz\Bigg]^{2}
		\nonumber \\
		&=h^{3}_{n}du
		\Bigg[\int_{w}K(z)K(z+u)(f_{y_{1}}+\mathcal{O}(h_{n}))dz\Bigg]^{2}
		\mbox{ \Bigg\{ Assumption \descref{(A3)} \Bigg\} }
		\nonumber \\
		\label{ch6:after variable transformation}
		&=f^{2}_{y_{1}}h^{3}_{n}du 
		\Bigg[\int_{w}K(z)K(z+u)dz\Bigg]^{2}+o(h^{3}_{n}).
	\end{align}
	This yields that
	\begin{gather}
		\mathds{E}(C^{2}_{n})
		=2n^{2}k^{4}h^{3}_{n}\int
		\Bigg[\sum_{x=0}^{1}(f_{x}f_{y_{1}})^{2k-2}
		(f_{x}f_{y_{1}+uh_{n}})^{2k-2}\phi''(f^{k}_{x}f^{k}_{y_{1}})
		\phi''(f^{k}_{x}f^{k}_{y_{1}+uh_{n}})
		f^{2}_{x}(1-f_{x})^{2}f^{2}_{y_{1}}
		\nonumber \\
		+2(f_{x_{0}}f_{y_{1}})^{2k-2}
		(f_{x_{1}}f_{y_{1}+uh_{n}})^{2k-2}
		\phi''(f^{k}_{x_{0}}f^{k}_{y_{1}})
		\phi''(f^{k}_{x_{1}}f^{k}_{y_{1}+uh_{n}})
		f^{2}_{x_{0}}f^{2}_{x_{1}}f^{2}_{y_{1}}\Bigg] dy_{1}
		\nonumber \\
		\times \int \Big( \int K(z)K(z+u)dz\Big)^{2}
		du+o(n^{2}h^{3}_{n}).
	\end{gather}
	Since $f_{y_{1}+uh_{h}}=f_{y_{1}}+\mathcal{O}(h_{n})$ by Assumption \descref{(A2)}, we can approximate $f_{y_{1}+uh_{n}} \approx f_{y_{1}}$ for sufficiently large $n$ and the remainder term will be $o(n^{2}h^{3}_{n})$. Using this approximation, we get
	\begin{align}
		\mathds{E}(C^{2}_{n})
		&=2n^{2}h^{3}_{n}
		\int \Bigg[
		\sum_{x=0}^{1}k^{2}(f_{x}f_{y})^{2k-1}\phi''\Big(f^{k}_{x}f^{k}_{y}\Big)(1-f_{x}) \Bigg]^{2} dy
		\int \Big( \int K(z)K(z+u)dz\Big)^{2}
		du+o(n^{2}h^{3}_{n})
		\nonumber \\
		&=4n^{2}h^{3}_{n}\sigma^{2}_{\phi}+o(n^{2}h^{3}_{n}).
	\end{align}
	Also, see that 
	\begin{align}
		\frac{1}{(nh_{n})^{2}}|Cov(B_{n}, C_{n})| 
		&\le \frac{1}{(nh_{n})^{2}} \sqrt{Var(B_{n}) Var(C_{n})}
		\nonumber \\
		&=\frac{1}{(nh_{n})^{2}}\sqrt{ \mathcal{O}(nh^{2}_{n}) \times \mathcal{O}(n^{2}h^{3}_{n})}
		\nonumber \\
		&=\mathcal{O}(n^{-1/2}h^{1/2}_{n}) \to 0
		\mbox{ as } n \to \infty.
	\end{align}
	So it follows that 
	\begin{gather}
		\mathds{E}(\widetilde{I}_{n})
		\approx \frac{\mu_{\phi}}{nh_{n}}
		\mbox{ and }
		Var(\widetilde{I}_{n})\approx \frac{\sigma^{2}_{\phi}}{n^{2}h_{n}}.
	\end{gather}
	We see that the asymptotic distribution of $\widetilde{I}_{n}$ is determined by $C_{n}$. We already know that  
	$S_{n}=\sum_{i=1}^{n}T_{i}$ where 
	\begin{align}
		T_{i}=
		\frac{2}{nh^{3/2}_{n}}\sum_{j< i}\sum_{x=0}^{1}
		\int_{f_{y}>0}
		k^{2}(f_{x}f_{y})^{2k-2}\phi''(f^{k}_{x}f^{k}_{y})
		Z_{ix}Z_{jx}
		K_{h_{n}i}(y)K_{h_{n}j}(y)dy. 
	\end{align}
	Lemma \ref{conditions of hall and heyde} verifies the condition of Hall and Heyde (2014) \cite*{hall2014martingale} (Theorem 3.2) which gives 
	\begin{align}
		\frac{C_{n}}{nh^{3/2}_{n}}=S_{n}=\sum_{i=1}^{n}T_{i} \overset{\mathcal{L}}{\longrightarrow}
		\mathcal{N}\Big(0, 4\sigma^{2}_{\phi}\Big)
		\mbox{ as } n \to \infty.
	\end{align}
	Substituting $C_{n}=nh^{3/2}_{n}S_{n}$ in the expression of $\widetilde{I}_{n}$, we get  
	\begin{align}
		nh^{1/2}_{n}
		\Big(\widetilde{I}_{n}-\frac{\mu_{\phi}}{nh_{n}}\Big)
		\overset{\mathcal{L}}{=}
		nh^{1/2}_{n}
		\Big(\widetilde{I}_{n}-\frac{B_{n}}{2(nh_{n})^{2}}\Big)
		=\frac{S_{n}}{2}
		\overset{\mathcal{L}}{\longrightarrow}
		\mathcal{N}\Big(0, \sigma^{2}_{\phi}\Big)
		\mbox{ as } n \to \infty.
	\end{align}
	This completes the proof.

\subsection{Theorem \ref{theorem: consistency of tests}}

\begin{itemize}
		\item[(i)] 
		As the Kernel density estimates are consistent so are $\widehat{\mu}_{\phi}, \widehat{\sigma}_{\phi}$. 
		Define 
		\begin{align}
			T(\lambda)=\frac{nh^{1/2}_{n} \Big(\widehat{I}_{D^{(k)}_{\phi}}-\frac{\mu_{\phi}+\lambda(\hat{\mu}_{\phi}-\mu_{\phi})}{nh_{n}} \Big) }{\sigma_{\phi}+\lambda(\widehat{\sigma}_{\phi}-\sigma_{\phi})}
			\mbox{ for } \lambda \in [0, 1].
		\end{align}
		See that $T(1)=\widehat{T}_{D^{(k)}_{\phi}}$ and $T(0)=T_{D^{(k)}_{\phi}}$. Expanding $T(1)$ around $\lambda=0$ we get 
		\begin{align}
			T(1)=T(0)+\frac{1}{2}T'(\lambda^{**})
			\mbox{ where } \lambda^{**} \in (0, 1),  
		\end{align}
		where 
		\begin{align}
			T'(\lambda^{**})=\frac{\mu_{\phi}- \hat{\mu}_{\phi}}{h^{1/2}_{n}(\sigma_{\phi}+\lambda^{**}(\widehat{\sigma}_{\phi}-\sigma_{\phi}))}
			-\underbrace{T(\lambda^{**})}_{\mathcal{O}_{\mathds{P}}(1)} \times \underbrace{\frac{(\widehat{\sigma}_{\phi}-\sigma_{\phi})}{\sigma_{\phi}+\lambda^{**}(\widehat{\sigma}_{\phi}-\sigma_{\phi})}}
			_{o_{\mathds{P}}(1)}.
		\end{align}
		Define
		\begin{align}
			\mu_{\phi}(\lambda)
			=\frac{1}{2}\int_{u}K^{2}(u)du
			\int_{f_{y}>0}\sum_{x=0}^{1}k^{2}\xi_{xy}^{2k-1} 
			\phi''\Big( \xi^{k}_{xy}\Big)
			(1-f_{x}-\lambda h_{x})dy,
		\end{align}
		where $\xi_{xy}=(f_{x}+\lambda h_{x})(f_{y}+\lambda h_{y})$. Observe that $\mu_{\phi}(0)=\mu_{\phi}$ and $\mu_{\phi}(1)=\hat{\mu}_{\phi}$. Expanding $\mu_{\phi}(1)$ around $\lambda=0$ up to first-order term we get 
		$\mu_{\phi}(1)=\mu_{\phi}(0)+\frac{\mu_{\phi}'(\lambda^{*})}{2}$ where $\lambda^{*} \in (0,1)$. See that   
		\begin{align}
			\Bigg|\frac{\mu_{\phi}'(\lambda^{*})}{0.5 \int_{u}K^{2}(u)du }\Bigg|
			&=\Bigg|\int \sum_{x=0}^{1} k^{2}
			\xi_{xy}^{2k-1}
			\Bigg\{-\phi''(\xi^{k}_{xy})h_{x} \nonumber \\
			&+ k\xi^{k}\phi'''(\xi_{xy}^{k})
			\Big(\frac{h_{y}}{f_{y}+\lambda^{*}h_{y}}+ \frac{h_{x}}{f_{x}+\lambda^{*}h_{x}} \Big)(1-f_{x}-\lambda^{*}h_{x}) 
			\nonumber \\
			&+(2k-1)\Big(\frac{h_{y}}{f_{y}+\lambda^{*}h_{y}}+ \frac{h_{x}}{f_{x}+\lambda^{*}h_{x}} \Big) \phi''(\xi_{xy}^{k})(1-f_{x}-\lambda^{*}h_{x})
			\Bigg\}dy\Bigg|
			\nonumber \\
			&\le \sup_{x}|h_{x}|
			\int \sum_{x=0}^{1}
			\Bigg|3k^{2} \xi_{xy}^{2k-1}
			\Bigg\{-\frac{\phi''(\xi^{k}_{xy})}{3}
			+  \frac{k\xi^{k}\phi'''(\xi_{xy}^{k})+(2k-1)\phi''(\xi_{xy}^{k})}
			{f_{x}+\lambda^{*}h_{x}} 
			\Bigg\}\Bigg|dy \nonumber \\
			&+ \sup_{y}|h_{y}|
			\int \sum_{x=0}^{1}
			\Bigg|3k^{2} \xi_{xy}^{2k-1}
			\Bigg\{
			\frac{k\xi^{k}\phi'''(\xi_{xy}^{k})+(2k-1)\phi''(\xi_{xy}^{k})}
			{f_{y}+\lambda^{*}h_{y}}  
			\Bigg\}\Bigg|dy \nonumber \\
			& \le \mathcal{O}_{\mathds{P}}\Big(\sup_{x}|h_{x}| \Big)
			+\mathcal{O}_{\mathds{P}}\Big(\sup_{y}|h_{y}| \Big). 
		\end{align}
		As $f_{x}$ is trapped between $[0, 1]$, the term $|1-f_{x}-\lambda^{*}h_{x}|$ is bounded by $3$, which together with Assumptions \descref{(A1)} and \descref{(A2)} imply that the above integrations are bounded. We also know that $\sup_{x}|h_{x}|$ and $\sup_{y}|h_{y}|$ are $o_{\mathds{P}}\Big(\frac{1}{(nh^{1/2}_{n})^{1/3}}\Big)$, so  $|\mu_{\phi}'(\lambda)|=o_{\mathds{P}}\Big(\frac{1}{(nh^{1/2}_{n})^{1/3}}\Big)$ uniformly in $x,y$. This gives  
		$\hat{\mu}_{\phi}=\mu_{\phi}+o_{\mathds{P}}\Big(\frac{1}{(nh^{1/2}_{n})^{1/3}}\Big)$. Thus we get  
		\begin{align}
			| \widehat{T}_{D^{(k)}_{\phi}}-T_{D^{(k)}_{\phi}}|= \frac{1}{2}|T'(\lambda^{**})|
			\le \frac{1}{|\sigma_{\phi}+\lambda^{**}(\widehat{\sigma}_{\phi}-\sigma_{\phi})|}
			o_{\mathds{P}}\Big(\frac{1}{(nh^{2}_{n})^{1/3}}\Big)+
			o_{\mathds{P}}(1)
		\end{align}
		by Assumption \descref{(A4)}. Hence both $\widehat{T}^{(n)}_{D^{(k)}_{\phi}}$ and $T_{D^{(k)}_{\phi}}$ converge to the same limit in distribution. Consequently, $\widehat{T}^{(n)}_{D^{(k)}_{\phi}} \overset{\mathcal{L}}{\longrightarrow} \mathcal{N}(0, 1)$ under the null hypothesis $\mathds{H}$.
		
		\item [(ii)] Lemma \ref{Lemma: consistency of MI} along with the consistency of $\widehat{\mu_{\phi}},\widehat{\sigma}_{\phi}$ imply that  
		\begin{align}
			(nh^{1/2}_{n})^{-1}\widehat{T}_{D^{(k)}_{\phi}}
			=\frac{\widehat{I}_{D^{(k)}_{\phi}}-\frac{\widehat{\mu_{\phi}}}{nh_{n}}}
			{\widehat{\sigma}_{\phi}}
			\overset{\mathds{P}}{\longrightarrow} 
			\frac{I_{D^{(k)}_{\phi}}}{\sigma_{\phi}}
			=\begin{cases}
				0  &\mbox{ if null hypothesis is true },  \\
				t  &\mbox{ if alternative hypothesis is true }
			\end{cases}
		\end{align}
		for some fixed $t >0$. Therefore $\widehat{T}_{D^{(k)}_{\phi}}$ diverges in probability when the alternative hypothesis $\mathds{K}$ is true. This proves the consistency of $\widehat{T}_{D^{(k)}_{\phi}}$. 
	\end{itemize}

\subsection*{Lemma \ref{lemma: contiguous alternatives}}

	Let us consider 
	\begin{align}
		\zeta(t)
		=\sum_{x=0}^{1}\int_{f_{y}>0}
		\Bigg[\phi\Big((f_{x}f_{y}+t\Delta_{x,y})^{k}\Big)-
		\phi\Big(g_{x,y}(t)^{k}\Big)
		-\Big\{(f_{x}f_{y}+t\Delta_{x,y})^{k} -g_{x,y}(t)^{k}\Big\}
		\phi'\Big(g_{x,y}(t)^{k}\Big)
		\Bigg]dy 
	\end{align}
	where $g_{x,y}(t)=(f_{x}+t\Delta_{x})(f_{y}+t\Delta_{y})$ with $t=\frac{d}{\sqrt{nh^{1/2}_{n}}}$. See that   
	\begin{align}
		f^{(n)}_{x,y} =f_{x}f_{y}+t \Delta_{x,y} ,
		\mbox{   }
		f^{(n)}_{x} =f_{x}+t\Delta_{x}     
		\mbox{  and  }
		f^{(n)}_{y} =f_{y}+t\Delta_{y}.
	\end{align}
	Notice that $\zeta(0)=\zeta'(0)=0$. 
	Expanding $\zeta(t)$ around $t=0$ up to second-order gives 
	\begin{align}
		\label{ch6:expansion of zeta(t)}
		\zeta(t)=\zeta(0)+t\zeta'(0)
		+\frac{t^{2}}{2}\zeta''(0)
		+\frac{t^{3}}{6}\zeta'''(\theta) 
		\mbox{ for } 0< \theta < t.
	\end{align}
	Simple calculations give  
	\begin{align}
		\label{ch6:second-derivative of zeta(t)}
		\zeta''(t)
		=\sum_{x=0}^{1}\int_{f_{y}>0}
		&\Bigg[ 
		k^{2}(f_{x}f_{y}+t\Delta_{x,y})^{2k-2}\Delta_{x,y}^{2}
		\phi''\Big( (f_{x}f_{y}+t\Delta_{x,y})^{k} \Big)
		\nonumber \\
		&+k(k-1)(f_{x}f_{y}+t\Delta_{x,y})^{k-2}\Delta_{x,y}^{2}
		\phi'\Big( (f_{x}f_{y}+t\Delta_{x,y})^{k}\Big)
		\nonumber \\
		&-k^{2}(f_{x}f_{y}+t\Delta_{x,y})^{k-1}\Delta_{x,y}
		g^{k-1}_{x,y}(t)g'_{x,y}(t)
		\phi''\Big( g^{k}_{x,y}(t) \Big)
		\nonumber \\
		&-k(k-1)(f_{x}f_{y}+t\Delta_{x,y})^{k-2}\Delta^{2}_{x,y}
		\phi'\Big( g^{k}_{x,y}(t)\Big)
		\nonumber \\
		&-\Big\{(f_{x}f_{y}+t\Delta_{x,y})^{k}-g^{k}_{x,y}(t) \Big\}
		k^{2}g^{2k-2}_{x,y}(t)(g'_{x,y}(t))^{2}\phi'''\Big( g^{k}_{x, y}(t) \Big)
		\nonumber \\
		&-\Big\{(f_{x}f_{y}+t\Delta_{x,y})^{k} -g^{k}_{x,y}(t) \Big\}
		kg^{k-1}_{x,y}(t)g''_{x,y}(t)\phi''\Big(g^{k}_{x,y}(t) \Big)
		\nonumber \\
		&-\Big\{(f_{x}f_{y}+t\Delta_{x,y})^{k} -g^{k}_{x,y}(t) \Big\}
		k(k-1)g^{k-2}_{x,y}(t) (g'_{x,y}(t))^{2} 
		\phi''\Big( g^{k}_{x,y}(t) \Big)
		\nonumber \\
		&-\Big\{k(f_{x}f_{y}+t\Delta_{x,y})^{k-1}\Delta_{x,y}
		-k g^{k-1}_{x,y}(t)g'_{x,y}(t)\Big\} 
		k g^{k-1}_{x,y}(t) g'_{x,y}(t) 
		\phi'' \Big( g^{k}_{x,y}(t) \Big)
		\Bigg] dy
		\nonumber \\
		&=\sum_{x=0}^{1}\int_{f_{y}>0}
		k^{2}(f_{x}f_{y})^{2k}\phi''(f^{k}_{x}f^{k}_{y})
		\Bigg(\frac{\Delta_{x}}{f_{x}}+\frac{\Delta_{y}}{f_{y}}
		-\frac{\Delta_{x,y}}{f_{x}f_{y}} \Bigg)^{2}dy
		\mbox{ at } t=0, 
	\end{align}
	\begin{singlespace}
		\begin{align}
			\label{eq:zeta3}
			\zeta'''(\theta)
			=\sum_{x=0}^{1}\int_{f_{y}>0}
			&\Bigg[k^{3}(f_{x}f_{y}+\theta \Delta_{x,y})^{3k-3}\Delta^{3}_{x,y}
			\phi'''\Big( (f_{x}f_{y}+\theta \Delta_{x,y})^{k} \Big)
			\nonumber \\
			&+k^{2}(2k-2)(f_{x}f_{y}+\theta \Delta_{x,y})^{2k-3}\Delta^{3}_{x,y}
			\phi''\Big( (f_{x}f_{y}+\theta \Delta_{x,y})^{k}\Big)
			\nonumber \\
			&+ k^{2}(k-1)(f_{x}f_{y}+\theta \Delta_{x,y})^{2k-3}\Delta^{3}_{x,y}
			\phi''\Big( (f_{x}f_{y}+\theta \Delta_{x,y})^{k} \Big)
			\nonumber \\
			&+k(k-1)(k-2)(f_{x}f_{y}+\theta \Delta_{x,y})^{k-3}\Delta^{3}_{x,y}
			\phi'\Big((f_{x}f_{y}+\theta \Delta_{x,y})^{k} \Big)
			\nonumber \\
			&-k^{3}(f_{x}f_{y}+\theta \Delta_{x,y})^{k-1}\Delta_{x,y}g^{2k-2}_{x,y}(\theta)
			(g'_{x,y}(\theta))^{2}\phi'''\Big( g^{k}_{x,y}(\theta)\Big)
			\nonumber \\
			&-k^{2}(f_{x}f_{y}+\theta \Delta_{x,y})^{k-1}
			\Delta_{x,y}g^{k-1}_{x,y}(\theta)g''_{x,y}(\theta)\phi''\Big(g^{k}_{x,y}(\theta)\Big)
			\nonumber \\
			&-k^{2}(k-1)(f_{x}f_{y}+\theta\Delta_{x,y})^{k-1}\Delta_{x,y} g^{k-2}_{x,y}(\theta)(g'_{x,y}(\theta))^{2}\phi''\Big( g^{k}_{x,y}(\theta)\Big)
			\nonumber \\
			&-k^{2}(k-1)(f_{x}f_{y}+\theta \Delta_{x,y})^{k-2}\Delta^{2}_{x,y} g^{k-1}_{x,y}(\theta)g'_{x,y}(\theta)\phi''\Big(g^{k}_{x,y}(\theta) \Big)
			\nonumber \\
			&-k^{2}(k-1)(f_{x}f_{y}+\theta\Delta_{x,y})^{k-2}\Delta^{2}_{x,y}
			g^{k-1}_{x,y}(\theta) g'_{x,y}(\theta)\phi''\Big( g^{k}_{x,y}(\theta)\Big)
			\nonumber \\
			&-k(k-1)(k-2)(f_{x}f_{y}+\theta \Delta_{x,y})^{k-3}\Delta^{3}_{x,y}
			\phi'\Big( g^{k}_{x,y}(\theta)\Big)
			\nonumber \\
			&-\Big\{ (f_{x}f_{y}+\theta \Delta_{x,y})^{k}-g^{k}_{x,y}(\theta) \Big\}
			k^{3}g^{3k-3}_{x,y}(\theta)(g'_{x,y}(\theta))^{3}
			\phi''''\Big(g^{k}_{x,y}(\theta) \Big)
			\nonumber \\
			&-\Big\{ (f_{x}f_{y}+\theta \Delta_{x,y})^{k}-g^{k}_{x,y}(\theta) \Big\}
			k^{2}g^{2k-2}_{x,y}(\theta)2g'_{x,y}(\theta)g''_{x,y}(\theta)
			\phi'''\Big( g^{k}_{x,y}(\theta)\Big)
			\nonumber \\
			&-\Big\{ (f_{x}f_{y}+\theta \Delta_{x,y})^{k}-g^{k}_{x,y}(\theta) \Big\}
			k^{2}(2k-2)g^{2k-3}_{x,y}(\theta)(g'_{x,y}(\theta))^{3}
			\phi'''\Big( g^{k}_{x,y}(\theta)\Big)
			\nonumber \\
			&-\Big\{k(f_{x}f_{y}+\theta\Delta_{x,y})^{k-1}\Delta_{x,y}
			-kg^{k-1}_{x,y}(\theta)g'_{x,y}(\theta) \Big\}
			k^{2}g^{2k-2}_{x,y}(\theta)(g'_{x,y}(\theta))^{2}
			\phi'''\Big( g^{k}_{x,y}(\theta)\Big)
			\nonumber \\
			&-\Big\{(f_{x}f_{y}+\theta \Delta_{x,y})^{k}-g^{k}_{x,y}(\theta) \Big\}
			k^{2}g^{2k-2}_{x,y}(\theta)g'_{x,y}(\theta)g''_{x,y}(\theta)
			\phi'''\Big( g^{k}_{x,y}(\theta)\Big)
			\nonumber \\
			&-\Big\{(f_{x}f_{y}+\theta \Delta_{x,y})^{k}-g^{k}_{x,y}(\theta) \Big\}
			k(k-1)g^{k-2}_{x,y}(\theta)
			g'_{x,y}(\theta)
			g''_{x,y}(\theta)\phi''\Big(g^{k}_{x,y}(\theta) \Big)
			\nonumber \\
			&-\Big\{ k(f_{x}f_{y}+\theta\Delta_{x,y})^{k-1}\Delta_{x,y}-
			k g^{k-1}_{x,y}(\theta)g'_{x,y}(\theta)\Big\}
			kg^{k-1}_{x,y}(\theta)g''_{x,y}(\theta)\phi''\Big(g^{k}_{x,y}(\theta)\Big)
			\nonumber \\
			&- \Big\{ (f_{x}f_{y}+\theta \Delta_{x,y})^{k}-g^{k}_{x,y}(\theta) \Big\}
			k^{2}(k-1)g^{2k-3}_{x,y}(\theta)(g'_{x,y}(\theta))^{3}
			\phi'''\Big( g^{k}_{x,y}(\theta) \Big) 
			\nonumber \\
			&- \Big\{ (f_{x}f_{y}+\theta \Delta_{x,y})^{k}-g^{k}_{x,y}(\theta) \Big\}
			k(k-1)g^{k-2}_{x,y}(\theta)2g'_{x,y}(\theta)g''_{x,y}(\theta)
			\phi''\Big(g^{k}_{x,y}(\theta) \Big)
			\nonumber \\
			&- \Big\{ (f_{x}f_{y}+\theta \Delta_{x,y})^{k}-g^{k}_{x,y}(\theta) \Big\}
			k(k-1)(k-2)g^{k-3}_{x,y}(\theta)(g'_{x,y}(\theta))^{3}
			\phi \Big( g^{k}_{x,y}(\theta)\Big)
			\nonumber \\
			&-\Big\{k(f_{x}f_{y}+\theta\Delta_{x,y})^{k-1}\Delta_{x,y} 
			-kg^{k-1}_{x,y}(\theta)g'_{x,y}(\theta)  \Big\}
			k(k-1)g^{k-2}_{x,y}(\theta)(g'_{x,y}(\theta))^{2}
			\phi''\Big( g^{k}_{x,y}(\theta) \Big)
			\nonumber \\
			&-\Big\{k(f_{x}f_{y}+\theta\Delta_{x,y})^{k-1}\Delta_{x,y}
			-kg^{k-1}_{x,y}(\theta)g'_{x,y}(\theta)\Big\} 
			k^{2}g^{2k-2}_{x,y}(\theta)(g'_{x,y}(\theta))^{2}\phi\Big( g^{k}_{x,y}(\theta)\Big)
			\nonumber \\
			&-\Big\{k(f_{x}f_{y}+\theta\Delta_{x,y})^{k-1}\Delta_{x,y}
			-kg^{k-1}_{x,y}(\theta)g'_{x,y}(\theta)\Big\}
			kg^{k-1}_{x,y}(\theta)g''_{x,y}(\theta)
			\phi''\Big( g^{k}_{x,y}(\theta)\Big)
			\nonumber \\
			&-\Big\{k(f_{x}f_{y}+\theta\Delta_{x,y})^{k-1}\Delta_{x,y}
			-kg^{k-1}_{x,y}(\theta)g'_{x,y}(\theta)\Big\}
			k(k-1)g^{k-2}_{x,y}(\theta)(g'_{x,y}(\theta))^{2}
			\phi''\Big( g^{k}_{x,y}(\theta)\Big)
			\nonumber \\
			&-\Big\{k(k-1)(f_{x}f_{y}+\theta\Delta_{x,y})^{k-2}\Delta^{2}_{x,y}
			-kg^{k-1}_{x,y}(\theta)g''_{x,y}(\theta)
			-k(k-1)g^{k-2}_{x,y}(\theta)(g'_{x,y}(\theta))^{2}
			\Big\}
			\nonumber \\
			& \times kg^{k-1}_{x,y}(\theta)g'_{x,y}(\theta)
			\phi''\Big( g^{k}_{x,y}(\theta)\Big)
			\Bigg]dy.
		\end{align}    
	\end{singlespace}
	Assumption \descref{(A1)} ensures that $|\zeta'''(\theta)|=\mathcal{O}(1)$. Therefore the remainder in the expansion of $\zeta(t)$ in (\ref{ch6:expansion of zeta(t)}) will be $\mathcal{O}(t^{3})=o(t^{2})$ when $t \to 0$. Substituting $t=\frac{d}{\sqrt{nh^{1/2}_{n}}}$ in (\ref{ch6:expansion of zeta(t)}) gives  
	\begin{align}
		I^{(n)}_{D^{(k)}_{\phi}}
		=\zeta(t)=\frac{d^{2}}{2(nh^{1/2}_{n})}
		\sum_{x=0}^{1}\int_{f_{y}>0}
		k^{2}(f_{x}f_{y})^{2k}\phi''(f^{k}_{x}f^{k}_{y})
		\Bigg(\frac{\Delta_{x}}{f_{x}}
		+\frac{\Delta_{y}}{f_{y}}-\frac{\Delta_{x,y}}{f_{x}f_{y}} \Bigg)^{2}dy
		+o\Bigg(\frac{d^{2}}{nh^{1/2}_{n}}\Bigg).
	\end{align}
	This completes the proof.

\subsection*{Theorem \ref{Theorem: Asymptotic normality of MI under contiguous alternative}}

Now see that
	\begin{align}
		T_{D^{(k)}_{\phi}}
		&=nh^{1/2}_{n}\sigma^{-1}_{\phi}\Big(\widehat{I}_{D^{(k)}_{\phi}}-\frac{\mu_{\phi}}{nh_{n}}
		\Big)
		\nonumber \\
		&=nh^{1/2}_{n}\sigma^{-1}_{\phi}\Big(\widehat{I}_{D^{(k)}_{\phi}}
		-I^{(n)}_{D^{(k)}_{\phi}}-\frac{\mu_{\phi}}{nh_{n}}\Big)
		+nh^{1/2}_{n}\sigma^{-1}_{\phi}I^{(n)}_{D^{(k)}_{\phi}}
		\nonumber \\
		&=nh^{1/2}_{n}\sigma^{-1}_{\phi}\Big(\widehat{I}_{D^{(k)}_{\phi}}
		-I^{(n)}_{D^{(k)}_{\phi}}-\frac{\mu_{\phi}}{nh_{n}}\Big)
		\nonumber \\
		\label{ch6:decompose under K_n}
		&+\frac{d^{2}}{2\sigma_{\phi}}
		\sum_{x=0}^{1}\int_{f_{y}>0}
		k^{2}(f_{x}f_{y})^{2k}\phi''(f^{k}_{x}f^{k}_{y})
		\Bigg(\frac{\Delta_{x}}{f_{x}}+\frac{\Delta_{y}}{f_{y}}
		-\frac{\Delta_{x,y}}{f_{x}f_{y}}\Bigg)^{2}dy+o(1).
	\end{align}
	Let us define 
	\begin{align}
		U_{n}=nh^{1/2}_{n}\sigma^{-1}_{\phi}\Big(\widehat{I}_{D^{(k)}_{\phi}}
		-I^{(n)}_{D^{(k)}_{\phi}}-\frac{\mu_{\phi}}{nh_{n}}\Big), 
		\mbox{  }
		W_{n}(t)=\mathds{1}\big\{U_{n} \le t \big\}
		\mbox{ and }
		L_{n}=\prod_{i=1}^{n}\frac{f^{(n)}_{X_{i},Y_{i}}(x_{i}, y_{i} )}{f_{X_{i}}(x_{i})f_{Y_{i}}(y_{i})}
	\end{align}
	for any fixed $t \in \mathds{R}$. The probability density function under the null hypothesis is $f_{x,y}=f_{x}f_{y}$ for all $x,y$. Let $F^{(n)}$ and $F_{0}$ be the distribution functions associated with the contiguous alternatives and null hypothesis. Since $f^{(n)}_{x,y}  \to f_{x}f_{y}$ pointwise, it follows from the theorem of  Scheff\`e that $F^{(n)} \overset{\mathcal{L}}{\to} F_{0}$. Then an application of the Portmanteau lemma implies that 
	\begin{align}    
		\underset{n \to \infty}{\limsup}
		\mathds{P}_{\mathds{K}_{n}} \big\{U_{n} \le t \big\}
		\le \underset{n \to \infty}{\limsup}
		\mathds{P}_{\mathds{H}}\big\{U_{n} \le t \big\}
		= \Phi_{1}(t),
	\end{align}
	as $d=0$ under the null hypothesis $\mathds{H}$. In the other direction, see that 
	\begin{align}
		\mathds{P}_{\mathds{K}_{n}}\big\{ U_{n} \le t\big\}
		=\mathds{E}_{\mathds{K}_{n}}\Big[W_{n}(t)\Big]
		&\ge \underset{\prod_{i=1}^{n}f_{x_{i}}f_{y_{i}}>0}{\sum\int}
		W_{n}(t) \prod_{i=1}^{n}f^{(n)}_{X_{i}Y_{i}}(x_{i}, y_{i})dy_{i} \nonumber \\\
		&=\mathds{E}_{\mathds{H}}\Big[W_{n}(t)L_{n} \Big]
		\mbox{ for all } n.    
	\end{align}
	Thus we have $\underset{n \to \infty}{\liminf}\mathds{P}_{\mathds{K}_{n}}\big\{ U_{n} \le t\big\}\ge \underset{n \to \infty}{\liminf}\mathds{E}_{\mathds{H}}\Big[W_{n}(t)L_{n} \Big]$. Under the null hypothesis $\mathds{H}$, see that 
	\begin{align}
		L_{n}=1 
		\mbox{ and }
		W_{n}(t) =
		\mathds{1}\Bigg\{nh^{1/2}_{n}\sigma^{-1}_{\phi}\Bigg(\widehat{I}_{D^{(k)}_{\phi}}-
		\frac{\mu_{\phi}}{nh^{1/2}_{n}}\Bigg) \le t\Bigg\}
		\overset{\mathcal{L}}{\longrightarrow} T,
	\end{align}
	where $T \sim Bernoulli(\Phi_{1}(t))$. This gives $W_{n}(t)L_{n}\overset{\mathcal{L}}{\longrightarrow} T$ for any $t \in \mathds{R}$ under the null hypothesis. Since $W_{n}$ is bounded, it is uniformly integrable. This results into  $\mathds{E}_{\mathds{H}}\Big(W_{n}(t)L_{n}\Big)\longrightarrow \mathds{E}(T)=\Phi_{1}(t)$, and consequently we obtain 
	\begin{align}
		\mathds{P}_{\mathds{K}_{n}}
		\Bigg\{ nh^{1/2}_{n}\sigma^{-1}_{\phi}\Big(\widehat{I}_{D^{(k)}_{\phi}}
		-I^{(n)}_{D^{(k)}_{\phi}}-\frac{\mu_{\phi}}{nh_{n}}\Big)\le t
		\Bigg\}
		\longrightarrow \Phi_{1}(t)
		\mbox{ for all } t \in \mathds{R}
	\end{align}
	when $n \to \infty$. Putting all these pieces together results in
	
	$T_{D^{(k)}_{\phi}} \overset{\mathcal{L}}{\longrightarrow}
	\frac{d^{2}}{2\sigma_{\phi}}\sum_{x=0}^{1}\int_{f_{y}>0}k^{2}
	(f_{x}f_{y})^{2k}\phi''(f_{x}^{k}f^{k}_{y})\Big(\frac{\Delta_{x}}{f_{x}}+\frac{\Delta_{y}}{f_{y}}-
	\frac{\Delta_{x,y}}{f_{x}f_{y}}\Big)^{2}dy+\mathcal{N}(0,1 )$ under $\mathds{K}_{n}$. This completes the proof.

\subsection*{Theorem \ref{Theorem: Asymptotic normality of MI under contiguous alternative}}

We know that 
\begin{align}
    T_{D^{(k)}_{\phi}}
    &=nh^{1/2}_{n}\sigma^{-1}_{\phi}\Big(\widehat{I}_{D^{(k)}_{\phi}}-\frac{\mu_{\phi}}{nh_{n}}
    \Big)
    \nonumber \\
    &=nh^{1/2}_{n}\sigma^{-1}_{\phi}\Big(\widehat{I}_{D^{(k)}_{\phi}}
    -I^{(n)}_{D^{(k)}_{\phi}}-\frac{\mu_{\phi}}{nh_{n}}\Big)
    +nh^{1/2}_{n}\sigma^{-1}_{\phi}I^{(n)}_{D^{(k)}_{\phi}}
    \nonumber \\
    &=nh^{1/2}_{n}\sigma^{-1}_{\phi}\Big(\widehat{I}_{D^{(k)}_{\phi}}
    -I^{(n)}_{D^{(k)}_{\phi}}-\frac{\mu_{\phi}}{nh_{n}}\Big)+\frac{d^{2}}{2\sigma_{\phi}}
    \sum_{x=0}^{1}\int_{f_{y}>0}
    k^{2}(f_{x}f_{y})^{2k}\phi''(f^{k}_{x}f^{k}_{y})
    \Bigg(\frac{\Delta_{x}}{f_{x}}+\frac{\Delta_{y}}{f_{y}}
    -\frac{\Delta_{x,y}}{f_{x}f_{y}}\Bigg)^{2}dy+o(1).
\end{align}
For fixed $t \in \mathds{R}$, define 
\begin{align}
    W_{n}(t)=\mathds{1}\Bigg\{
    nh^{1/2}_{n}\sigma^{-1}_{\phi}\Big(\widehat{I}_{D^{(k)}_{\phi}}
    -I^{(n)}_{D^{(k)}_{\phi}}-\frac{\mu_{\phi}}{nh_{n}}\Big) \le t \Bigg\}
    \mbox{ and }
    L_{n}=\prod_{i=1}^{n}\frac{f^{(n)}_{X_{i},Y_{i}}}{f_{X_{i}}f_{Y_{i}}}.
\end{align}
See that 
\begin{align}
    \mathds{P}_{\mathds{K}_{n}}
    \Bigg[nh^{1/2}_{n}\sigma^{-1}_{\phi}\Big(\widehat{I}_{D^{(k)}_{\phi}}
    -I^{(n)}_{D^{(k)}_{\phi}}-\frac{\mu_{\phi}}{nh_{n}}\Big) \le t\Bigg]
    &=\mathds{E}_{\mathds{K}_{n}}\Big(W_{n}(t)\Big)
    \nonumber \\
    &=\mathds{E}_{\mathds{H}}\Big(W_{n}(t)L_{n}\Big).
\end{align}
The above expression holds as the densities under $\mathds{K}_{n}$ are contiguous to the probability density under $\mathds{H}$ ( formally denoted as $\mathds{K}_{n} \triangleleft \mathds{H}$; see Van der vaart (2000) \cite{van2000asymptotic}, pp. 87). If $\mathds{H}$ is true, then $d=0$ and $I^{(n)}_{D^{(k)}_{\phi}}=0$. This implies that
\begin{align}
    L_{n}\overset{\mathds{H}}{=}1 
    \mbox{ and }
   W_{n}(t) \overset{\mathds{H}}{=}
   \mathds{1}\Bigg\{nh^{1/2}_{n}\sigma^{-1}_{\phi}\Big(\widehat{I}_{D^{(k)}_{\phi}}-
   \frac{\mu_{\phi}}{nh^{1/2}_{n}}\Big) \le t\Bigg\}
   \overset{\mathcal{L}}{\longrightarrow} T,
\end{align}
where $T \sim Bernoulli(\Phi(t))$. So $W_{n}(t)L_{n}\overset{\mathcal{L}}{\longrightarrow} T$ for any $t \in \mathds{R}$, under the null hypothesis $\mathds{H}$. Since $W_{n}$ is bounded, it is uniformly intergrable, thus $\mathds{E}_{\mathds{H}}\Big(W_{n}(t)L_{n}\Big)\longrightarrow \mathds{E}(T)=\Phi(t)$. Therefore for $n \to \infty$,
\begin{align}
    \mathds{P}_{\mathds{K}_{n}}
    \Bigg[nh^{1/2}_{n}\sigma^{-1}_{\phi}\Big(\widehat{I}_{D^{(k)}_{\phi}}
    -I^{(n)}_{D^{(k)}_{\phi}}-\frac{\mu_{\phi}}{nh_{n}}\Big)\le t\Bigg]
    =\mathds{E}_{\mathds{H}}\Big(W_{n}(t)L_{n}\Big)
    \longrightarrow \Phi(t),
    \mbox{ for all } t \in \mathds{R}.
\end{align}
Thus $T_{D^{(k)}_{\phi}} \overset{\mathcal{L}}{\longrightarrow}
\frac{d^{2}}{2\sigma_{\phi}}\sum_{x=0}^{1}\int_{f_{y}>0}k^{2}
(f_{x}f_{y})^{2k}\phi''(f_{x}^{k}f^{k}_{y})\Big(\frac{\Delta_{x}}{f_{x}}+\frac{\Delta_{y}}{f_{y}}-
\frac{\Delta_{x,y}}{f_{x}f_{y}}\Big)^{2}dy+\mathcal{N}(0,1 )$ under $\mathds{K}_{n}$. This completes the proof.

\subsection*{Lemma \ref{ch6:MI under contaminated contiguous}}
  
  	Write 
	\begin{align*}
		\Gamma(s, t)
		&=\sum_{x=0}^{1}\int
		\Bigg\{\phi \Big((f^{(P_{n})}_{x,y})^{k}\Big) 
		- \phi\Big((f^{(P_{n})}_{x})^{k}(f^{(P_{n})}_{y})^{k}
		\Big) \nonumber\\
		&-\Big((f^{(P_{n})}_{x,y})^{k} - 
		(f^{(P_{n})}_{x})^{k}(f^{(P_{n})}_{y})^{k}
		\Big)\phi'\Big((f^{(P_{n})}_{x})^{k}(f^{(P_{n})}_{y})^{k}\Big)
		\Bigg\}dy	
	\end{align*}
	where $s=\frac{\epsilon}{\sqrt{nh^{1/2}_{n}}}$ and $t=\frac{d}{\sqrt{nh^{1/2}_{n}}}$. Expanding $\Gamma(s, t)$ around $s=0$ upto second-order gives 
	\begin{align}
		\Gamma(s, t)=\Gamma(0, t)+s \Gamma^{(10)}(0, t)+ \frac{s^{2}}{2}\Gamma^{(20)}(0, t)
		+\frac{s^{3}}{3}\Gamma^{(30)}(s^{*}, t)
		\mbox{ with } 0 < s^{*} < s. 
	\end{align}
	Here, the partial derivatives are denoted as 
	$\Gamma^{(r_{1}r_{2})}(s_{1}, t_{1})=\frac{\partial^{r_{1}+r_{2}}}{\partial s^{r_{1}}\partial s^{r_{2}}}\Gamma(s, t)\Bigg|_{s=s_{1}, t=t_{1}}$, $r_{1}, r_{2}=1, 2, \ldots$. We get $f^{(P_{n})}_{x,y}=f^{(n)}_{x,y}$ for all $x,y$ when $s=0$. This gives $\Gamma(0, t)=I^{(n)}_{D^{(k)}_{\phi}}$ as in (\ref{I(n)}). From Lemma \ref{lemma: contiguous alternatives} we know that 
	\begin{align}
		\label{ch6: expansion eq1}	
		\Gamma(0, t)
		=I^{(n)}_{D^{(k)}_{\phi}} 
		=\frac{d^{2}}{2nh^{1/2}_{n}}
		\sum_{x=0}^{1}\int
		k^{2}(f_{x}f_{y})^{2k}
		\Bigg( \frac{\Delta_{x,y}}{f_{x}f_{y}} -
		\frac{\Delta_{x}}{f_{x}} - \frac{\Delta_{y}}{f_{y}}\Bigg)^{2}
		\phi''\Big(f^{k}_{x}f^{k}_{y}\Big)dy+o\Bigg(\frac{d^{2}}{nh^{1/2}_{n}}\Bigg).
	\end{align}	
	A simple calculation yields that 
	\begin{align}
		\Gamma^{(10)}(0, t)
		&=\sum_{x=0}^{1}\int_{f_{y}>0}
		\Bigg[k \Big(f^{(n)}_{x,y}\Big)^{k}
		\frac{\Delta'_{x,y}}{f^{(n)}_{x,y}}
		\Big\{\phi'\Big((f^{(n)}_{x,y})^{k}\Big) - 
		\phi' \Big( (f^{(n)}_{x} f^{(n)}_{y})^{k}\Big)
		\Big\}
		\nonumber \\
		&- k
		\Bigg( \frac{\Delta'_{x}}{f^{(n)}_{x}}
		+ \frac{\Delta'_{y}}{f^{(n)}_{y}} \Bigg)
		\Big( (f^{(n)}_{x,y})^{k} - (f^{(n)}_{x}f^{(n)}_{y})^{k} \Big)
		\Big(f^{(n)}_{x}f^{(n)}_{y}\Big)^{k}
		\phi''\Big((f^{(n)}_{x}f^{(n)}_{y})^{k}\Big)
		\Bigg]dy.
	\end{align}	
	Expanding $\Gamma^{(10)}(0, t)$ around $t=0$ upto gives 
	\begin{align}
		\Gamma^{(10)}(0, t)=\Gamma^{(10)}(0, 0)+t\Gamma^{(11)}(0, 0)+\frac{t^{2}}{2}\Gamma^{(12)}(0, t^{*})
		\mbox{ where } 0 < t^{*} < t. 
	\end{align}	
	Simple calculations show that $\Gamma^{(10)}(0, 0)=0$ and 	
	\begin{align}
		\Gamma^{(11)}(0, 0)
		=\sum_{x=0}^{1}\int k^{2}(f_{x}f_{y})^{2k}
		\Bigg( \frac{\Delta_{x,y}}{f_{x}f_{y}} -
		\frac{\Delta_{x}}{f_{x}} - \frac{\Delta_{y}}{f_{y}}\Bigg)
		\Bigg( \frac{\Delta'_{x,y}}{f_{x}f_{y}} -
		\frac{\Delta'_{x}}{f_{x}} - \frac{\Delta'_{y}}{f_{y}}\Bigg)
		\phi''\Big(f^{k}_{x}f^{k}_{y}\Big)dy.
	\end{align}
	Using the same argument as before, one can show that the remainder term is $o(t)$. Thus we obtain
	\begin{align}
		\label{ch6: expansion eq2}		
		s\Gamma^{(10)}(0, t)
		=\frac{d\epsilon}{nh^{1/2}_{n}}\sum_{x=0}^{1}\int k^{2}(f_{x}f_{y})^{2k}
		\Bigg( \frac{\Delta_{x,y}}{f_{x}f_{y}} -
		\frac{\Delta_{x}}{f_{x}} - \frac{\Delta_{y}}{f_{y}}\Bigg)
		\Bigg( \frac{\Delta'_{x,y}}{f_{x}f_{y}} -
		\frac{\Delta'_{x}}{f_{x}} - \frac{\Delta'_{y}}{f_{y}}\Bigg)
		\phi''\Big(f^{k}_{x}f^{k}_{y}\Big)dy+
		o\Bigg( \frac{d\epsilon}{nh^{1/2}_{n}}\Bigg).
	\end{align}
	Next, see that 
	\begin{align}
		\Gamma^{(20)}(0,t)
		&=\sum_{x=0}^{1}
		\int \Bigg\{ 
		k(k-1)(f^{(n)}_{x,y})^{k}\Bigg(\frac{\Delta'_{x,y}}
		{f^{(n)}_{x,y}}\Bigg)^{2}
		\Big\{\phi' \Big(f^{(n)}_{x,y}\Big)^{k}
		- \phi' \Big(f^{(n)}_{x}f^{(n)}_{y}\Big)^{k}\Big\}
		\nonumber \\
		&+k^{2}(f^{(n)}_{x,y})^{2k}
		\Bigg(\frac{\Delta'_{x,y}}{f^{(n)}_{x,y}}\Bigg)^{2}
		\phi''\Big((f^{(n)}_{x,y})^{k}\Big)  \nonumber \\
		&-k^{2}(f^{(n)}_{x,y}f^{(n)}_{x}f^{(n)}_{y})^{k}\frac{\Delta'_{x,y}}{f^{(n)}_{x,y}}
		\Bigg(\frac{\Delta'_{x}}{f^{(n)}_{x}}
		+ \frac{\Delta'_{y}}{f^{(n)}_{y}}
		\Bigg)
		\phi''\Big((f^{(n)}_{x}f^{(n)}_{y})^{k}\Big) \nonumber \\
		&-k^{2}(f^{(n)}_{x}f^{(n)}_{y})^{2k}
		\Bigg\{
		\Bigg(\frac{f^{(n)}_{x,y}}{f^{(n)}_{x}f^{(n)}_{y}} \Bigg)^{k}
		\frac{\Delta'_{x,y}}{f^{(n)}_{x,y}}
		- 
		\frac{\Delta'_{x}}{f^{(n)}_{x}}
		- 
		\frac{\Delta'_{y}}{f^{(n)}_{y}}
		\Bigg\} 
		\Bigg(\frac{\Delta'_{x}}{f^{(n)}_{x}} 
		+ \frac{\Delta'_{y}}{f^{(n)}_{y}}  \Bigg) 
		\phi''\Big((f^{(n)}_{x}f^{(n)}_{y})^{k}\Big)
		\nonumber \\
		&-\Big( (f^{(n)}_{x,y})^{k} - (f^{(n)}_{x}f^{(n)}_{y})^{k}\Big)
		(f^{(n)}_{x}f^{(n)}_{y})^{k}
		\Bigg\{k(k-1)
		\Bigg(\frac{\Delta'_{x}}{f^{(n)}_{x}}\Bigg)^{2}
		+2k^{2}\Bigg(\frac{\Delta'_{x}\Delta'_{y}}{f^{(n)}_{x}f^{(n)}_{y}}\Bigg) \nonumber \\
		&+k(k-1) \Bigg(\frac{\Delta'_{y}}{f_{y}}\Bigg)^{2}
		\Bigg\}\phi''\Bigg( \Big(f^{(n)}_{x}f^{(n)}_{y}\Big)^{k} \Bigg) \nonumber \\
		&-k\Bigg\{(f^{(n)}_{x,y})^{k} - (f^{(n)}_{x}f^{(n)}_{y})^{k}
		\Bigg\}\Bigg(\frac{\Delta'_{x}}{f^{(n)}_{x}} + \frac{\Delta'_{y}}{f^{(n)}_{y}} \Bigg)^{2}
		\phi'''\Bigg( (f^{(n)}_{x}f^{(n)}_{y})^{k} \Bigg)
		\Bigg\} dy.
	\end{align}	
	As before, we expand $\Gamma^{(20)}(0, t)$ around $t=0$ and get
	\begin{align}
		\Gamma^{(20)}(0, t)=\Gamma^{(20)}(0, 0)+t\Gamma^{(21)}(0, t^{**})
		\mbox{ where } 0 < t^{**} < t,
	\end{align}	 
	where  
	\begin{align}
		\Gamma^{(20)}(0,0)
		&=\sum_{x=0}^{1}
		\int 
		k^{2}(f_{x}f_{y})^{2k}
		\Bigg(\frac{\Delta'_{x,y}}{f_{x}f_{y}}
		-
		\frac{\Delta'_{x}}{f_{x}}
		- \frac{\Delta'_{y}}{f_{y}}
		\Bigg)^{2}
		\phi''\Big(f^{k}_{x}f^{k}_{y}\Big) dy.
	\end{align}	
	Thus we obtain 
	\begin{align}
		\label{ch6: expansion eq3}		
		\frac{s^{2}}{2}\Gamma^{(20)}(0, t)
		=\frac{\epsilon^{2}}{2nh^{1/2}_{n}}\sum_{x=0}^{1}
		\int 
		k^{2}(f_{x}f_{y})^{2k}
		\Bigg(\frac{\Delta'_{x,y}}{f_{x}f_{y}}
		-
		\frac{\Delta'_{x}}{f_{x}}
		- \frac{\Delta'_{y}}{f_{y}}
		\Bigg)^{2}
		\phi''\Big(f^{k}_{x}f^{k}_{y}\Big) dy
		+O\Bigg(\frac{s^{2}t}{2}\Bigg).
	\end{align}	
	Combining (\ref{ch6: expansion eq1}), (\ref{ch6: expansion eq2}) and (\ref{ch6: expansion eq3}) we get 
	\begin{align*}
		I^{(P_{n})}_{D^{(k)}_{\phi}}
		&=\frac{k^{2}}{2nh^{1/2}_{n}}
		\sum_{x=0}^{1} \int (f_{x}f_{y})^{2k}
		\Bigg\{
		d\Bigg(\frac{\Delta_{x,y}}{f_{x}f_{y}}
		-
		\frac{\Delta_{x}}{f_{x}}
		- \frac{\Delta_{y}}{f_{y}}
		\Bigg)
		+
		\epsilon
		\Bigg(\frac{\Delta'_{x,y}}{f_{x}f_{y}}
		-
		\frac{\Delta'_{x}}{f_{x}}
		- \frac{\Delta'_{y}}{f_{y}}
		\Bigg)\Bigg\}^{2}\phi''\Big(f^{k}_{x}f^{k}_{y}\Big)dy
		+o\Bigg(\frac{(d+\epsilon)^{2}}{nh^{1/2}_{n}}\Bigg).
	\end{align*}	
	This completes the proof.

\subsection*{Theorem \ref{theorem: asymptotic dist under contaminated contiguous}}
    
    Using Lemma \ref{ch6:MI under contaminated contiguous} we see that  
	\begin{align*}
		T_{D^{(k)}_{\phi}}
		&=nh^{1/2}_{n}\sigma^{-1}_{\phi}\Big(\widehat{I}_{D^{(k)}_{\phi}}-\frac{\mu_{\phi}}{nh^{1/2}_{n}}-
		I^{(P_{n})}_{D^{(k)}_{\phi}}\Big)
		+nh^{1/2}_{n}\sigma^{-1}_{\phi} I^{(P_{n})}_{D^{(k)}_{\phi}}
		\nonumber \\
		&=U_{n}+\frac{k^{2}}{2\sigma_{\phi}}
		\sum_{x=0}^{1} \int (f_{x}f_{y})^{2k}
		\Bigg\{
		d\Bigg(\frac{\Delta_{x,y}}{f_{x}f_{y}}
		-
		\frac{\Delta_{x}}{f_{x}}
		- \frac{\Delta_{y}}{f_{y}}
		\Bigg)
		+
		\epsilon
		\Bigg(\frac{\Delta'_{x,y}}{f_{x}f_{y}}
		-
		\frac{\Delta'_{x}}{f_{x}}
		- \frac{\Delta'_{y}}{f_{y}}
		\Bigg)\Bigg\}^{2}\phi''\Big(f^{k}_{x}f^{k}_{y}\Big)dy+o(1) ,
	\end{align*}
	where $U_{n}=nh^{1/2}_{n}\sigma^{-1}_{\phi}\Big(\widehat{I}_{D^{(k)}_{\phi}}-\frac{\mu_{\phi}}{nh^{1/2}_{n}}-
	I^{(P_{n})}_{D^{(k)}_{\phi}}\Big)$. See that $f^{(P_{n})}_{x,y} \longrightarrow f_{x}f_{y}$ for all $x,y$, when $n \to \infty$. So $f^{(P_{n})}_{X, Y}$ is contiguous to the joint density under the null hypothesis. Applying the results of Portmanteau as in Theorem \ref{Theorem: Asymptotic normality of MI under contiguous alternative}, we find that
	\begin{align}
		\underset{n \to \infty}{\lim}
		\mathds{P}_{f^{(P_{n})}_{X,Y}}\Big[U_{n}\le t\Big]
		=\underset{n \to \infty}{\lim}
		\mathds{E}_{f^{(P_{n})}_{X,Y}}\Big(\mathds{1}\{U_{n}\le t\}\Big)
		=\underset{n \to \infty}{\lim}
		\mathds{E}_{\mathds{H}}\Big(\mathds{1}\{U_{n}\le t\}L_{n}\Big)
		\mbox{ for fixed } t \in \mathds{R},
	\end{align}
	where $L_{n}=\prod_{i=1}^{n}\frac{f^{(P_{n})}_{x,y}}{f_{x}
		f_{y}}$. From the condition $0 \le \max\{d, \epsilon\} \le C^{*} \sup_{x,y}|f_{x,y}-f_{x}f_{y}|$ we find that $d=\epsilon=0$, and consequently $f^{(P_{n})}_{x,y}=f_{x}f_{y}$ for all $x,y$ when the null hypothesis is true. Thus we get $L_{n}=1$ and 
	\begin{align}
		\mathds{1}\Big\{U_{n}\le t\Big\}=
		\mathds{1}\Bigg\{nh^{1/2}_{n}\Big(\widehat{I}_{D^{(k)}_{\phi}}-\frac{\mu_{\phi}}{nh_{n}}\Big)\le t\Bigg\}
		\overset{L}{\longrightarrow} T
		\mbox{ under }
		\mathds{H},
	\end{align}
	where $T \sim Bernoulli(\Phi_{1}(t))$. Since $\mathds{1}\Big\{U_{n}\le t\Big\}$ is bounded, it is uniformly intergrable. Then it follows that  
	\begin{align}
		\mathds{P}_{f^{(P_{n})}_{X,Y}}\Big[U_{n}\le t\Big]
		\longrightarrow \mathds{E}_{\mathds{H}}(T)=\Phi_{1}(t)
	\end{align}
	for all $t \in \mathds{R}$. Thus, it follows that
	\begin{align*}
		T_{D^{(k)}_{\phi}}
		- \frac{k^{2}}{2\sigma_{\phi}}
		\sum_{x=0}^{1} \int (f_{x}f_{y})^{2k}
		\Bigg\{
		d\Bigg(\frac{\Delta_{x,y}}{f_{x}f_{y}}
		-
		\frac{\Delta_{x}}{f_{x}}
		- \frac{\Delta_{y}}{f_{y}}
		\Bigg)
		+
		\epsilon
		\Bigg(\frac{\Delta'_{x,y}}{f_{x}f_{y}}
		-
		\frac{\Delta'_{x}}{f_{x}}
		- \frac{\Delta'_{y}}{f_{y}}
		\Bigg)\Bigg\}^{2}\phi''\Big(f^{k}_{x}f^{k}_{y}\Big)dy
		\overset{\mathcal{L}}{\to}
		\mathcal{N}(0, 1)
	\end{align*}
	under $f^{(P_{n})}_{X, Y}$ when $n \to \infty$. This completes the proof.

\subsection*{Theorem \ref{PIF and LIF theorem}}

    	\begin{itemize}              
		
		\item[(a)] Let us denote  
		\begin{align}
			r_{n}
			=\frac{k^{2}}{2\sigma_{\phi}}
			\sum_{x=0}^{1} \int (f_{x}f_{y})^{2k}
			\Bigg\{
			d\Bigg(\frac{\Delta_{x,y}}{f_{x}f_{y}}
			-
			\frac{\Delta_{x}}{f_{x}}
			- \frac{\Delta_{y}}{f_{y}}
			\Bigg)
			+
			\epsilon
			\Bigg(\frac{\Delta'_{x,y}}{f_{x}f_{y}}
			-
			\frac{\Delta'_{x}}{f_{x}}
			- \frac{\Delta'_{y}}{f_{y}}
			\Bigg)\Bigg\}^{2}\phi''\Big(f^{k}_{x}f^{k}_{y}\Big)dy.
		\end{align}
		See that $r_{n} \to r_{*}$ where 
		\begin{align}
			r_{*}
			=\frac{k^{2}}{2\sigma_{\phi}}
			\sum_{x=0}^{1} \int (f_{x}f_{y})^{2k}
			\Bigg\{
			d\Bigg(\frac{\Delta_{x,y}}{f_{x}f_{y}}
			-
			\frac{\Delta_{x}}{f_{x}}
			- \frac{\Delta_{y}}{f_{y}}
			\Bigg)
			+
			\epsilon
			\Bigg(\frac{\Delta^{*}_{x,y}}{f_{x}f_{y}}
			-
			\frac{\Delta^{*}_{x}}{f_{x}}
			- \frac{\Delta^{*}_{y}}{f_{y}}
			\Bigg)\Bigg\}^{2}\phi''\Big(f^{k}_{x}f^{k}_{y}\Big)dy.
		\end{align}
		It is known from Theorem \ref{theorem: asymptotic dist under contaminated contiguous} that
		\begin{gather}
			\Bigg|\mathds{P}_{f^{(P_{n})}_{X, Y}}\Big\{T_{D^{(k)}_{\phi}}  - t_{n}> \tau_{c} -r_{n} \Big\}
			- (1- \Phi_{1}(\tau_{c}-r_{n}))\Bigg|
			\longrightarrow 0
			\text{ when } n \to \infty. 
		\end{gather}
		A first-order Taylor series expansion gives 
		\begin{align}
			\Phi_{1}(\tau_{c}-r_{n})
			=	\Phi_{1}(\tau_{c}-r_{*}) + (r_{*}-r_{n})
			\phi_{1}(\tau_{c}-r'_{*}) 
		\end{align}
		where $r^{'}_{*}$ is an intermediate point between $r_{n}, r_{*}$. Since $\phi_{1}$ is bounded, 
		\begin{gather}
			\mathds{P}_{f^{(P_{n})}_{X, Y}}\Big\{T_{D^{(k)}_{\phi}}  - r_{n}> \tau_{c} -r_{n} \Big\}
			\longrightarrow 1- \Phi_{1}(\tau_{c}-r_{*}) 
			\text{ when } n \to \infty. 
		\end{gather}
		The assumption of uniformly convergence of $\Big\{ \frac{\partial}{\partial \epsilon } \pi(f^{(P_{n})}_{X,Y}, t_{1})\Big\}$ implies that differentiation and limit can be interchanged. This, in turn, gives the power influence function as 
		\begin{align}
			\mathcal{PIF}(\pi, t_{1} )=
			\Bigg[\frac{\partial }{\partial \epsilon}
			\Bigg\{\underset{n \longrightarrow \infty}{\lim} \pi(f^{(P_{n})}_{X,Y}, t_{1})\Bigg\}
			\Bigg]_{\epsilon=0}
			=\Bigg[\frac{\partial r_{*}}{\partial \epsilon}
			\cdot \phi_{1}(\tau_{c}-r_{*})\Bigg]_{\epsilon=0}.
		\end{align}	
		When $\epsilon=0$, we find that 
		$r_{*}=\frac{d^{2}}{2\sigma_{\phi}}
		\mathcal{IF}_{2}(I_{D^{(k)}_{\phi}}, f_{X}f_{Y}, t_{1})$  and  
		\begin{align}
			\Bigg[\frac{\partial r_{*}}{\partial \epsilon}\Bigg]_{\epsilon=0}
			=\frac{d}{\sigma_{\phi}}
			\sum_{x=0}^{1} \int k^{2}(f_{x}f_{y})^{2k}
			\Bigg(\frac{\Delta^{*}_{x,y}}{f_{x}f_{y}}
			-
			\frac{\Delta^{*}_{x}}{f_{x}}
			- \frac{\Delta^{*}_{y}}{f_{y}}
			\Bigg)
			\Bigg(\frac{\Delta_{x,y}}{f_{x}f_{y}}
			-
			\frac{\Delta_{x}}{f_{x}}
			- \frac{\Delta_{y}}{f_{y}}
			\Bigg)\phi''\Big(f^{k}_{x}f^{k}_{y}\Big)dy.	
		\end{align}
		This gives the desired result. 
		
		\item[(b)] When $t_{0}=t_{1}$, we get $\Delta_{x,y}=\Delta^{*}_{x,y}$ for all $x,y$. Thus, the result follows. 
		
		\item[(c)] See that $f^{(P_{n})}_{x,y}=f^{(n)}$ for all $x,y$ when $d=0$ and $t_{0}=t_{1}$. Substituting $d=0$ and $t_{0}=t_{1}$ in the expression of $\mathcal{PIF}$ as in (b) gives that $\mathcal{LIF}(\alpha, t_{0}) \equiv 0$.  
	\end{itemize}
	This completes the proof.

\subsection*{Theorem \ref{ch6:first BP theorem}} 
    
    See that 
	\begin{align}
		I_{m}=\int\sum\Big\{\phi((f^{m}_{x,y})^{k})-
		\phi((f^{m}_{x} f^{m}_{y})^{k})
		-\Big( (f^{m}_{x,y})^{k}
		- (f^{m}_{x} f^{m}_{y})^{k}  \Big)
		\phi'((f^{m}_{x} f^{m}_{y})^{k})
		\Big\}.
	\end{align}
	Define $A_{m}=\Big\{(x,y): f_{x,y} > \max\{k_{xy,m},
	f^{m}_{x}f^{m}_{y}\}\Big\}$. The summation over "$x$" and integration over "$y$" are implicit for simplicity. Note that, as $m \to \infty$, 
	\begin{align}
		\int \sum_{A_{m}}k_{xy, m}
		= \int \sum_{A_{m}} \min\{f_{x,y}, k_{xy,m} \}
		& \le \int \sum \min\{f_{x,y}, k_{xy,m} \} 
		\nonumber \\
		&= \int \sum \min\{f_{x}f_{y|x}, k_{x,m}k_{y|x,m} \}  
		\nonumber \\
		&\le \sum \max\{f_{x},k_{x,m} \}\int \min\{f_{y|x}, k_{y|x, m}\}
		\to 0  
	\end{align}
	by Assumption \descref{(BP1)} as $f_{x}, k_{x,m}$ are discrete and bounded in $[0,1]$. So we get $k_{xy,m} \to 0$, and subsequently $f^{m}_{x,y} \to (1-\epsilon)f_{x,y}$ on $A_{m}$ as $m \to \infty$. Similarly, we also have  
	\begin{align}
		\int \sum_{A_{m}}f^{m}_{x}f^{m}_{y}
		=\int \sum_{A_{m}} \min\{f_{x,y}, f^{m}_{x}f^{m}_{y} \}
		&\le \int \sum_{A_{m}} \max\{f_{x}, f^{m}_{x}\} 
		\min\{f_{y|x}, f^{m}_{y} \}
		\nonumber \\
		&\le \sum \max\{f_{x}, f^{m}_{x}\} 
		\int \min\{f_{y|x}, f^{m}_{y} \} 
		\to 0
		\mbox{ as } m \to \infty,
	\end{align}
	by Assumption \descref{(BP2)}. Thus we get $f^{m}_{x}f^{m}_{y} \to 0$ on $A_{m}$ for $m \to \infty$. Now see that 
	\begin{align}
		\max\{k_{xy,m}, f^{m}_{x}f^{m}_{y}\}
		& \ge \min\{k_{x,m}, f^{m}_{x}\} \min\{k_{y|x, m}, f^{m}_{y}\}
		\nonumber \\
		&\ge \min\{k_{x,m}, f^{m}_{x}\}
		\min\{k_{y|x, m}, f_{y|x}, f^{m}_{y}\} \nonumber \\
		&= \min\{k_{x,m}, f^{m}_{x}\}
		\min\Big\{ \min\{k_{y|x, m}, f_{y|x}\},
		\min\{ f_{y|x}, f^{m}_{y}\}  \Big\}.
	\end{align}
	The last factor tends to $0$ by Assumptions \descref{(BP1)} and \descref{(BP2)} when $m \to \infty$. This implies that $A_{m} \to \{(x,y) : f_{x,y} > 0\}$. 
	Therefore we get  
	\begin{gather}
		\Bigg|\int\sum_{A_{m}}\Big\{\phi((f^{m}_{x,y})^{k})-
		\phi((f^{m}_{x} f^{m}_{y})^{k})
		-\Big( (f^{m}_{x,y})^{k}
		- (f^{m}_{x} f^{m}_{y})^{k}  \Big)
		\phi'((f^{m}_{x} f^{m}_{y})^{k})
		\Big\}
		\nonumber \\
		-\int\sum\Big\{\phi \Big((1-\epsilon)^{k}f^{k}_{x,y}\Big)-
		\phi(0)
		-\Big( (1-\epsilon)^{k}f^{k}_{x,y} \Big) \phi'(0)
		\Big\} \Bigg| 
		\nonumber \\
		\label{ch6: A_m}
		=\Bigg|\int\sum_{A_{m}}\Big\{\phi((f^{m}_{x,y})^{k})-
		\phi((f^{m}_{x} f^{m}_{y})^{k})
		-\Big( (f^{m}_{x,y})^{k}
		- (f^{m}_{x} f^{m}_{y})^{k}  \Big)
		\phi'((f^{m}_{x} f^{m}_{y})^{k})
		\Big\}- D^{(k)}_{\phi}\Big((1-\epsilon)f_{X,Y}, 0\Big)\Bigg|
		\to 0
	\end{gather}
	as $m \to \infty$, because $\phi(0), \phi'(0)$ are assumed to be finite in Assumption \descref{(BP4)}. Next, see that 
	\begin{align}
		\int \sum_{A^{c}_{m}}f_{x,y}  
		&\le \int \sum \min\{ f_{x,y}, k_{xy,m}\} +
		\int \sum \min\{ f_{x,y},f^{m}_{x}f^{m}_{y}\}
		\nonumber \\
		&\le  \sum \max\{f_{x}, k_{x,m}\}  \int \min\{ f_{y|x}, k_{y|x,m}\} 
		\nonumber \\
		&+\sum  \max\{f_{x}, f^{m}_{x}\} \int \min\{ f_{y|x},f^{m}_{y}\}.
	\end{align}
	The first factor with each integral is bounded. So Assumptions \descref{(BP1)} and \descref{(BP2)} together imply that the above integration goes to $0$ as $m \to \infty$. So $A^{c}_{m}$ is asymptotically a null set under $f_{X,Y}$. Therefore we get 
	\begin{align}
		&\Bigg|\int\sum_{A^{c}_{m}}\Big\{\phi((f^{m}_{x,y})^{k})-
		\phi((f^{m}_{x} f^{m}_{y})^{k})
		-\Big( (f^{m}_{x,y})^{k}
		- (f^{m}_{x} f^{m}_{y})^{k}  \Big)
		\phi'((f^{m}_{x} f^{m}_{y})^{k})
		\Big\}
		\nonumber \\
		\label{ch6: A_m complement}
		&-\underbrace{\sum \int \Big\{\phi( (\epsilon k_{xy,m})^{k} )-
			\phi((f^{m}_{x} f^{m}_{y})^{k})
			-\Big( (\epsilon k_{xy,m})^{k}
			- (f^{m}_{x} f^{m}_{y})^{k}  \Big)
			\phi'((f^{m}_{x} f^{m}_{y})^{k})}_{D^{(k)}_{\phi}(\epsilon k_{XY,m}, f^{m}_{X}f^{m}_{Y})  }
		\Big\}
		\Bigg| \to 0
	\end{align}
	for $m \to \infty$. Combining (\ref{ch6: A_m}) and (\ref{ch6: A_m complement}) gives  
	\begin{align}
		\Bigg|I_{m}- 
		D^{(k)}_{\phi}\Big((1-\epsilon)f_{X,Y}, 0\Big)
		-D^{(k)}_{\phi}\Big(\epsilon k_{XY,m}, f^{m}_{X}f^{m}_{Y} \Big) \Bigg| \to 0
		\mbox{ as }
		m \to \infty. 
	\end{align}
	Define  
	\begin{align}
		a_{1}(\epsilon)
		=D^{(k)}_{\phi}\Big((1-\epsilon)f_{X,Y}, 0\Big)+
		\underset{m \to \infty}{\liminf}
		D^{(k)}_{\phi}\Big(\epsilon k_{XY,m}, f^{m}_{X}f^{m}_{Y}\Big).
	\end{align}
	Next, see that 
	\begin{align}
		\underset{m \to \infty}{\lim \inf} I_{m}
		&=\underset{m \to \infty}{\lim \inf} \Bigg[
		D^{(k)}_{\phi}\Big((1-\epsilon)f_{X,Y}, 0\Big)
		+D^{(k)}_{\phi}\Big(\epsilon k_{XY,m}, f^{m}_{X}f^{m}_{Y}\Big)
		\Bigg] \nonumber \\
		&\ge 
		D^{(k)}_{\phi}((1-\epsilon)f_{X,Y}, 0)+
		\underset{m \to \infty}{\lim \inf} 
		D^{(k)}_{\phi}(\epsilon k_{XY,m}, f^{m}_{X}f^{m}_{Y})  
		=a_{1}(\epsilon).
	\end{align}
	Also, see that $\underset{m \to \infty}{\limsup} I_{m} \le \underset{m \to \infty}{\limsup} I'_{m}$ as $I_{m} \le I'_{m}$ for all $m\ge 1$ by Assumption \descref{(BP5)}. Consider 
	\begin{align}
		B_{m}=\Big\{ (x,y): k_{xy,m} > \max\{f_{x,y}, f_{x}f_{y}\}\Big\}.
	\end{align}
	Similarly as before, $B_{m}$ is asymptotically null set under both $f_{X,Y}$ and $f_{X}f_{Y}$. Also note that $B^{c}_{m}$ is asymptotically null under $k_{XY,m}$. Using the same argument as before, we obtain 
	\begin{align}
		\Bigg|I'_{m} - D^{(k)}_{\phi}(\epsilon k_{XY,m}, 0)
		-D^{(k)}_{\phi}\Big((1-\epsilon)f_{X,Y}, f_{X}f_{Y}\Big)\Bigg|
		\to 0
		\mbox{ for } m \to \infty.
	\end{align}
	Next, see that  
	\begin{align}
		\underset{m \to \infty}{\limsup}I_{m}
		\le 
		\underset{m \to \infty}{\limsup}I'_{m}
		&=\underset{m \to \infty}{\limsup}\Bigg[D^{(k)}_{\phi}(\epsilon k_{XY,m}, 0)
		+D^{(k)}_{\phi}\Big((1-\epsilon)f_{X,Y}, f_{X}f_{Y}\Big) \Bigg]
		\nonumber \\
		&\le \underset{m \to \infty}{\limsup}
		D^{(k)}_{\phi}(\epsilon k_{XY,m}, 0)
		+D^{(k)}_{\phi}\Big((1-\epsilon)f_{X,Y}, f_{X}f_{Y}\Big)
		=a_{2}(\epsilon) \mbox{ (say)}.
	\end{align}
	
	Suppose $a_{2}(\epsilon) < a_{1}(\epsilon)$, then $\underset{m \to \infty}{\limsup}  I_{m} < \underset{m \to \infty}{\liminf} I_{m}$. So $\underset{m \to \infty}{\lim} I_{m}$ exists, i.e., asymptotically there will be no breakdown, when $a_{2}(\epsilon) < a_{1}(\epsilon)$, which is the same as 
	\begin{gather}
		D^{(k)}_{\phi}(\epsilon k_{XY,m}, 0)+D^{(k)}_{\phi}\Big((1-\epsilon)f_{X,Y}, f_{X}f_{Y}\Big) 
		< D^{(k)}_{\phi}(\epsilon k_{XY,m}, f^{m}_{X}f^{m}_{Y})
		+D^{(k)}_{\phi}\Big((1-\epsilon)f_{X,Y}, 0\Big)
	\end{gather} 
	for sufficiently large $m$. This condition is equivalent to 
	\begin{align}
		\label{ch6:BP: last eq}
		\underset{m \to \infty}{\liminf}D^{(k)}_{\phi}(\epsilon k_{XY,m}, f^{m}_{X}f^{m}_{Y}) 
		&> \underset{m \to \infty}{\limsup} D^{(k)}_{\phi}(\epsilon k_{XY,m}, 0)
		+ D^{(k)}_{\phi}\Big((1-\epsilon)f_{X,Y}, f_{X}f_{Y}\Big) -D^{(k)}_{\phi}\Big((1-\epsilon)f_{X,Y}, 0 \Big)
		\nonumber \\
		&=\underset{m \to \infty}{\limsup}\sum \int \Bigg[\phi( (\epsilon k_{xy,m})^{k}) -(\epsilon k_{xy,m})^{k}\phi'(0)\Bigg] \nonumber \\
		&+ \sum \int((1-\epsilon)f_{xy})^{k}\phi'(0)
		-\phi(f^{k}_{x}f^{k}_{y}) \nonumber \\
		&-\Big\{ ((1-\epsilon)f_{xy})^{k}-(f_{x}f_{y})^{k} \Big\}\phi'(f^{k}_{x}f^{k}_{y})\Bigg] dy.
	\end{align}
	Clearly, Assumption \descref{(BP6)} ensures that the (\ref{ch6:BP: last eq}) holds for $\epsilon < \tilde{\epsilon}$. This completes the proof.

\subsection*{Lemma \ref{ch6:lemma1: Breakdown point}} 
    
    Let us assume $M_{f,f} \ge M_{g,f}$. See that 
	\begin{align}
		&D^{(k)}_{\phi}(\epsilon g, f)-
		D^{(k)}_{\phi}(\epsilon g, g)
		\nonumber \\
		&=\sum \int\Big[ \phi((\epsilon g)^{k})-\phi(f^{k})-\Big\{(\epsilon g)^{k}-f^{k}\Big\} \phi'(f^{k})
		-
		\phi((\epsilon g)^{k})+\phi(g^{k})+\Big\{(\epsilon g)^{k}-g^{k}\Big\} \phi'(g^{k})\Big]
		\nonumber \\
		&= -\sum\int\phi(f^{k})-\epsilon^{k}M_{g,f}+M_{f,f}+\sum \int\phi(g^{k})
		+(\epsilon^{k}-1)M_{g,g}
		\nonumber \\
		&\ge  \sum\int [\phi(g^{k})-\phi(f^{k})]-\epsilon^{k}M_{f,f}+M_{f,f}
		+(\epsilon^{k}-1)M_{g,g}
		\nonumber \\
		&=\sum\int [\phi(g^{k})-\phi(f^{k})]-(\epsilon^{k}-1)M_{f,f}
		+(\epsilon^{k}-1)M_{g,g}
		\nonumber \\
		&=\sum\int [\phi(g^{k})-\phi(f^{k})]-(\epsilon^{k}-1)(M_{f,f}
		-M_{g,g}) \ge 0
	\end{align}
	when 
	\begin{align}
		\epsilon^{k} \le 
		1+\frac{\sum\int [\phi(g^{k})-\phi(f^{k})]}{(M_{f,f} - M_{g,g})}
		\mbox{ for } g \ne f.
	\end{align}
	The reverse inequality can be similarly proved when $M_{f,f} \le M_{g,f}$. This completes the proof. 

\subsection*{Theorem \ref{ch6:second BP theorem}}
    
    	An application of Lemma \ref{ch6:lemma1: Breakdown point} in combination of Assumption \descref{(BP7)} gives
	\begin{align}
		\label{ch6: BP th2 eq1}		
		D^{(k)}_{\phi}(\epsilon k_{m}, f^{m}_{X}f^{m}_{Y})  
		\ge D^{(k)}_{\phi}(\epsilon k_{m}, k_{m}) 
		\mbox{ for }
		\epsilon \le \Bigg[1+\frac{\sum\int \big[\phi[(k_{xy,m})^{k}\big]-\phi\big[(f^{m}_{x}f^{m}_{y})^{k}\Big]}
		{(M_{f^{m},f^{m}} - M_{k_{m},k_{m}})}\Bigg]^{1/k}
		\le \epsilon_{1}.
	\end{align}
	Similarly, we also find that  
	\begin{gather}
		\label{ch6: BP th2 eq2}		
		D^{(k)}_{\phi}((1-\epsilon) f_{X, Y}, f_{X,Y})
		\ge 
		D^{(k)}_{\phi}((1-\epsilon) f_{X,Y}, f_{X}f_{Y})
		\text{ for }
		\epsilon \le  1- \Bigg[1+\frac{\sum \int [\phi(f^{k}_{xy})- \phi(f^{k}_{x}f^{k}_{y})] }{M_{f_{X}f_{Y}, f_{X}f_{Y}} - M_{f_{X,Y}, f_{X,Y}}}\Bigg]^{1/k}=\epsilon_{2}.
	\end{gather}
	Combining these results, we further obtain that, for sufficiently large $m$,
	\begin{align}
		D^{(k)}_{\phi}(\epsilon k_{m}, f^{m}_{X}f^{m}_{Y})
		&\ge D^{(k)}_{\phi}(\epsilon k_{m}, k_{m})
		\mbox{ \Big[ by (\ref{ch6: BP th2 eq1}) \Big]}
		\nonumber\\
		&> D^{(k)}_{\phi}(\epsilon k_{m}, 0)
		+D^{(k)}_{\phi}((1-\epsilon)f_{X,Y}, f_{X,Y})
		-D^{(k)}_{\phi}((1-\epsilon)f_{X,Y}, 0)
		\nonumber \\
		& \text{ \Big[ by Assumption \descref{(BP8)} and then (\ref{ch6: BP th2 eq2}) \Big] } \nonumber\\
		\label{ch6: BP th2 eq3}
		&\ge D^{(k)}_{\phi}(\epsilon k_{m}, 0)
		+D^{(k)}_{\phi}((1-\epsilon)f_{X,Y}, f_{X}f_{Y})
		-D^{(k)}_{\phi}((1-\epsilon)f_{X,Y}, 0)
	\end{align}
	when $\epsilon \le \min\big\{\epsilon_{1}, \epsilon_{2}, \frac{1}{2}\big\}$. It is also clear from (\ref{ch6: BP th2 eq3}) that the Assumption \descref{(BP6)} is also satisfied along the way under such a choice of $\epsilon$. So the Assumptions \descref{(BP7)} and \descref{(BP8)} together work as a sufficient condition for Assumption \descref{(BP6)}, and therefore Theorem \ref{ch6:first BP theorem} is applicable. 
	Hence, the asymptotic breakdown point is atleast $\min\big\{\epsilon_{1}, \epsilon_{2}, \frac{1}{2}\big\}$.

\end{document}